\documentclass[%
preprint,
amsmath,amssymb,
aps,
prb,
floatfix,
]{revtex4-2}

\usepackage[T1]{fontenc}
\usepackage[utf8]{inputenc}

\usepackage{siunitx}
\usepackage{amsmath} 
\usepackage{bm}
\usepackage{mathtools}

\usepackage{xcolor}
\usepackage[pdfencoding=auto]{hyperref}
\hypersetup{colorlinks=true, citecolor=blue, urlcolor=blue, linkcolor=black}

\newcommand{\subfigref}[2][Fig.~]{#1\ref{#2}}

\newcommand{\figref}[2][Fig.~]{#1\ref{#2}}

\newcommand{\tabref}[2][Table~]{#1\ref{#2}}
\newcommand{\secref}[2][Sec.~]{#1\ref{#2}}
\newcommand{\suppfigref}[2][Fig.~]{#1\ref{#2}}
\newcommand{\supptabref}[2][Table~]{#1\ref{#2}}
\newcommand{\suppsecref}[2][Sec.~]{#1\ref{#2}}
\renewcommand{\eqref}[2][Eq.~]{#1(\ref{#2})}

\DeclareSIUnit\angstrom{\protect\text{Å}}

\usepackage{graphicx} 
\graphicspath{ {./}{DGraphics_main/}{DGraphics_supp/} }
\usepackage{caption}
\usepackage{subfig}
\usepackage{wrapfig}
\usepackage[export]{adjustbox}

\usepackage{relsize}
\usepackage{multirow}
\usepackage{dcolumn}
\usepackage{array}
\renewcommand{\arraystretch}{1.2}

\begin{document}

\title{High-efficiency solid-state electrochemical thermal switching through phonon engineering of 2D materials} 

\author{Harry Mclean}
\author{Steven Paul Hepplestone*}
\affiliation{%
  Department of Physics and Astronomy, University of Exeter, Stocker Road, Exeter, EX4 4QL, United Kingdom\\
  \text{*Corresponding Author:}~\texttt{s.p.hepplestone@exeter.ac.uk}%
}
\date{\today}

\begin{abstract}
Reversible solid-state thermal switches with large thermal contrasts remain rare, yet are essential for advanced thermal management. Using universal machine-learned interatomic potentials (uMLiPs) and density functional theory (DFT), we screen 57 transition metal dichalcogenides (TMDCs), spanning the 1T and 2H phases with Li and Mg intercalation. Among the candidates, Li$_a$ZrS$_2$ combines exceptional thermal conductivity switching with low heat capacity. Lithiation to Li$_{0.125}$ZrS$_2$ reduces the cross-plane thermal conductivity from approximately 3.7 to 0.05~W~m$^{-1}$~K$^{-1}$. The switching behaviour exceeds that of previously reported solid-state electrochemical thermal switches. Device-level modelling further demonstrates the potential for waste-heat recovery, with a pulsed system capturing 86\% of the energy recovered by an ideal switch. We attribute the switching behaviour to substantial softening of the Zr--Zr framework, with the mean force constant decreasing from approximately 4.2 to 1.2~eV~\AA$^{-2}$, alongside enhanced scattering of heat-carrying acoustic phonons. These findings establish electrochemical intercalation as a promising route to engineering thermal transport in layered materials.
\end{abstract}

\maketitle

\section{Introduction}

Heat is responsible for approximately 90\% of the world's energy generation~\cite{Henry2020}. Conversely, roughly 60\% of global primary energy consumption is lost as waste heat~\cite{Forman2016,Jia2024}. How heat transport is exploited is therefore important across a wide range of technologies and environments~\cite{Kumar2020,Liu2025_2,Huang2026,Chen2025,Cao2022,Sklan2015,Utaka2019,Klinar2020}. Critical to this is the development of materials and structures that manipulate thermal properties, from thermal metamaterials that enable heat cloaking~\cite{Fan2024,Mclean2025} and thermomechanically coupled lattice designs~\cite{Xiang2026} to architectures that spatially redistribute thermal currents~\cite{Yang2021,ThermalMetaPaper,ThermalMetaJAP,Mclean2025,Mclean2027}.
One category of high interest is thermal devices that seek to control heat flow in ways analogous to how electronic components control the flow of charge~\cite{Wang2007}. These consist of: thermal switches~\cite{Utaka2019,Cao2022,Kimling2015,Wehmeyer2017,Swoboda2020,Li2023}, thermal diodes~\cite{Li2004,Li2015,Wehmeyer2017,Swoboda2020}, thermal transistors~\cite{ChungLo2008,Swoboda2020,Yang2023,Sood2018}, and thermal logic gates~\cite{Wang2007}. For active control of the heat flux, the thermal switch is critical, providing an electrically controlled route to manipulate the flow of heat between components.

Thermal switches work by modulating heat flow reversibly between distinct on- and off-states by changing the thermal conductivity ($\kappa$) of the active material. Such controllable switching is of interest for applications including thermal management in space~\cite{Cao2022,Hengeveld2010,Beasley2004}, waste-heat recovery~\cite{McKay2013}, and thermal computing~\cite{Sklan2015,Li2012}. Thermal switches can be actuated through a broad range of physical mechanisms~\cite{Chen2025_2}, phase transitions~\cite{Wang2017}, magnetic~\cite{Zhao2016,Kimling2015} and electric fields~\cite{Cao2022,Liu2018,Li2023}, intercalation~\cite{Cho2014}, chemical reactions and changes in the surrounding environment~\cite{Bywaters1973}. Magnetic-field switches, for example, exploit field-dependent changes in thermal conductivity~\cite{Zhao2016,Kimling2015}, while phase-transition switches rely on the change in thermal transport that accompanies a structural or electronic transition~\cite{Wang2015}. Other approaches exploit atom insertion into a host structure~\cite{Cho2014}, changes in oxygen stoichiometry~\cite{Jeong2025}, liquid motion or redistribution~\cite{Cho2008}, electric-field control of ferroelectric materials~\cite{Cao2022,Liu2018,Li2023} or variation of gas pressure across a thermal gap~\cite{Bywaters1973}.  Key to all these approaches is the magnitude and range of temperatures for thermal conductivity modulation.

Solid-state electrochemical thermal switches, which exploit intercalation or oxidation~\cite{Bian2024}, are particularly attractive as the insertion and extraction of ions modifies the thermal conductivity of the bulk active material itself, rather than a contact or a gap. Such devices operate at modest voltages, can be readily miniaturised, and require neither a narrow temperature window nor large changes in external pressure. Despite these advantages, the thermal-conductivity modulation achieved in demonstrated electrochemical switches remains limited. LaNiO$_3$-based solid-state devices, for example, have been reported with $\Delta\kappa \approx 4.3$~W~m$^{-1}$~K$^{-1}$ and $R \approx 3.5$ ($S = 0.72$)~\cite{Bian2024}, while CeO$_2$-based switches reach $R \approx 5.8$~\cite{Jeong2025}. Thermal switching behaviour is also demonstrated in battery electrodes, such as Graphite~\cite{Qian2016}, MoS$_2$\cite{MoS2Kinter} and others~\cite{McleanArxiv}.

Layered van der Waals crystals are a natural platform for electrochemical thermal switching and are ubiquitous in battery electrode materials. Their thermal transport is intrinsically anisotropic, and the cross-plane direction, which is the one that must be modulated in a thin-film or heterostructure device, possesses weak interlayer bonding that intercalation perturbs directly. Graphite~\cite{Qian2016} and MoS$_2$~\cite{MoS2Kinter} exhibit large and reversible changes in thermal conductivity upon electrochemical Li insertion, and Li intercalation in transition-metal dichalcogenides (TMDCs) has been used to realise an electrochemical thermal transistor~\cite{Sood2018}. In MoS$_2$, the reported modulation has been attributed to a combination of Li-induced rattler modes, $c$-axis strain and stacking disorder~\cite{MoS2Kinter}. Establishing the microscopic origin of such changes nevertheless remains difficult, requiring detailed atomic-scale studies. However, what is required is a high-throughput multi-scale study of the many different 2D materials and their potential for thermal switching.
This is challenging, however, and requires the careful selection of a representative material because of several factors.  These include: intercalation hosts require prohibitively large supercells because they either exhibit chemical disorder or non-uniform intercalation; intercalation resulting in phase changes or chemical conversion of the host; Li insertion modifies the anharmonic phonon scattering processes at each stage of intercalation; high system anisotropy; and other factors. These factors prevent the isolation of the key driving mechanisms that govern lattice thermal transport in thermal switching devices.

Here we screen 57 TMDC MX$_2$ materials spanning the octahedrally coordinated (1T) and trigonal-prismatic (2H) phases with Li/Mg intercalation to identify the optimal thermal switching material (where M is the metal, and X is Sulphur or Selenium). We combine \textit{ab initio} density functional theory (DFT) with the universal machine-learned interatomic potential (uMLiP) to calculate phonon spectra, phonon lifetimes and lattice thermal conductivity by finite-displacement methods~\cite{Phonopy,Phono3py}. 
To perform a rapid screening, we employed the NequIP-OAM-L uMLiP~\cite{NEQUIP} to assess phonon stability and thermal conductivity of 57 sulphur- and selenium-based TMDCs. Our initial screening criteria are based on phonon calculations for both the bulk structure and the first stages of intercalation (12.5\% filling). If either structure produced imaginary phonon frequencies with the NequIP uMLiP, the result was automatically discarded (see Data Repository~\cite{Figshare}), reducing our search to twenty-seven dynamically stable systems. The thermal conductivity, both cross-plane ($\kappa_\perp$) and in-plane ($\kappa_\parallel$), is evaluated using finite-displacement methods~\cite{Phonopy,Phono3py}.
From this set, we select the best switching material, Li$_a$ZrS$_2$, to be studied in detail with first-principles DFT. We then follow Li$_a$ZrS$_2$ across the full intercalation range with both methods, resolving how phonon frequencies, group velocities and scattering rates produce the switching; and finally we propagate the resulting on/off conductivities through a pulsed waste-heat-recovery device to obtain a device-level metric. The outcome is a microscopic connection between electrochemical Li intercalation, phonon scattering and thermal-conductivity switching in a layered vdW host, together with the design rules that the screen extracts for solid-state electrochemical thermal switches.
To assess the performance of these structures, three complementary metrics are used throughout this work: the switching width $\Delta\kappa = \kappa^{\mathrm{on}}-\kappa^{\mathrm{off}}$, which measures the absolute change; the on/off ratio $R=\kappa^{\mathrm{on}}/\kappa^{\mathrm{off}}$, which measures the relative modulation; and the switch score $S = |\kappa^{\mathrm{on}}-\kappa^{\mathrm{off}}|/\kappa^{\mathrm{on}}$, which for any switch that lowers the conductivity on actuation equals $1-1/R$ and lies in $[0,1]$, and is therefore convenient for comparing switches whose on-state conductivities differ by orders of magnitude (a score above unity marks an inverted switch, as found for a few screened hosts below). Switching speed, cyclability and compatibility with solid-state device architectures are equally important but are not computed here, though details can be found here~\cite{Price2023,Kim2021}.

\section{Results}

\subsection{Screening of transition-metal dichalcogenides}
\label{subsec:screening}

Our screening identified twenty-seven MX$_2$ TMDC host--intercalant (Li or Mg) systems that are dynamically stable between 0 and 12.5\% filling ([Li/Mg]$_a$MX$_2$, $a=0\rightarrow0.125$). These consist of twelve Li systems and fifteen Mg systems. \subfigref{subfig:SwitchScreen} compares the cross-plane switch score $S_\perp$ at 300~K for the twelve Li systems. Eleven switch positively, i.e.\ the pristine host has the higher thermal conductivity, with $S$ spanning 0.11 (Li$_{0.125}$CrSe$_2$) to 0.96 (Li$_{0.125}$ZrS$_2$, shown in gold). Li$_{0.125}$VSe$_2$ is the exception: the intercalated system is more conductive than the pristine host, with $S =-0.34$. 
From this result, the leading Li candidates are Li$_{0.125}$ZrS$_2$ (0.96), Li$_{0.125}$TiS$_2$ (0.90) and Li$_{0.125}$VS$_2$ (0.90).  
To further narrow the selection, we apply a second criterion, heat capacity, since it is connected to how quickly the switch layer equilibrates with its surroundings.
Li$_{0.125}$ZrS$_2$ has one of the lowest mean heat capacities in the set, $c_v \approx$1.7~J~K$^{-1}$~cm$^{-3}$ averaged over its two states, behind only ZrSe$_2$ (1.58~J~K$^{-1}$~cm$^{-3}$). Using this we can obtain the state-averaged cross-plane thermal diffusivities, $\bar{\alpha}_\perp=(\alpha^{\mathrm{on}}_{\perp}+\alpha^{\mathrm{off}}_{\perp})/2$, where $\alpha_\perp=\kappa_\perp/C_v$, of Li$_a$TiS$_2$ and Li$_a$VS$_2$ are 0.55 and 0.17~mm$^2$~s$^{-1}$, compared with 1.24~mm$^2$~s$^{-1}$ for Li$_{0.125}$ZrS$_2$. Hence, a 5~$\mu$m layer has an estimated equilibration time of $\approx$45 and $\approx$145~$\mu$s for the TiS$_2$ and VS$_2$ systems, respectively, compared with $\approx$20~$\mu$s for ZrS$_2$, indicating that ZrS$_2$ is the lead candidate.

The Mg-intercalated TMDCs (\suppfigref{suppfig:DTMDCKappaMg}) show similar behaviour to the Li intercalation. The mean switch score over the systems that switch positively is $S = 0.71$, against $0.70$ for Mg and Li, respectively. However, the best Mg performer, Mg$_{0.125}$HfS$_2$ ($S = 0.83$), falls short of Li$_{0.125}$ZrS$_2$ (0.96). The set of calculated properties for all the materials studied, such as the thermal conductivities, switch scores, heat capacities, and thermal diffusivities, can be found in \suppsecref{suppsec:dtmdc}.  Five systems have negative switch scores, i.e.\ the intercalated cross-plane conductivity, $\kappa_\perp$ exceeds the bulk non-intercalated value, $\kappa_{zz}^{\mathrm{off}}>\kappa_{zz}^{\mathrm{on}}$. Together with Li$_{0.125}$VSe$_2$, the four inverted Mg systems are Mg$_{0.125}$VSe$_2$, Mg$_{0.125}$CrS$_2$, Mg$_{0.125}$CrSe$_2$ and Mg$_{0.125}$WS$_2$. Their switching widths are 0.07, 0.44, 0.09, 0.03, and 0.23~W~m$^{-1}$~K$^{-1}$, and their scores are $S = -0.34$, -2.15, -0.10, -0.06, and -0.26, respectively. These systems are presented in \subfigref{subfig:SwitchScreen} and \suppfigref{suppfig:DTMDCKappaMg} (for Li and Mg, respectively). All the inverted systems are 2H phase and, in general, make poor thermal switches because they have low conductivities (mean $\kappa_{\perp}\approx0.9$ vs $\approx2.6$~W~m$^{-1}$~K$^{-1}$ for 1T), rising a little upon intercalation, resulting in a much lower mean switch score (0.79 for 1T systems against 0.55 for 2H). It is important to note that in these negative switch score systems, the ratio is highly influenced by the low conductivity results, meaning that inaccuracies in the model could result in changes in the switching score. Our search through the total set of TMDC structures has suggested Li$_{0.125}$ZrS$_2$ is the choice material to explore further, with a high switch score of 0.96. 


\subsection{Electrochemistry: voltage and capacity}
\label{subsec:voltage}

Having selected Li$_{0.125}$ZrS$_2$ from the search, we now perform a full first-principles study, assessing its electrochemistry.  For our first-principles calculations, we use VASP with the PBE and HSE06 functionals. Firstly, we compare our modelled results for the intercalation voltage of  Li$_a$ZrS$_2$ with the experimental discharge curve of Kim \textit{et al.}~\cite{Kim2021} to validate our approach. Our HSE06 voltage profile captures the high initial discharge ($\approx$3.7~V against $\approx$3.3~V), while PBE and the uMLiP underestimate it at $\approx$2.1 and $\approx$2.0~V (\subfigref{subfig:VoltageProfile}). This is consistent with the well-documented tendency of PBE to underestimate intercalation voltages in transition-metal hosts through self-interaction error, which the hybrid functional largely corrects~\cite{Chevrier2010}. Regardless of functional, however, our results indicate a relatively uniform filling for these systems at low filling factors, rather than a clustering of the Li at each stage of filling. Across the operating plateau ($0.25 \lesssim a \lesssim 1$) the measured voltage is $\approx$1.6--1.8~V: HSE06 overestimates it by $\approx$0.4--0.6~V and PBE by $\approx$0.2~V, while the uMLiP underestimates it by $\approx$0.1~V. No single functional is accurate over the whole curve, but all three place the first Li insertion, which is the switching step used here, within 1--2~V of the measured onset. The switch can be driven at voltages compatible with a solid electrolyte. For both PBE and the uMLiP the convex-hull and sequential profiles are nearly coincident (\suppfigref{suppfig:SequenceVoltage}), an expected consequence of the functional used in the training dataset.  

Our results indicate that the filling behaviour at higher levels of intercalation is highly dependent on the choice of functional.  For each DFT functional (PBE~\cite{Perdew1996} and HSE06~\cite{Heyd2003,Heyd2006}) and for the uMLiP~\cite{NEQUIP}, the convex-hull profile can be found in \suppfigref{suppfig:SequenceVoltage}, which connects only the chemically stable states. With PBE the intermediate compositions $a = 0.375$, $0.625$ and $0.875$ lie 0.8, 2.1 and 2.4~meV per formula unit (f.u.) above the hull, and the uMLiP places the same states 0.6, 2.6 and 2.2~meV per f.u.\ above it. Both profiles comprise a staircase of short plateaus that broadly follows the sloping experimental discharge. With HSE06 only $a = 0$, $0.125$, $0.25$ and $1$ are hull-stable, the remaining states lying $\approx$22--61~meV per f.u.\ above the hull. Each intermediate composition is represented here by a single Li ordering in the $2\times2\times1$ supercell of the unit cell of the pristine host, and the true ground-state ordering, which is not exhaustively sampled, could bring some of these states onto the hull. The few-meV PBE and uMLiP excursions lie within the uncertainty introduced by this limited sampling and well below the thermal energy $k_BT \approx 26$~meV at 300~K, so they should not be read as evidence of phase separation. The HSE06 excursions exceed $k_BT$ but carry the same caveat. The voltage staircase is approximate, although the switching step at $a=0.125$ is hull-stable in all three descriptions. Regardless of which functional is chosen, the thermal switching performance is dominated by the early regime of filling, strongly supporting ZrS$_2$ as an ideal model platform for thermal switching.


\subsection{Thermal conductivity under intercalation}
\label{subsec:thermal_props}

\begin{figure*}
    \centering
    \subfloat[]{\includegraphics[width=0.5\linewidth]{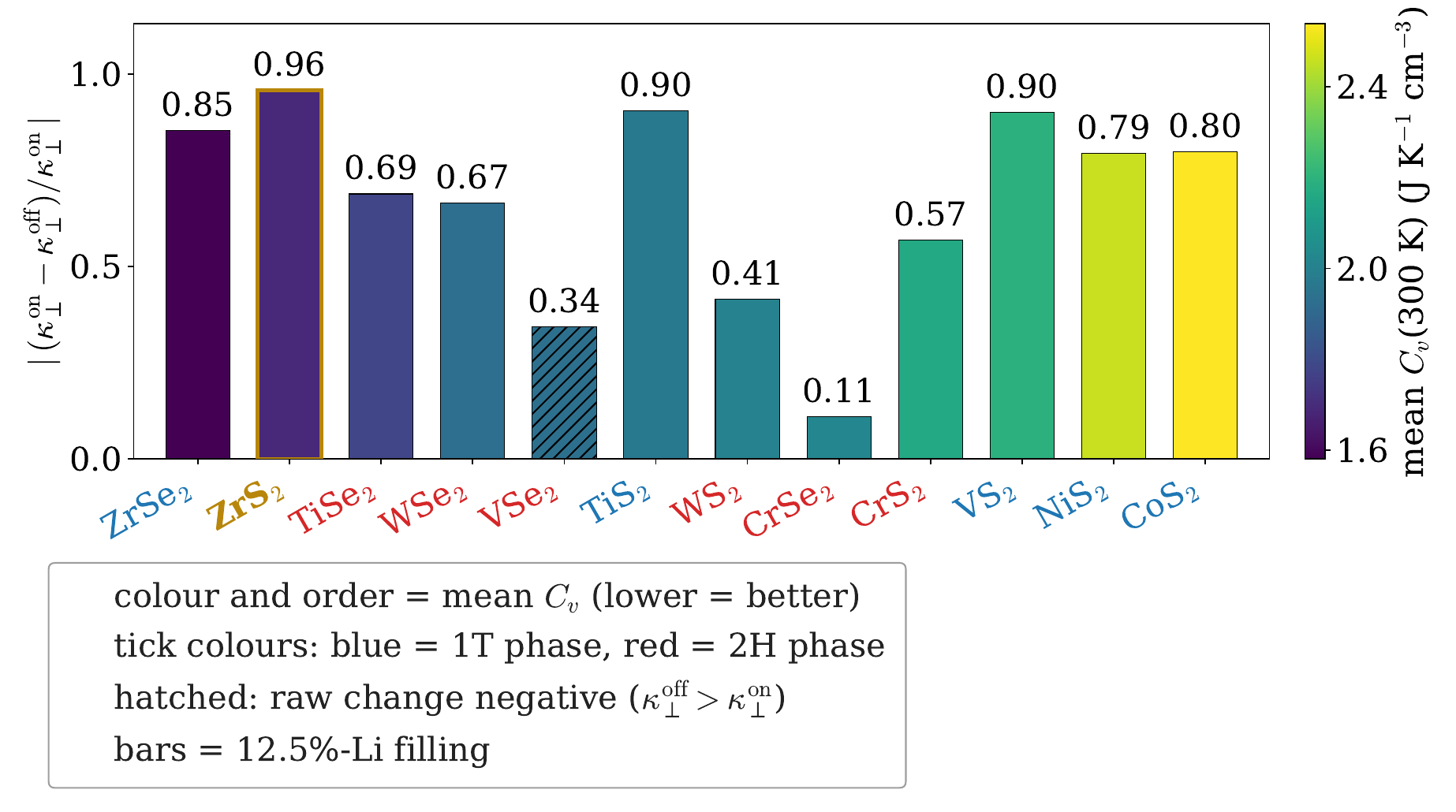}\label{subfig:SwitchScreen}}
    \subfloat[]{\includegraphics[width=0.5\linewidth]{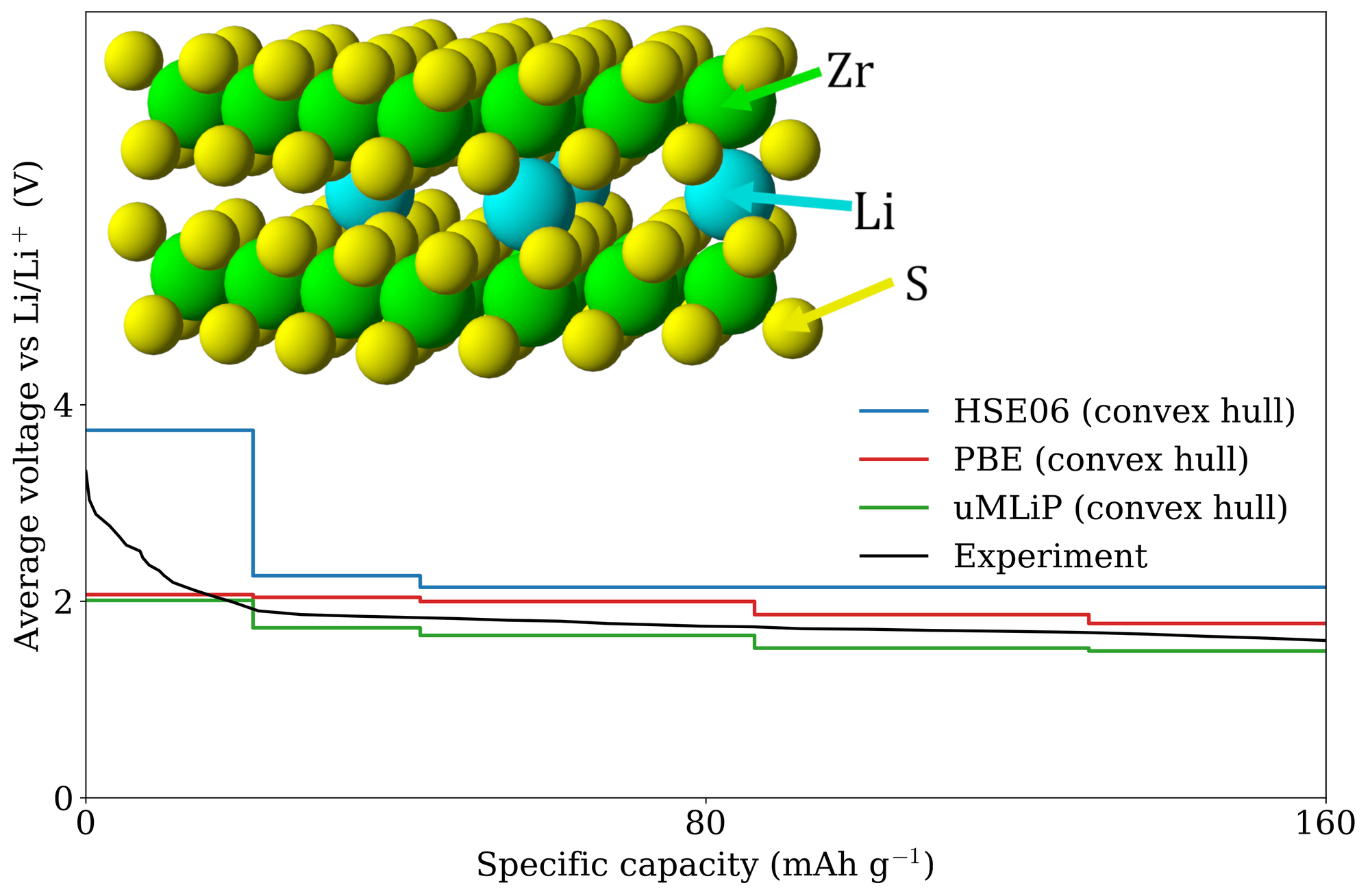}\label{subfig:VoltageProfile}}
    \hfill
    \subfloat[]{\includegraphics[width=0.5\linewidth]{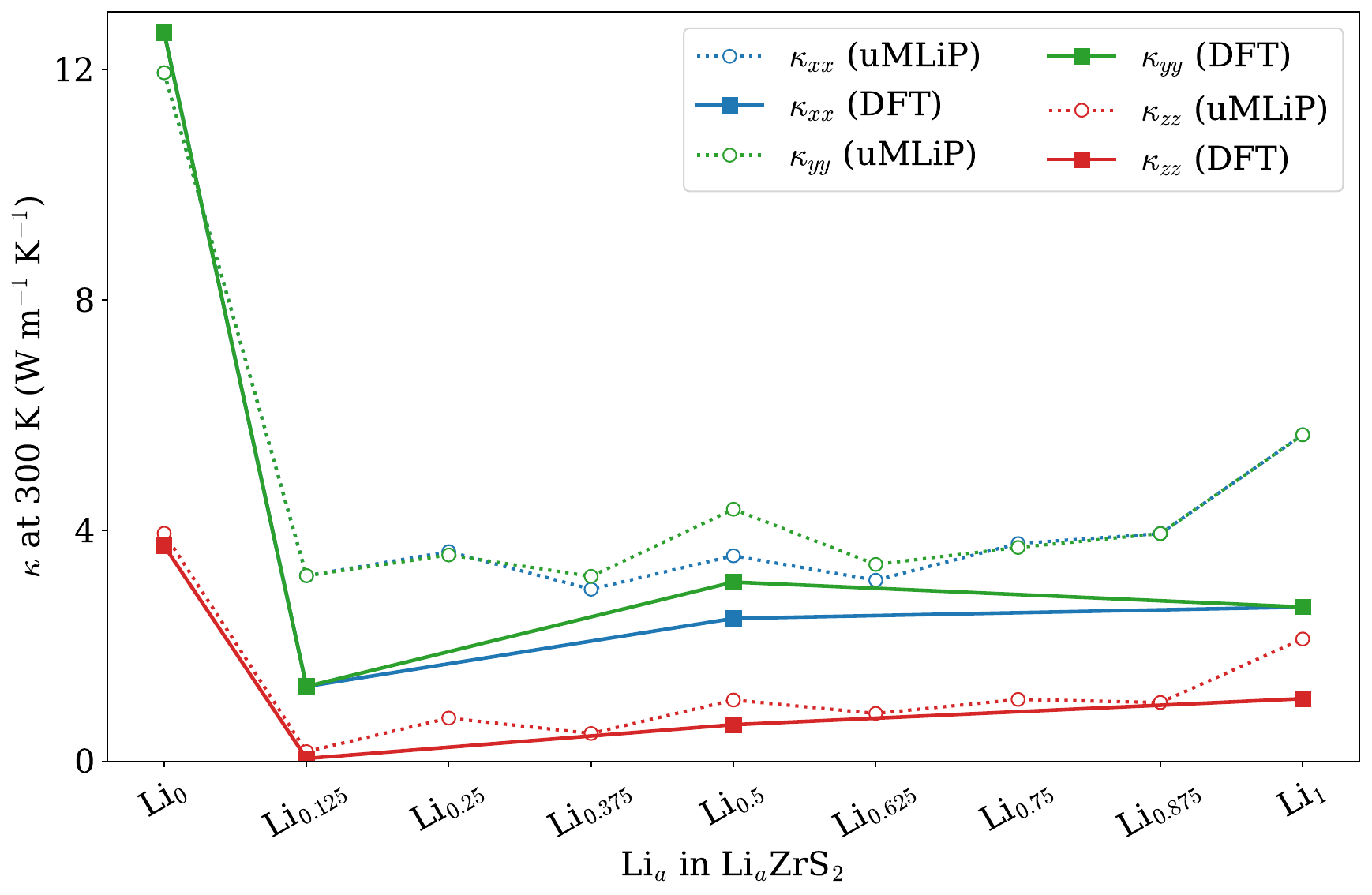}\label{subfig:ThermalConductivityVLi}}
    \caption{Screening, electrochemistry and thermal response of Li$_a$ZrS$_2$. (a) Cross-plane switch score, $S_\perp$, at 300~K for the twelve Li systems of the screen (uMLiP, BTE-RTA); bars are ordered by the volumetric heat capacity encoded by the colour scale, hatched bars mark hosts where $\kappa_{zz}=\kappa_\perp$ rises on intercalation, and blue (red) tick labels denote 1T (2H) hosts; Li$_{0.125}$ZrS$_2$ is highlighted in gold. (b) Convex-hull voltage profiles from PBE, HSE06 and the uMLiP against the experimental discharge curve~\cite{Kim2021}, with an inset of $2\times2\times1$ supercell of Li$_{0.125}$ZrS$_2$. (c) Cartesian thermal-conductivity components at 300~K versus Li content from DFT and the uMLiP.}
    \label{fig:VoltageVsCapacity}
\end{figure*}

To firmly establish Li$_a$ZrS$_2$ as a thermal switch, we evaluate how the thermal properties such as thermal conductivity (\subfigref{subfig:ThermalConductivityVLi}) and heat capacity (\suppfigref{suppfig:HeatCapacity}) evolve under intercalation, evaluating with both DFT and the uMLiP.
Our results show that the cross-plane thermal conductivity, $\kappa_\perp$ drops from $\approx3.7$ to $\approx0.05$  as we intercalate Li.  
The in-plane components of thermal conductivity $\kappa_{xx}$ and $\kappa_{yy}$ are equal in the high-symmetry compositions but split when the Li ordering breaks the in-plane three-fold symmetry. Conversely, the cross-plane component $\kappa_\perp=\kappa_{zz}$ is substantially smaller than the in-plane components and displays the largest relative drop upon initial lithiation.
DFT and the uMLiP show agreement to within $\approx$4\% for the pristine host ($a=0$) but demonstrate divergence as lithiation proceeds: the orientation-averaged conductivity, $\kappa_{ave}$ differs by $\approx$0.9~W~m$^{-1}$~K$^{-1}$ at $a=0.5$ and by $\approx$2.3~W~m$^{-1}$~K$^{-1}$ at $a=1$.
Both descriptions are evaluated on the same PBE-relaxed geometries, so this divergence reflects the force constants of the potential rather than different structures.  
The uMLiP captures the qualitative trend and the low-lithium regime but underestimates the magnitude of the suppression at intermediate and full lithiation. 
This is expected as the range of data for which the baseline uMLiP is trained upon lacks this particular region of structures, indicating that whilst they provide a good initial estimate, further data is needed to ensure accuracy.  However, to hone the entire data set would require substantial fine-tuning and additional first-principles data, which undermines the purpose of any initial screening methodology, and emphasises that the screening process is about qualitative comparison, not quantitative.

The heat capacity of the switch system sets how much energy is stored per unit temperature change, and hence is related to how fast the device responds.  Heat capacity rises monotonically with temperature towards the Dulong--Petit limit at approximately 350-400~K. At 300~K the heat capacity increases with Li content from $\approx$437 to $\approx$550~J~kg$^{-1}$~K$^{-1}$  between $a = 0$ and $a = 1$, reflecting the light Li atoms added per formula unit. DFT and the uMLiP agree to within $\approx$1~J~kg$^{-1}$~K$^{-1}$ on average over all temperatures and compositions (\suppfigref{suppfig:HeatCapacity}). Perhaps most importantly, we can estimate for any electrode the room temperature heat capacity change with lithiation as $\Delta c_v=3a k_B N_A/M_r$ where $a$ is the fraction of intercalation and $M_r$ is the unintercalated molar mass. This arises from the original Dulong-Petit model for high temperatures and is a consequence of adding additional modes of vibration.

Reported room-temperature thermal conductivities of ZrS$_2$ (non-intercalated) span $\approx$2--13~W~m$^{-1}$~K$^{-1}$ in calculations~\cite{Glebko2018,Abdulsalam2020,Tang2023,Glebko2019,Li2025PhysScr}, while experimental values cluster around 2.1--2.5~W~m$^{-1}$~K$^{-1}$~\cite{Herninda2020,Li2023CuS,Li2025PhysScr}. The calculated range reflects differences in structure, exchange-correlation treatment, scattering order, Brillouin-zone sampling and force-field or molecular-dynamics methodology. The experimental values are sample-dependent: surface oxidation and chalcogen deficiency introduce mass and bonding disorder that adds phonon scattering~\cite{Herninda2020}, while defects, sample size and measurement method contribute further spread~\cite{Li2023CuS,Li2025PhysScr}. This makes absolute comparison difficult, and the presented values are estimates in the same sense, given the dependence on boundary scattering.  The robust quantity is the relative suppression on lithiation, which is qualitatively consistent between DFT and the uMLiP even though its magnitude is not. By imposing a 1~$\mu$m boundary mean free path on every composition, our results allow a robust comparison between lithiated and unlithiated systems (\secref{subsec:phonon_method}). From this, we obtain a cross-plane on/off ratio of $\approx$81 ($\approx$24) from our DFT (uMLiP) calculations, which provides an estimate of an upper (lower) bound of the phenomenon.

Stacking-disordered WSe$_2$ provides an independent benchmark for validating our screening method and assessing the physical plausibility of ultra-low cross-plane thermal conductivity, $\kappa_\perp$ in layered dichalcogenides. At 300~K, the measured cross-plane thermal conductivity, $\kappa_\perp$ decreases from $\approx 1.44$~W~m$^{-1}$~K$^{-1}$ in single-crystal WSe$_2$ to $\approx 0.048$~W~m$^{-1}$~K$^{-1}$ in stacking-disordered WSe$_2$~\cite{Chiritescu2007}. Our screening method yields $1.65$~W~m$^{-1}$~K$^{-1}$ for single-crystal WSe$_2$ (\supptabref{supptab:DTMDCscreen}), in close agreement with experiment. The substantial reduction in $\kappa_\perp$ upon stacking disorder has been attributed to the localisation of lattice vibrations caused by random stacking of otherwise crystalline layers. The predicted off-state conductivity of Li$_{0.125}$ZrS$_2$ is comparable to this value, supporting the physical plausibility of the ultra-low $\kappa_\perp$ predicted for intercalated TMDCs. Within our set of screened candidates, Li$_{0.125}$VS$_2$ (0.069~W~m$^{-1}$~K$^{-1}$) reaches an equally low off-state (evaluated at the uMLiP level), against 0.17~W~m$^{-1}$~K$^{-1}$ for Li$_{0.125}$ZrS$_2$ from the same potential). This indicates that ultra-low cross-plane conduction is a property of cross-plane defects of TMDCs rather than an anomalous result of ZrS$_2$.

\subsection{Origin of the switching: force constants and phonon scattering}

\begin{figure*}
    \centering
    \subfloat[]{\includegraphics[width=0.5\linewidth]{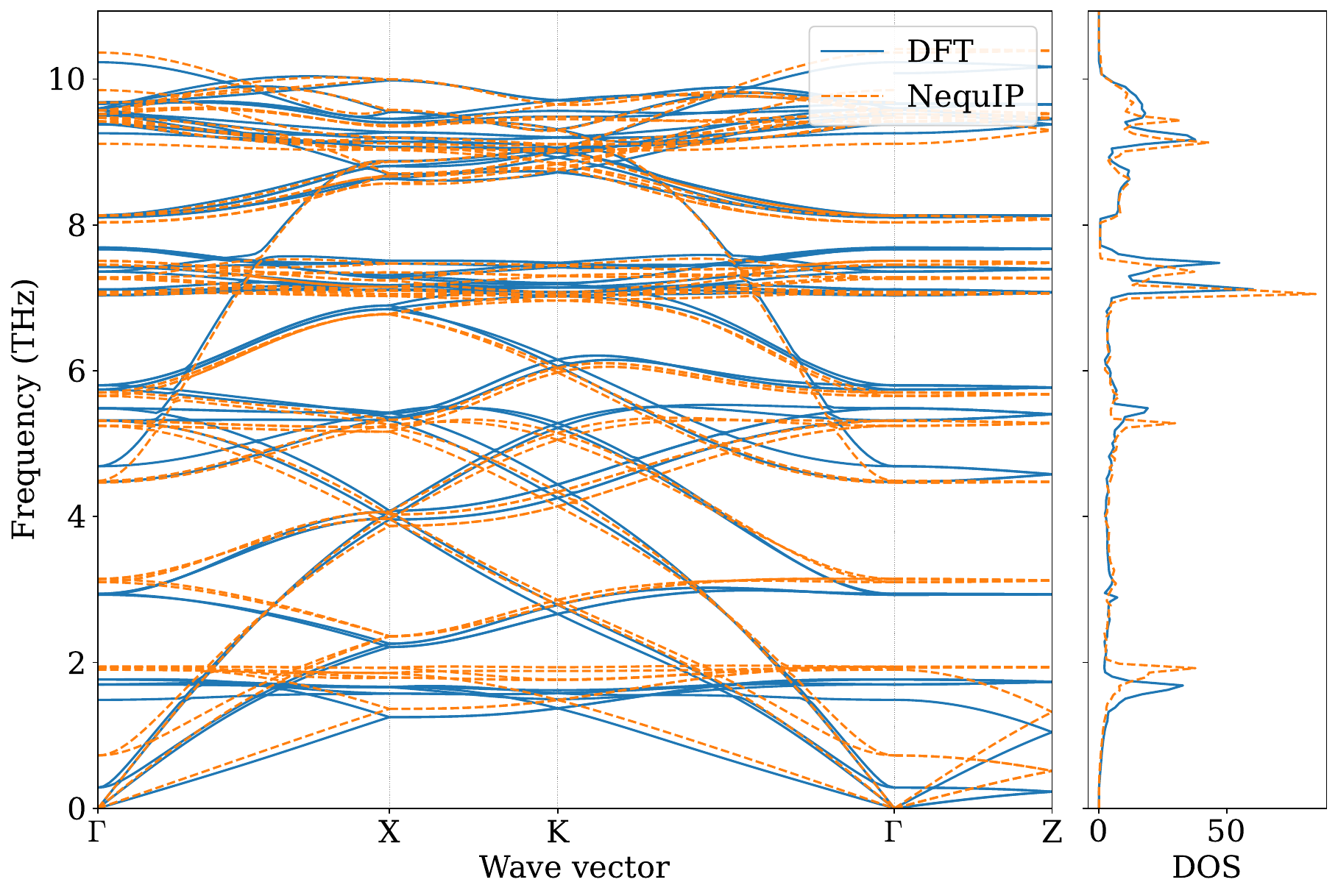}\label{subfig:bandx=0}}
    \subfloat[]{\includegraphics[width=0.5\linewidth]{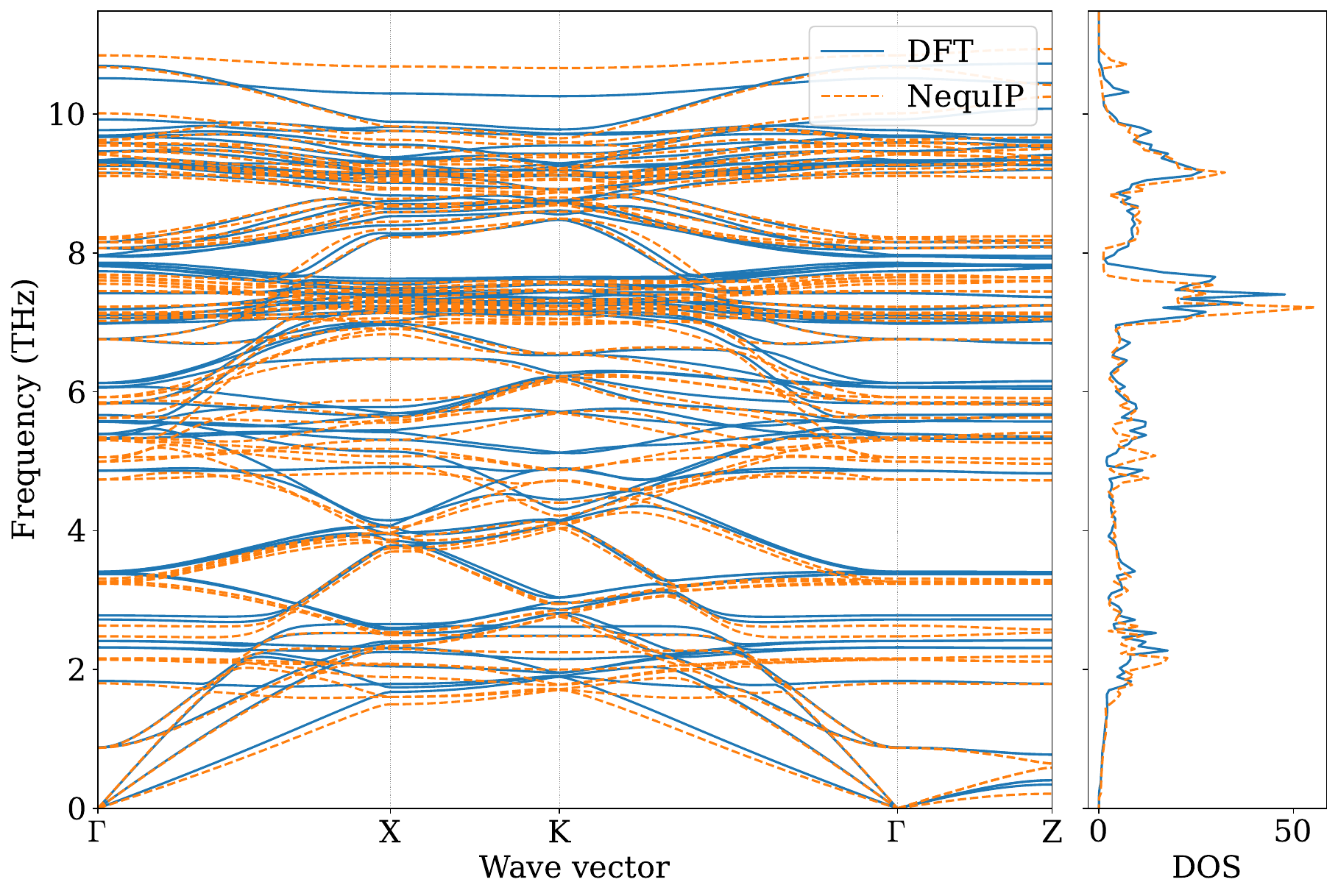}\label{subfig:bandx=0.125}}
    \hfill
    \subfloat[]{\includegraphics[width=0.5\linewidth]{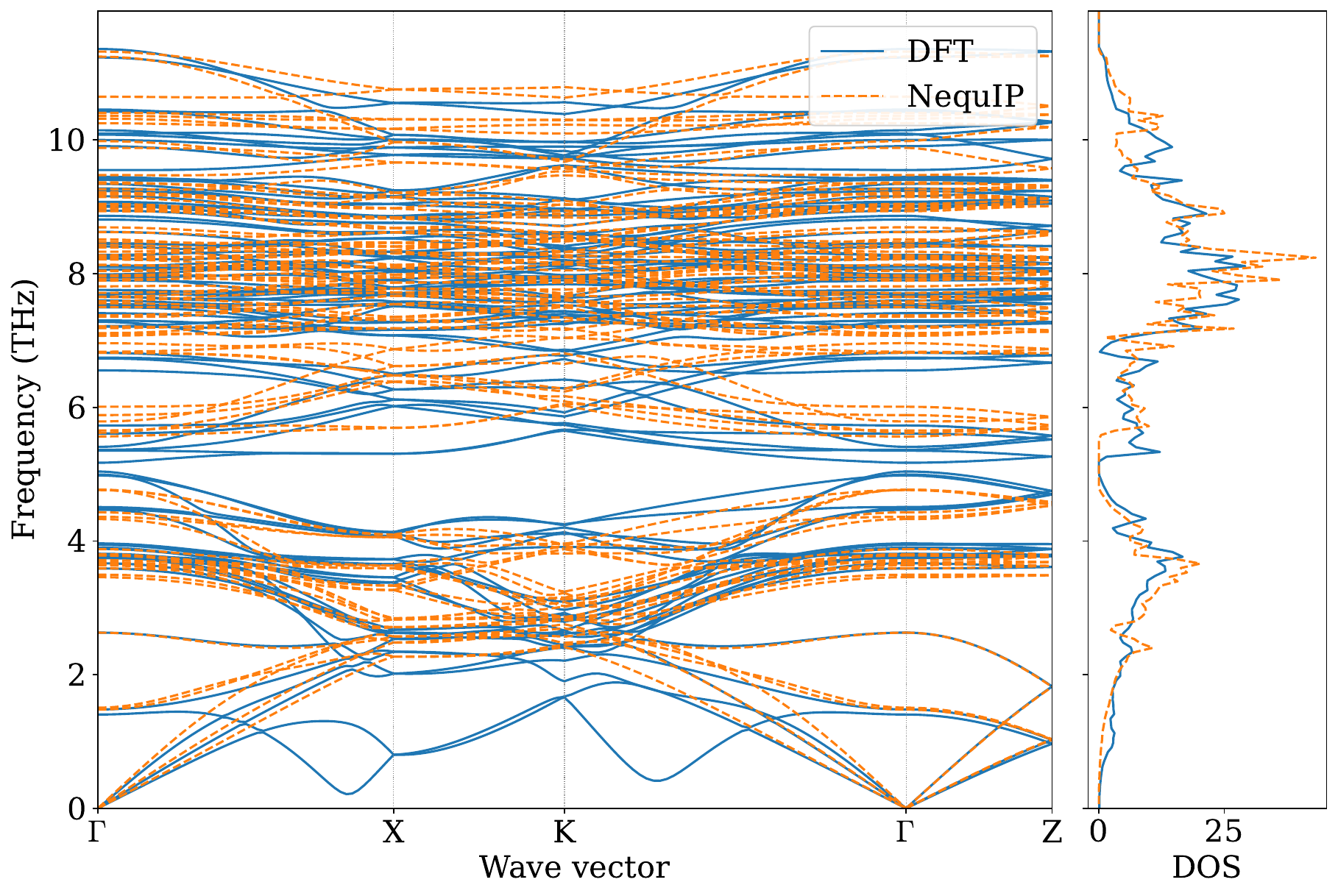}\label{subfig:bandx=0.75}}
    \subfloat[]{\includegraphics[width=0.5\linewidth]{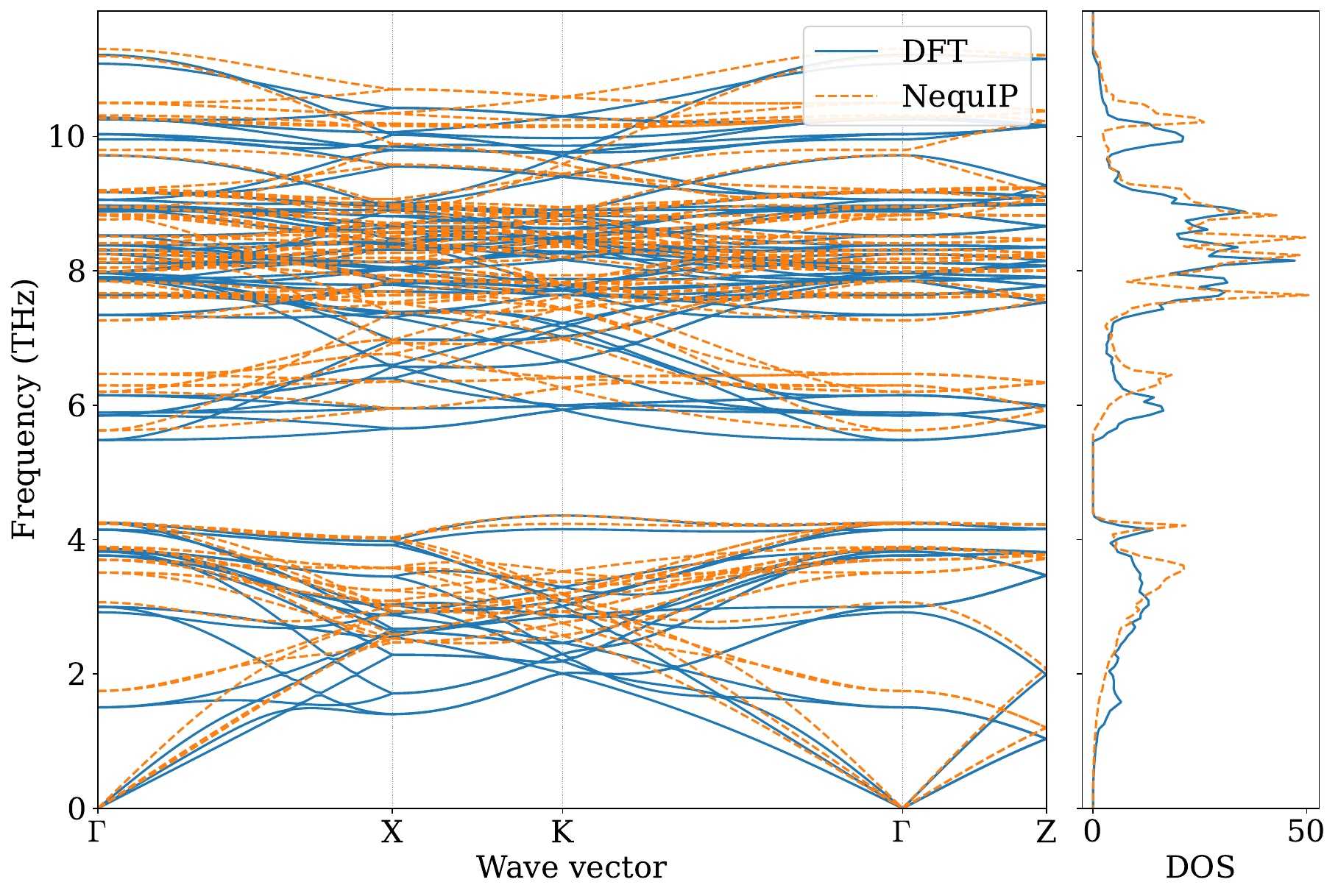}\label{subfig:bandx=1}}
    \caption{Phonon dispersions and densities of states of Li$_a$ZrS$_2$. Band structure and total density of states calculated with DFT (solid lines) and with the uMLiP (dashed lines) for (a) pristine ZrS$_2$ (the unintercalated $2\times2\times1$ supercell), (b) Li$_{0.125}$ZrS$_2$, (c) Li$_{0.75}$ZrS$_2$ and (d) LiZrS$_2$. Dispersions for the remaining compositions are given in \suppsecref{suppsec:band_dispersions}.}
    \label{fig:main_phonon}
\end{figure*}

To fully describe the origin of the change in thermal conductivity that makes Li$_{0\rightarrow0.125}$ZrS$_2$ a solid-state electrochemical thermal switch, we need to examine the constituents of the thermal conductivity.  As $\kappa \propto \nu \otimes \nu ~\tau$, we focus on phonon group velocities, $\nu$, and lifetimes, $\tau$, by investigating the behaviour of the phonon dispersions, the atomic force constants, and the phonon scattering rates. 
The key features shown in the phonon dispersions are the out-of-plane acoustic modes softening and an acoustic--optical gap widening with lithiation (\figref{fig:main_phonon}). Between the pristine host and 12.5\% filling (\subfigref{subfig:bandx=0} and \subfigref{subfig:bandx=0.125}) the median group velocity of the acoustic modes below 3~THz falls from $\approx$1.0 to $\approx$0.7~km~s$^{-1}$ ($\approx$1.2 to $\approx$0.8~km~s$^{-1}$ from the uMLiP), the mean out-of-plane component falls from $\approx$0.25 to $\approx$0.14~km~s$^{-1}$ ($\approx$0.30 to $\approx$0.23~km~s$^{-1}$), and the high-frequency optical branches flatten, their mean bandwidth above 7~THz narrowing from $\approx$0.45 to $\approx$0.39~THz ($\approx$0.46 to $\approx$0.36~THz). 
The phonon dispersion relations show (\suppsecref{suppsec:band_dispersions}) a narrow acoustic--optical gap is already present at $a=0.375$ ($\approx$0.05~THz near 5.6~THz) and $a=0.5$ ($\approx$0.13~THz between 5.71 and 5.85~THz).
It shifts to lower frequency and widens rapidly beyond $a=0.5$, reaching $\approx$0.44~THz (5.02--5.46~THz) at $a=0.625$, $\approx$0.80~THz (4.77--5.57~THz) at $a=0.75$ (\subfigref{subfig:bandx=0.75}) and $\approx$1.28~THz (4.35--5.63~THz) at LiZrS$_2$ (\subfigref{subfig:bandx=1}). DFT reproduces the endpoint, with a $\approx$1.10~THz gap (4.39--5.49~THz) at $a=1$. The softening of the out-of-plane acoustic modes and the opening of the acoustic--optical gap reflect the same weakening of the in-plane Zr--Zr bonding identified below.
The phonon dispersion DFT and the uMLiP agree closely for ZrS$_2$ (\subfigref{subfig:bandx=0}) and Li$_{0.125}$ZrS$_2$ (\subfigref{subfig:bandx=0.125}), with a mean absolute frequency difference of $\approx$0.11~THz over the full $\Gamma$--X--K--$\Gamma$--Z path at both compositions. For Li$_{0.75}$ZrS$_2$ (\subfigref{subfig:bandx=0.75}) and LiZrS$_2$ (\subfigref{subfig:bandx=1}), however, the uMLiP does not reproduce the low-frequency acoustic branches (most clearly shown in \subfigref{subfig:bandx=0.75}): the mean difference below 3~THz rises to $\approx$0.28 and $\approx$0.40~THz, and the lowest acoustic branch at X, which softens in DFT from $\approx$2.07~THz at $a=0.5$ to $\approx$0.80~THz at $a=0.75$ and hardens again to $\approx$1.40~THz at $a=1$, remains at $\approx$2.3 and $\approx$2.5~THz in the uMLiP. 

\begin{figure*}[h!]
    \centering
    \subfloat[]{\includegraphics[width=0.5\linewidth]{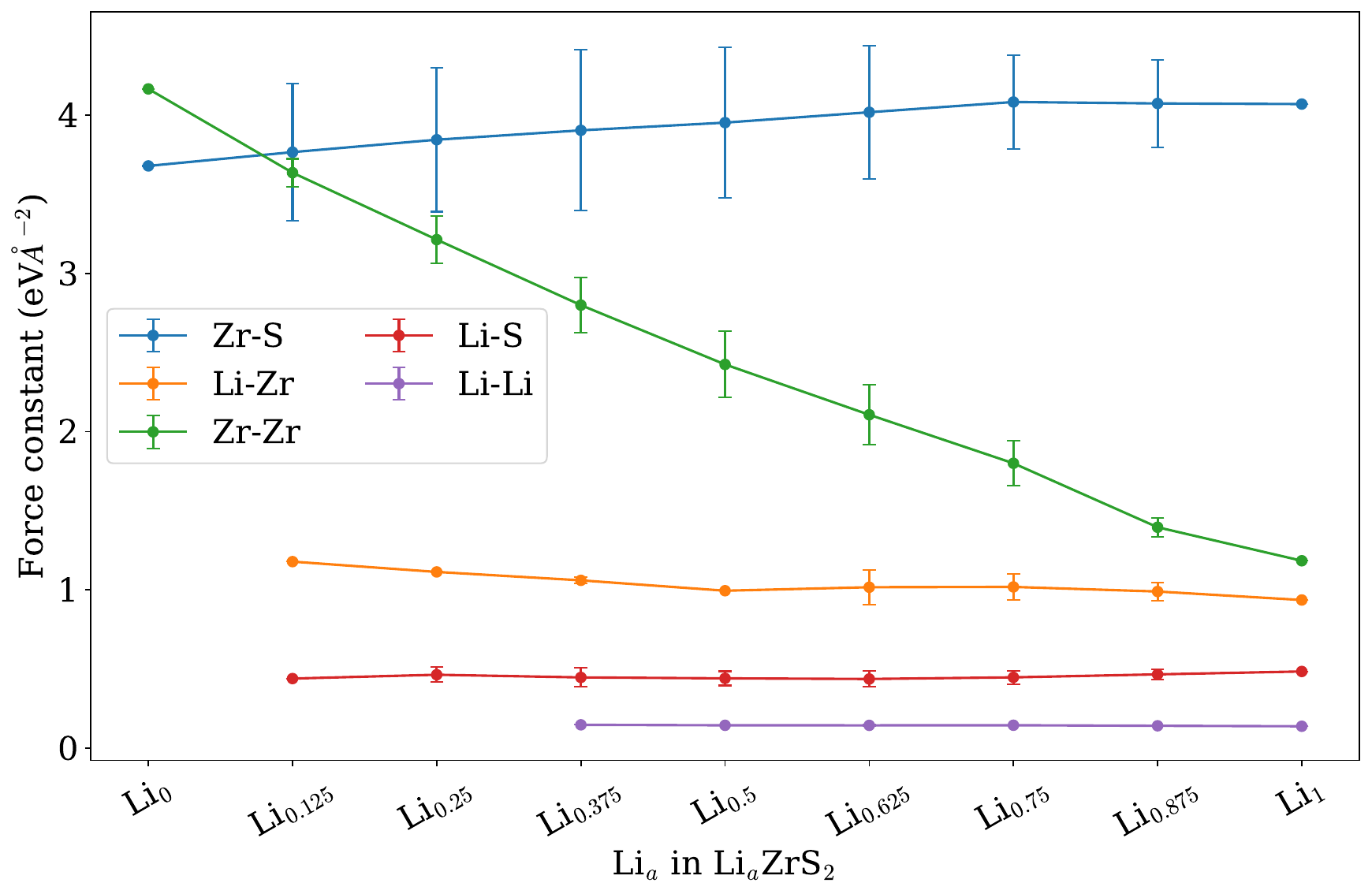}\label{subfig:BondFCVLiDFT}}
    \subfloat[]{\includegraphics[width=0.5\linewidth]{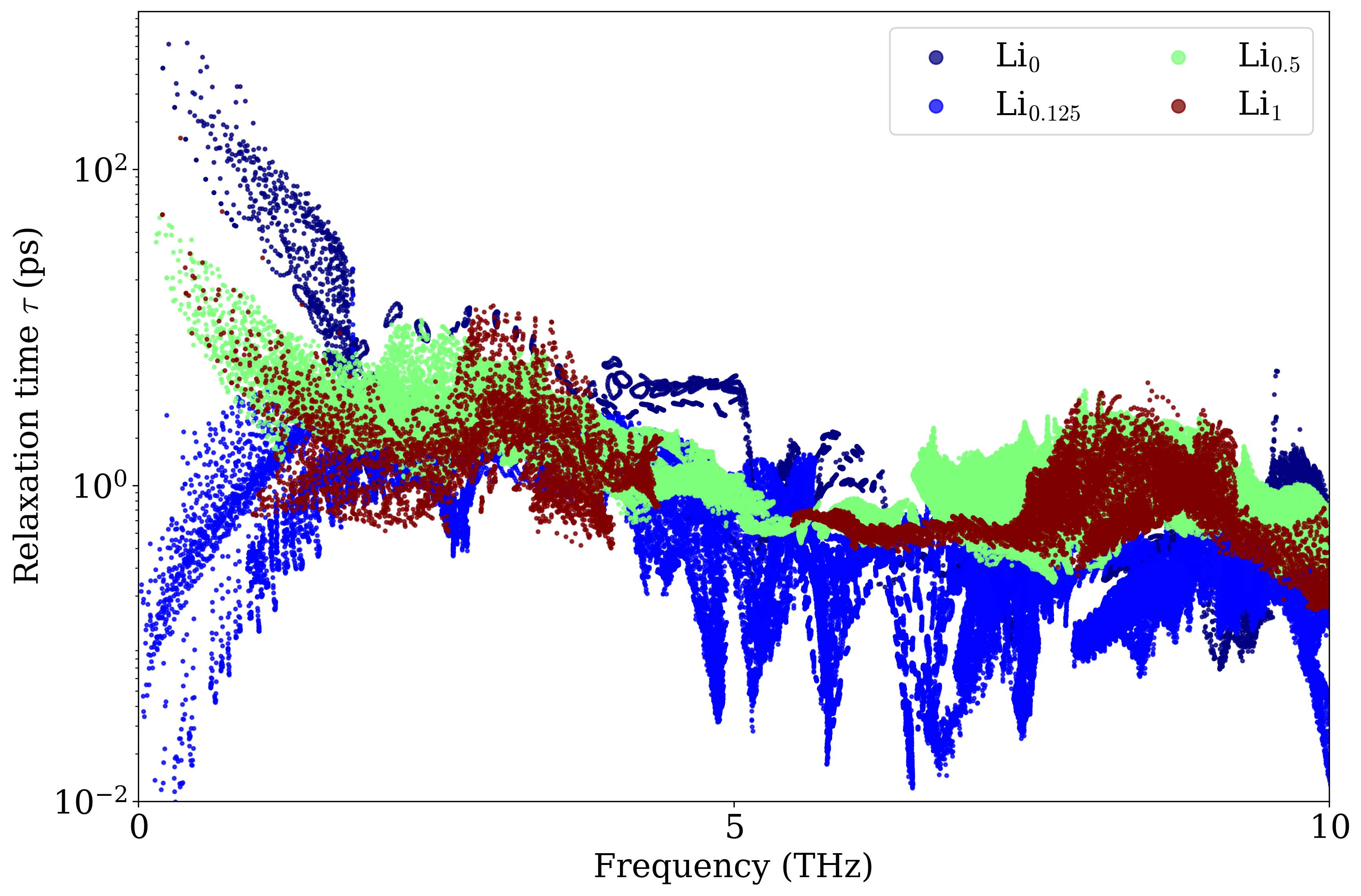}\label{subfig:RelaxationTimesVLiDFT}}
    \hfill
    \subfloat[]{\includegraphics[width=0.5\linewidth]{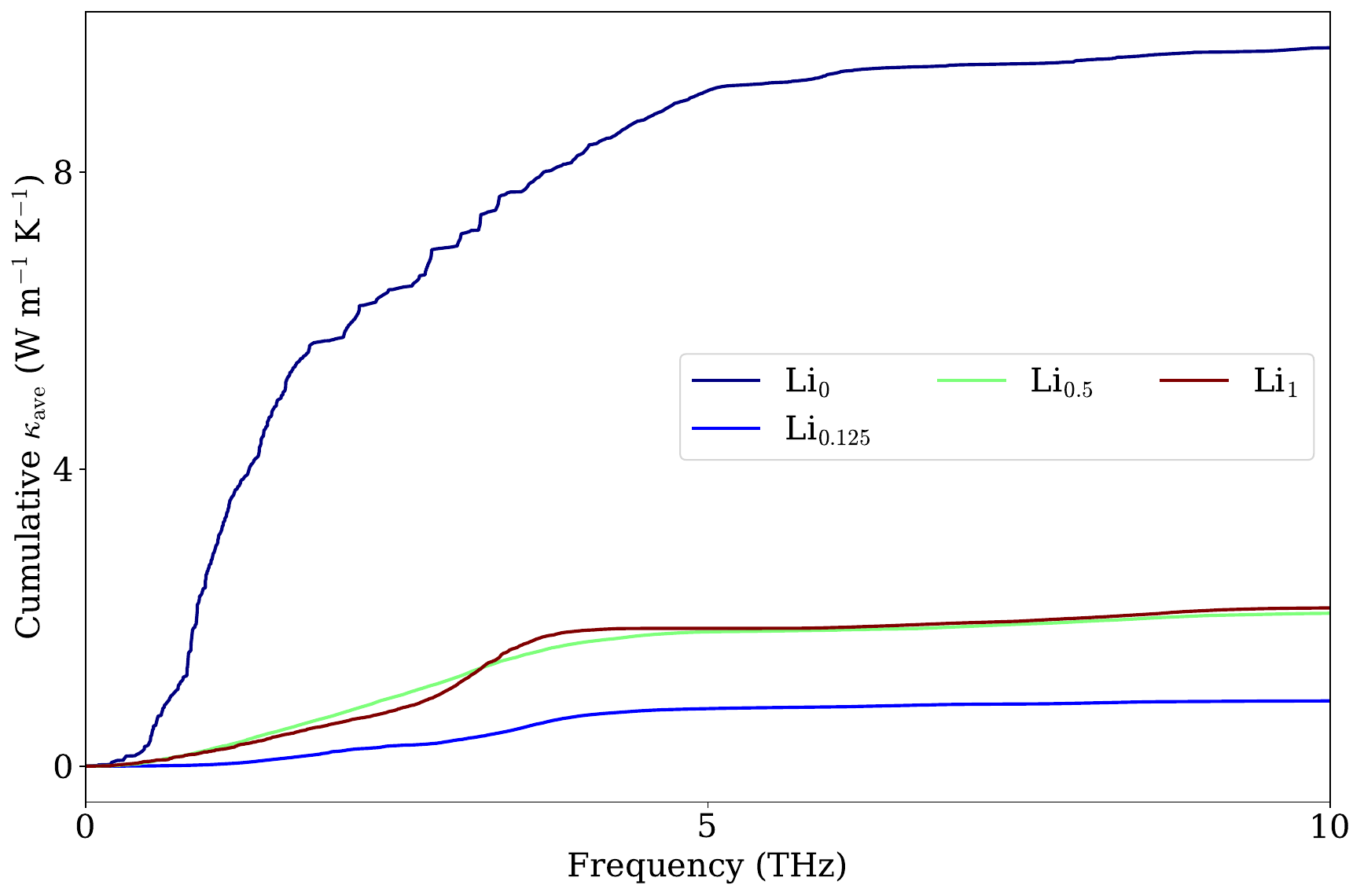}\label{subfig:CumulativeKappaDFT}}
    \caption{Microscopic origin of the switching in Li$_a$ZrS$_2$. (a) Bond-resolved second-order force constants from DFT. (b) RTA relaxation times $\tau$ at 300~K versus frequency for $a = 0$, $0.125$, $0.5$ and $1$; the uMLiP results are given in \suppfigref{subfig:RelaxationTimesVLiNequIP}, and the mode-free window at $a=1$ is the acoustic--optical gap of \subfigref{subfig:bandx=1}. (c) Cumulative lattice thermal conductivity versus frequency at 300~K from the DFT RTA calculation.}
    \label{fig:Mechanism}
\end{figure*}

Examining the second-order force constants (FCs) resolved by bond type shows how the interatomic forces change with intercalation.  We show these results calculated with  DFT in \subfigref{subfig:BondFCVLiDFT} and with the uMLiP in \suppfigref{subfig:BondFCVLiNequIP}. Intercalation weakens the Zr--Zr intra-layer bonding and slightly hardens the Zr--S bonds, whose force-constant spread is negligible at $a=0$ and $a=1$ but substantial at intermediate compositions. The mean Zr--Zr force constant falls from $\approx$4.2~eV\,\AA$^{-2}$ (DFT, $a=0$) to $\approx$1.2~eV\,\AA$^{-2}$ at $a=1$, and the uMLiP reproduces the trend almost quantitatively ($4.0 \rightarrow 1.2$~eV\,\AA$^{-2}$). The newly formed Li--Zr ($\approx$0.9--1.2~eV\,\AA$^{-2}$) and Li--S ($\approx$0.5~eV\,\AA$^{-2}$) bonds are softer than the host Zr--S bonds. The small Li--Li force constant ($\approx$0.14~eV\,\AA$^{-2}$) reflects the weak coupling of the Li sublattice, close to the $\approx$0.13~eV\,\AA$^{-2}$ obtained for bulk Li at 293~K from lattice-dynamics models~\cite{Beg1976}. The inserted Li introduces both a mass contrast and a set of soft bonds into the interlayer region.  The reduction in the Zr--Zr bond strength is evidenced by the softening of the in-plane Zr framework, which suppresses the low-frequency group velocities that dominate heat transport and is the primary structural driver of the reduction in $\kappa$.

We apply the single-mode relaxation-time approximation to obtain the phonon relaxation times for the compositions $a=0$, $0.125$, $0.5$ and $1$ (\subfigref{subfig:RelaxationTimesVLiDFT}) using DFT (the uMLiP are shown in \suppfigref{subfig:RelaxationTimesVLiNequIP}), and find that the overall distribution changes little along the series: the median lifetime at $a=0$ is $\approx$0.6~ps for both methodologies. For the heat-carrying acoustic modes, however, $\tau$ changes markedly as a function of composition. The mean of $\log_{10}\tau$ over all modes at 300~K, $\langle\log_{10}\tau\rangle$, falls from $-0.07$ at $a=0$ to $-0.44$ at $a=0.125$ in DFT, i.e.\ the geometric-mean lifetime drops from 0.84 to 0.36~ps. The mean lifetime of the low-frequency acoustic branches (below $\approx$3~THz) falls by $\approx$4.0$\times$ between those two compositions ($\approx$2.1$\times$ from the uMLiP). The population of long-lived acoustic modes is progressively reduced as Li is inserted. Since each mode contributes to $\kappa$ as $C_v v^2 \tau$, a factor of four in $\tau$ cannot by itself account for the $\approx$81-fold drop in cross-plane conductivity, $\kappa_\perp$. The remainder comes from the softened acoustic group velocities (\suppfigref{suppfig:GroupVelocity}) implied by the Zr--Zr force-constant reduction.


As expected from the phonon dispersion relations, we observe the same mode-free window in the relaxation-time distribution, i.e. no modes exist between 4.39 and 5.49~THz at $a=1$ as seen in \subfigref{subfig:RelaxationTimesVLiDFT} (the uMLiP results can be found in \suppfigref{subfig:RelaxationTimesVLiNequIP}). 
Effectively, the conductivity associated with this frequency interval in the bulk system collapses upon intercalation.   This range contributes 0.54~W~m$^{-1}$~K$^{-1}$ ($\approx$6\% of the total) at $a=0$ and 0.05--0.07~W~m$^{-1}$~K$^{-1}$ at $a=0.125$--$0.5$, then falls to $\approx$0.04, $\approx$0.01 and $<$0.001~W~m$^{-1}$~K$^{-1}$ at $a=0.625$, $0.75$ and $a\ge0.875$ (uMLiP) and to zero at $a=1$.
This effectively pushes states into the adjacent frequency regions, creating flatter phonon branches which in turn result in higher scattering rates.
Consistent with this, $\kappa$ has a non-linear dependence on filling fraction, $a$. After the initial collapse at $a=0.125$ it partially recovers at the 50\% Li content before reducing and rising again at fully lithiated.  This is to be expected, as the more disordered the systems ($a$=0.125, 0.625), the greater the number of allowed modes for scattering.  This disorder means that whilst the number of states remains similar, the number of states with different momentum and energy increases (states with similar energy and momenta decrease), which means more scattering events satisfy momentum and energy conservation conditions than in the bulk, resulting in a higher scattering rate. 

The change in group velocity and phonon relaxation time is reflected in the cumulative thermal conductivity $\kappa(\omega)$ (\subfigref{subfig:CumulativeKappaDFT} and \suppfigref{subfig:CumulativeKappaNequIP}). We find that acoustic modes below $\approx$5~THz carry $\approx$94\% of the total $\kappa$ in pristine ZrS$_2$, $\approx$88\% in Li$_{0.125}$ZrS$_2$ and $\approx$86\% at full lithiation. Our uMLiP calculations give $\approx$94\%, $\approx$93\% and $\approx$84\%. The spectral midpoint of the cumulative curve, the frequency below which half of $\kappa$ is carried, rises from $\approx$1.5~THz to $\approx$3.3~THz on lithiation (uMLiP: $\approx$1.6 to $\approx$2.3~THz), which is the spectral signature of acoustic softening. The residual contribution above $\approx$5~THz is small in every phase ($\lesssim$14\%, and largest at full lithiation), so the intercalation-induced suppression is concentrated in the acoustic branches.

We attribute the high-lithiation recovery in thermal conductivity to a second factor: the symmetry of the Li ordering. The in-plane components $\kappa_{xx}$ and $\kappa_{yy}$ are equal ($\kappa_{xx}=\kappa_{yy}=\kappa_{\parallel}$) whenever the ordering preserves the in-plane three-fold symmetry, and split when it does not ($\kappa_{xx}\neq\kappa_{yy}$). This splitting is largest at $a=0.5$ (uMLiP 3.56 versus 4.37~W~m$^{-1}$~K$^{-1}$; DFT 2.48 versus 3.11~W~m$^{-1}$~K$^{-1}$), is a few percent over $0.25\le a\le0.625$, and vanishes at $a=0.125$, $0.875$ and $1$, where the ordering retains the symmetry. At $a=0.75$ the components are again nearly degenerate (within $\approx$2\%). The near-isotropic orderings at $a=0.75$--$1$ thus coincide with the recovery of the thermal conductivity at high lithiation. The composition dependence of $\kappa$ has two distinct parts. The first is a sharp collapse at low Li content, driven by the force-constant softening and the enhanced acoustic scattering described above. The second is a partial recovery towards $a=1$, associated with the acoustic--optical gap and the restoration of in-plane isotropy. Li$_{0.125}$ZrS$_2$ sits at the minimum of that curve, which is why this composition, rather than the fully lithiated phase, is the optimal off state.

\subsection{Switching performance}
\label{subsec:switch_score}

\begin{figure*}
    \centering
    \vspace{-2cm}
    \subfloat[]{\includegraphics[width=0.5\linewidth]{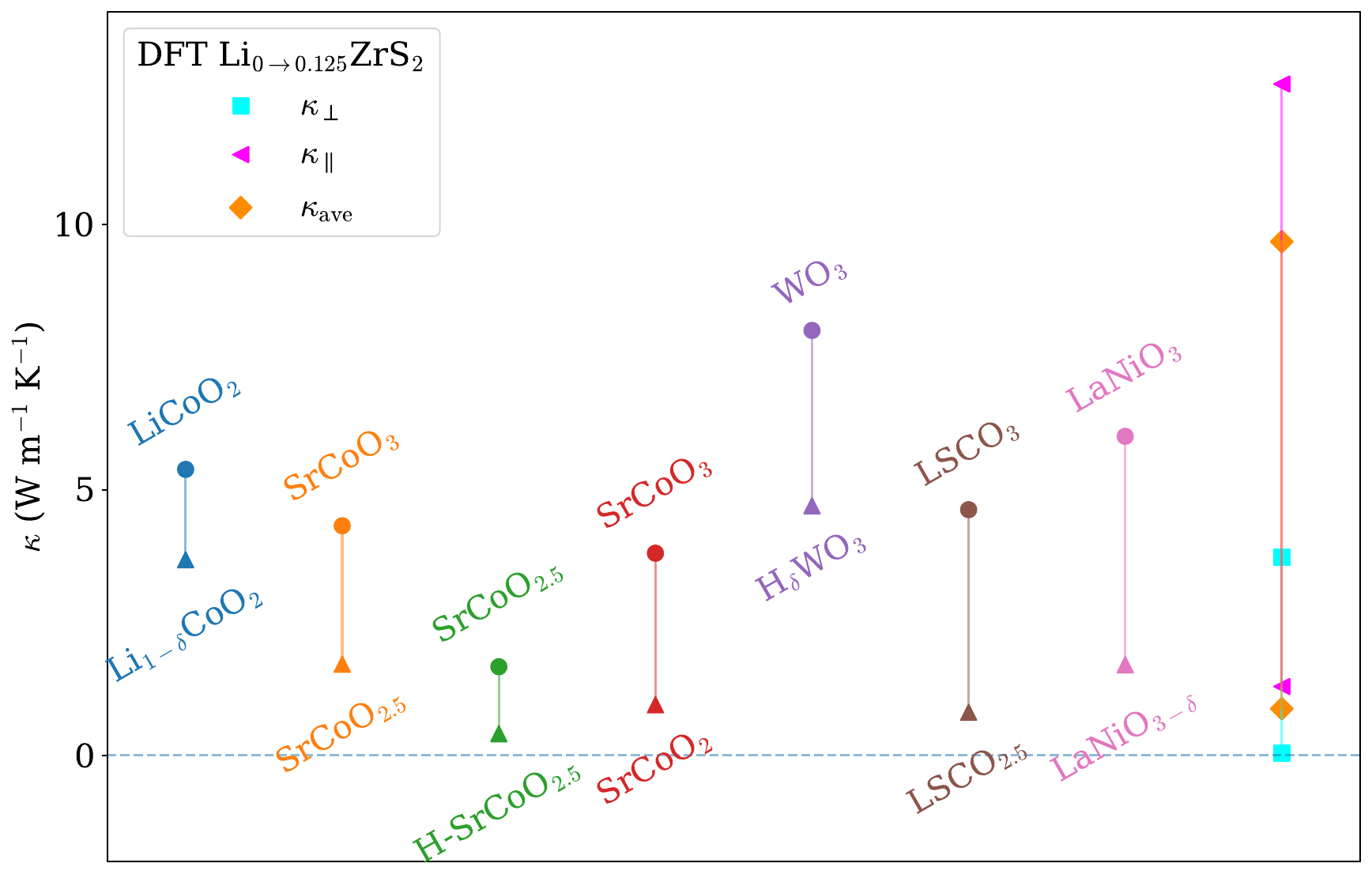}\label{subfig:KappaComparisonExp}}
    \subfloat[]{\includegraphics[width=0.5\linewidth]{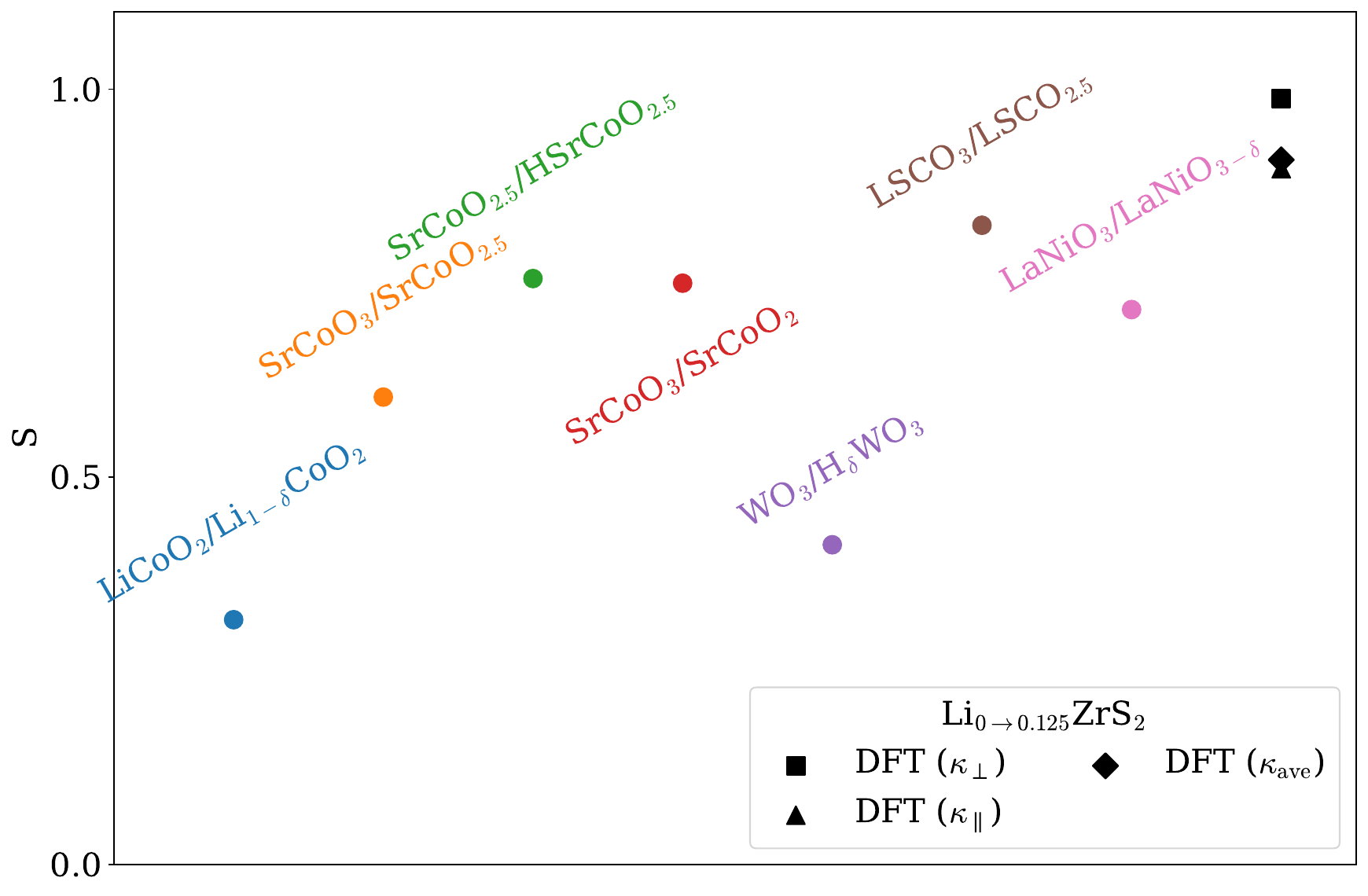}\label{subfig:SwitchScoreComparison}}
    \hfill
    \subfloat[]{\includegraphics[width=0.5\linewidth]{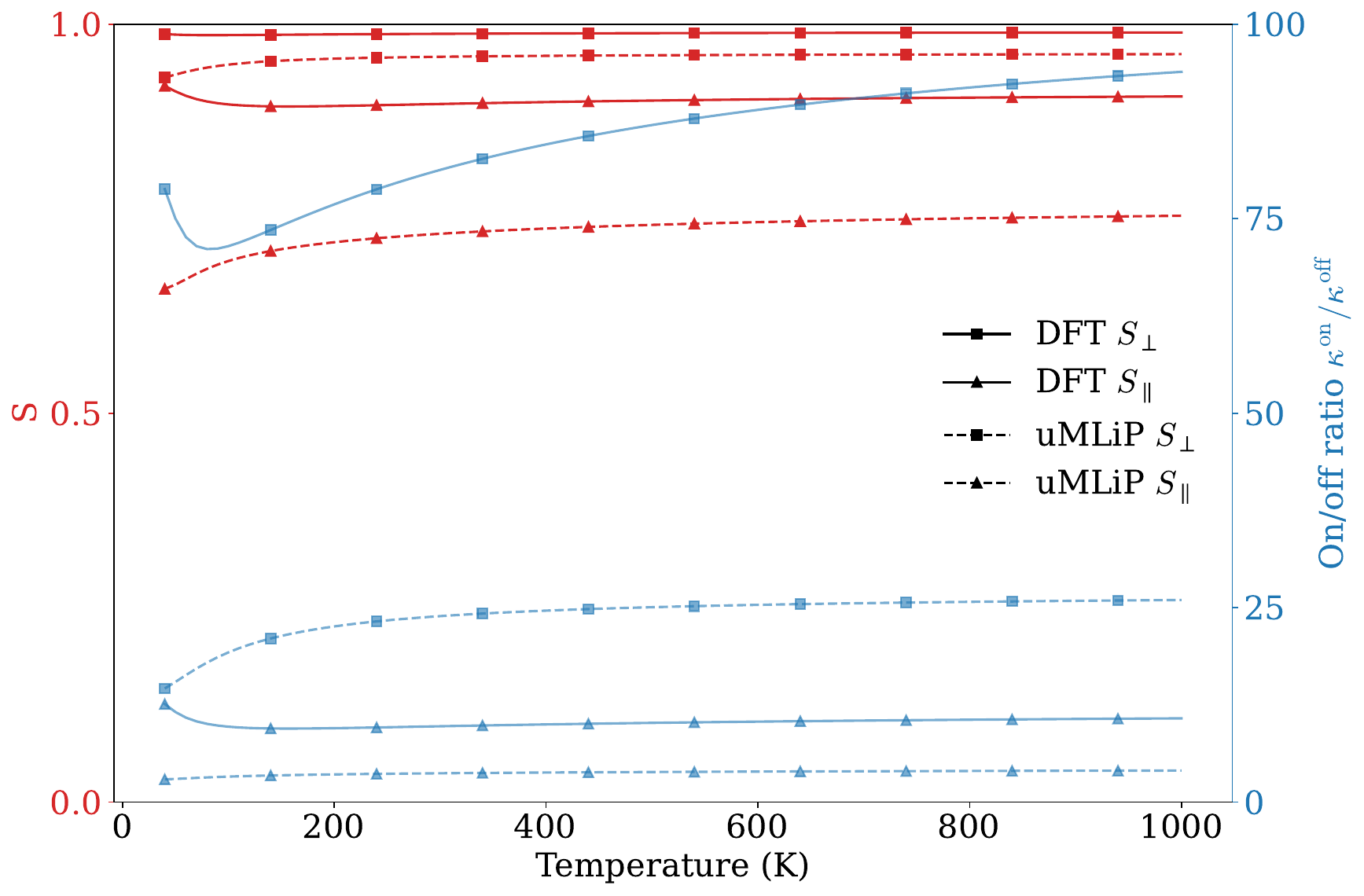}\label{subfig:SwitchScoreVsT}}
    \subfloat[]{\includegraphics[width=0.5\linewidth]{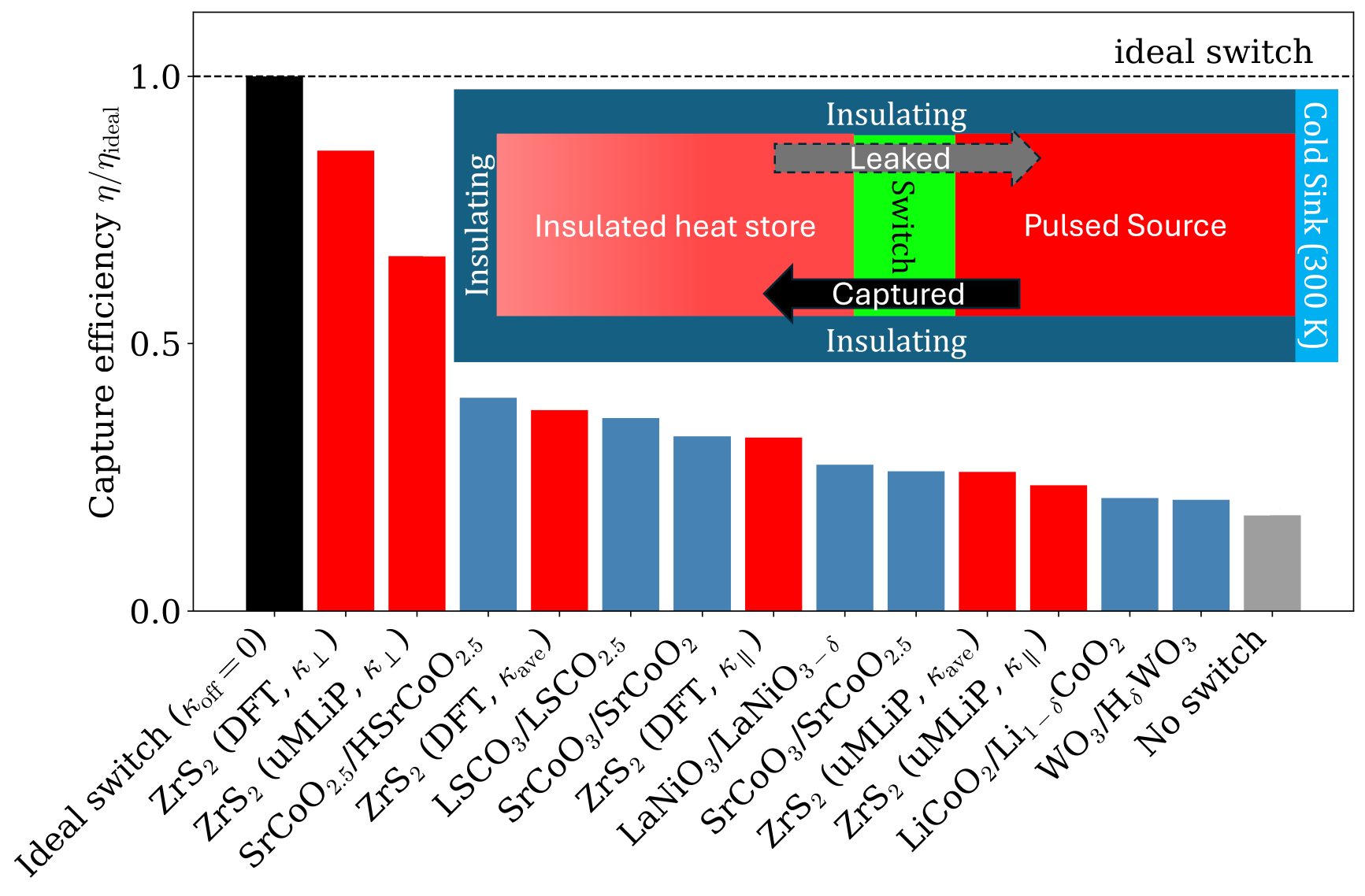}\label{subfig:WHR_eta}}
    \caption{Waste-heat recovery through an electrochemical thermal switch. A pulsing heat source is coupled through the switch layer to an insulated sink, with identical device geometry and pulse schedule for each material (inset of d). (a) On- and off-state thermal conductivity and (b) switch score of Li$_a$ZrS$_2$ (DFT RTA; $S_\perp$:cross-plane, $S_\parallel$:in-plane and $S_{ave}$ orientation-averaged) compared with other electrochemical thermal switches~\cite{Cho2014,Lu2020,Yang2023,Ning2019,Zhang2023,Bian2024}. (c) Switch score (left axis) and on/off ratio (right axis) versus temperature for the cross-plane (solid) and in-plane (dashed) channels. (d) Capture efficiency normalised to the ideal switch, marked by the dashed horizontal line at unity; the bar labelled ``no switch'' is the unswitched host conductor.}
    \label{fig:WasteHeatRecovery}
\end{figure*}

Having established that Li$_a$ZrS$_2$ has an electrochemically tuneable thermal conductivity, we compare its thermal-switching capability with other known thermal switches.
The cross-plane switch score, $S_\perp$,  reaches $S = 0.99$ (0.96) based on the DFT (uMLiP) results with switching ratios of $R \approx 81$ ($R \approx 24$).  These results are higher than previously studied solid-state electrochemical thermal switches (\subfigref{subfig:SwitchScoreComparison}): LiCoO$_2$ $\rightarrow$ Li$_{1-\delta}$CoO$_2$~\cite{Cho2014}, SrCoO$_3$ $\rightarrow$ SrCoO$_{2.5}$ and SrCoO$_{2.5}$ $\rightarrow$ HSrCoO$_{2.5}$~\cite{Lu2020}, SrCoO$_3$ $\rightarrow$ SrCoO$_2$~\cite{Yang2023}, WO$_3$ $\rightarrow$ H$_{\delta}$WO$_3$~\cite{Ning2019}, La$_{0.5}$Sr$_{0.5}$CoO$_3$ $\rightarrow$ La$_{0.5}$Sr$_{0.5}$CoO$_{2.5}$ (LSCO)~\cite{Zhang2023}, and LaNiO$_3$ $\rightarrow$ LaNiO$_{3-\delta}$~\cite{Bian2024}. 

The comparators reach LSCO~\cite{Zhang2023} $0.82$, SrCoO$_{2.5}$/HSrCoO$_{2.5}$~\cite{Lu2020} $0.76$, LaNiO$_3$~\cite{Bian2024} $0.72$, WO$_3$~\cite{Ning2019} $0.41$ and LiCoO$_2$~\cite{Cho2014} $0.32$. Our in-plane score, $S_\parallel$ is calculated from the mean of $\kappa_{xx}$ and $\kappa_{yy}$, and provides $S_\parallel = 0.90$ ($R \approx 9.8$), and the orientation average, relevant to a randomly oriented powder or a polycrystalline film, is $S_{ave} = 0.91$ ($R \approx 11$). Both also exceed every comparator. The cross-plane channel is thus the strongest by a wide margin, while the in-plane and orientation-averaged channels give a weaker but still substantial contrast. The absolute switching width, $\Delta\kappa_\perp = 3.69$~W~m$^{-1}$~K$^{-1}$ cross-plane, is comparable to the $\approx$4.3~W~m$^{-1}$~K$^{-1}$ reported for LaNiO$_3$-based devices~\cite{Bian2024}, but is achieved with a ratio more than twenty times larger and without the risk of the oxygen being lost to the environment. In addition, Li$_a$ZrS$_2$ switching performance is essentially temperature-independent (\subfigref{subfig:SwitchScoreVsT}, ignoring additional anharmonic effects) as between 100 and 1000~K the cross-plane $S_\perp$ score varies only from 0.986 to 0.989 (DFT) and the in-plane score, $S_\parallel$ from 0.897 to 0.907 (DFT). The contrast persists across the whole operating window. 
In addition, the cross-plane ratio grows from $\approx$71 at 100~K to $\approx$94 at 1000~K, which matters for any device that must hold its contrast under thermal load.

\subsection{Waste-heat recovery}
\label{subsec:WasteHeat}

To demonstrate and compare, we model using a continuum-level model~\cite{Mclean2026} the switching contrast in a one-dimensional waste-heat-recovery system (\subfigref{subfig:WHR_eta}, inset), in which rectangular heat pulses of fixed energy per unit area are injected into a source region. The source is anchored to a fixed-temperature reservoir and is coupled through the switch layer to an insulated sink in which the recovered heat is stored. The switch opens only while the temperature drop across it exceeds a threshold large enough to drive heat into the sink, and remains off otherwise, so any heat that returns from the sink through the off-state conductivity is counted as leakage. We assign every material a temperature-independent thermal conductivity and a common volumetric heat capacity of 1.69$\times10^{6}$~J~m$^{-3}$~K$^{-1}$, the average of the calculated ZrS$_2$ and Li$_{0.125}$ZrS$_2$ values at 300~K. This isolates the switching contrast from differences in heat storage. 
We then simulate the identical device, geometry and pulse schedule for the on/off conductivity pair of every electrochemical switch compared in \secref{subsec:switch_score} based on the uMLiP results. Each is benchmarked against an ideal switch with zero off-state conductivity and the cross-plane DFT on-state conductivity, $\kappa_\perp$, of this work, and against an unswitched bar of the source/sink host conductor (model and parameters in \secref{subsec:WHRmodel}; complete results in \suppsecref{suppsec:WHRresults}).

Normalized to the ideal-switch benchmark, we recover 0.86 (DFT) and 0.66 (uMLiP) of the ideal-switch energy after 20 heat pulses, against $\approx$0.40 for the best previously reported electrochemical switch (SrCoO$_{2.5}$/HSrCoO$_{2.5}$), $\approx$0.36 for LSCO, $\approx$0.33 for SrCoO$_3$/SrCoO$_2$, $\approx$0.27 for LaNiO$_3$ and $\approx$0.21 for both LiCoO$_2$ and WO$_3$ (\figref{fig:WasteHeatRecovery}). An unswitched conductor retains only $\approx$0.18. The orientation-averaged and in-plane ZrS$_2$ switches, which are the relevant limits for a randomly oriented powder, retain 0.32--0.38 (DFT) and 0.24--0.26 (uMLiP). In absolute terms, the cross-plane switch captures $\eta = 5.65\%$ of the injected pulse energy, against 6.56\% for the ideal switch. The cycle is deliberately demanding, and in this idealistic model of thermal switches 93--99\% of the injected energy drains to the reservoir (is wasted) for every material.  Hence, the meaningful comparison is the relative one, which indicates that the Li$_a$ZrS$_2$ is still the best-performing candidate presented here.   The full set of values, including the leakage and the on-time fraction of the gate, is given in \supptabref{supptab:WHRresults}.

We find that recovery is governed by the on/off contrast rather than by the on-state conductance (\suppfigref{suppfig:WHR_stored}). Materials with a small on/off ratio cycle their storage, capturing each pulse and then returning a large fraction of it through the off-state between pulses, which produces the sawtooth profiles in \suppfigref{suppfig:WHR_stored}. The cross-plane ZrS$_2$ switch instead retains each capture and accumulates almost monotonically. Off-state leakage drains 12.1--15.5\% of the pulse energy in the four weakest comparators, against $\approx$1.3\% through the cross-plane ZrS$_2$ switch and $\approx$19\% when the device is left unswitched. The ideal switch shares the low on-state conductivity of cross-plane ZrS$_2$ (3.73~W~m$^{-1}$~K$^{-1}$) and still outperforms the other presented materials. The cross-plane ZrS$_2$ switch approaches that bound because its off-state conductivity ($\approx$0.05~W~m$^{-1}$~K$^{-1}$) nearly closes the gate.

\clearpage
\section{Discussion}

Combining first-principles phonon calculations with a universal machine-learned interatomic potential identifies Li$_a$ZrS$_2$ as a high-contrast solid-state electrochemical thermal switch from a consideration of an additional 56 TMDC candidates. Our results also identify the mechanism behind the thermal switching behaviour and indicate how this can be applied in other systems. Furthermore, our high-throughput uMLiP-based search shows that this switching behaviour is common in the broader material class of TMDCs.

Our screening process shows that cross-plane drop in thermal conductivity, $\kappa_\perp$, is phase-dependent. All fifteen 1T systems, including Li$_{0.125}$ZrS$_2$, show decreases in the thermal conductivity upon intercalation, i.e. they switch positively (scores 0.48--0.96). Conversely, 2H systems show a mixture of responses, with five systems in which intercalation raises the $\kappa_\perp$, and four of these are Mg-intercalated. In these systems, the 2H hosts are already close to the low conductivities that a switch must reach (mean $\kappa_{\perp}\approx0.9$ vs $\approx2.6$~W~m$^{-1}$~K$^{-1}$ for 1T) and strongly anisotropic ($\kappa_{\parallel}/\kappa_{\perp}=10$--84), leaving little room for suppression. We believe this is driven by the reduced symmetry in 2H systems compared to 1T.  These results highlight the power of using uMLiPs to perform rapid screening of material systems to find desirable properties.

For our identified target, Li$_{0 \rightarrow0.125}$ZrS$_2$, we have applied a higher level of theoretical modelling, DFT, to validate the uMLiP result.  Our results have shown that lithiation reduces the cross-plane thermal conductivity, $\kappa_\perp$, from $\approx$3.73 to $\approx$ 0.046~W~m$^{-1}$~K$^{-1}$, resulting in an on/off ratio of $\approx$81. The in-plane and orientation-averaged contrasts are weaker, $\approx$9.8 and $\approx$11. The mechanism for reduction is twofold as is a result of intercalation softening the interaction between Zr--Zr atoms, which results in a drop in the acoustic group velocities. 
In addition, the break in symmetry created by intercalation shortens the lifetime of the heat-carrying acoustic phonons by approximately a factor of $4$ between $a=0$ and $a=0.125$.  Unlike previous discussions, these results do not support the concept of rattler modes dropping the thermal conductivity~\cite{Qian2016} in intercalation materials, but instead point clearly to a decrease in symmetry and low bond strengths being key.

Whilst we have focused on Li$_{0\rightarrow0.125}$ZrS$_2$, we note that Li$_{0\rightarrow0.125}$TiS$_2$ and Li$_{0\rightarrow0.125}$VS$_2$ are the closest competitors based on switching score alone ($S\approx$0.90), with off-state cross-plane conductivities, $\kappa_\perp$, of 0.19 and 0.069~W~m$^{-1}$~K$^{-1}$, respectively, but have longer switching times. Li$_{0\rightarrow0.125}$ZrS$_2$ thus combines a large conductivity contrast with a faster thermal response among the leading candidates. Unlike previous approaches exploiting $c$-axis strain, stacking disorder, and rattler modes in MoS$_2$~\cite{MoS2Kinter} and stacking disorder in WSe$_2$~\cite{Chiritescu2007}, the chemo-mechanical strain in Li$_a$ZrS$_2$ is small ($\approx$0.4\% between $a=0$ and $a=0.125$ as shown in \suppfigref{suppfig:VolumeExpansion}). The switching is therefore consistent with intercalation-induced changes in the host force constants and phonon scattering, rather than a substantial structural transformation such as a 1T-to-2H phase change or c-axis strain.

Device-level modelling demonstrates the potential of this thermal contrast for pulsed waste-heat recovery. Under an identical pulse schedule, the cross-plane switch recovers 86\% of the energy recovered by an ideal switch using DFT-derived conductivities, compared with 66\% using the uMLiP results. The DFT-based performance exceeds the $\approx$40\% reported for the best previously studied solid-state electrochemical thermal switches and recovers nearly five times the energy of the unswitched host. This improvement arises from reduced thermal backflow between pulses in the low-conductivity off state. However, the absolute capture efficiency remains modest ($\eta=5.65\%$, compared with 6.56\% for the ideal switch), so the principal benefit is the increased fraction of recoverable energy captured rather than high absolute energy-conversion efficiency. The estimated equilibration time of $\approx$20~$\mu$s for a 5~$\mu$m layer, falling to $\approx$11~$\mu$s in the on state, is shorter than the 50~$\mu$s pulse duration of the model heater. However, the switching kinetics of the thermal switch are not included and must be considered when assessing practical device performance.

Overall, this study establishes intercalation-induced phonon engineering as a promising route to controlling thermal transport in layered materials. By combining high-throughput screening with microscopic analysis and device-level modelling, we identify both the physical origin of the switching and the criteria governing its performance. Future work should extend first-principles validation to defects, structural disorder, and phase transitions, using these data to refine machine-learned potentials and explore more challenging materials spaces. Incorporating electrochemical switching kinetics and optimising pulse schedules will also be essential for translating the predicted thermal contrast into practical thermal-management and waste-heat-recovery technologies.

\section{Methods}

\subsection{Density functional theory}
\label{subsec:dft_method}

The intercalated 1T and 2H structures are those of the electronic-structure study of Price \textit{et al.}~\cite{Price2023}.
All DFT calculations were performed with VASP~\cite{Kresse1993,Kresse1996,Kresse1996_2} within the projector-augmented-wave (PAW) formalism. The PBE generalised gradient approximation was used for the structures, force constants and thermal transport, and both PBE and the HSE06 hybrid functional~\cite{Heyd2003,Heyd2006} were used for the voltage calculation. A plane-wave cutoff of 700~eV, collinear spin polarisation, and DFT-D3 dispersion corrections were used throughout, with a self-consistent-field (SCF) convergence of $10^{-8}$~eV for PBE and $10^{-6}$~eV for HSE06. The voltage calculations used $\Gamma$-centred Monkhorst--Pack meshes of $6\times6\times6$ for Li$_a$ZrS$_2$ ($a=0$ to $1$ in steps of $0.125$) and $24\times24\times24$ for body-centred-cubic bulk Li with PBE, and $1\times1\times1$ and $4\times4\times4$ respectively with HSE06; the HSE06 lithiated cells are therefore $\Gamma$-point calculations. The force-constant calculations used a $\Gamma$-centred $3\times3\times3$ mesh for the phonopy \& phono3py calculations discussed in \secref{subsec:phonon_method}.

\subsection{Universal machine-learned interatomic potential}

Our screening search was evaluated with the NequIP-OAM-L foundation potential~\cite{NEQUIP} (model NequIP-OAM-L-0.1) through the Atomic Simulation Environment interface. The uMLiP is used only to evaluate forces on the PBE-relaxed geometries obtained previously, so the DFT and uMLiP results differ solely through the potential energy surface and not through structure.

\subsection{Intercalation voltage and capacity}
\label{subsec:voltage_method}

The formation energy per formula unit (f.u.) of each lithiated state Li$_a$ZrS$_2$ is computed relative to the host ZrS$_2$ and a metallic Li reference,
\begin{equation}
E_{\mathrm{form}}(a) = \frac{E_{\mathrm{tot}}(a) - E_{\mathrm{host}} - n_{\mathrm{Li}}\,\mu_{\mathrm{Li}}}{N_{\mathrm{ZrS}_2}},
\label{eq:formation_energy}
\end{equation}
where $E_{\mathrm{tot}}(a)$ is the DFT or uMLiP total energy of the lithiated cell, $E_{\mathrm{host}}$ the total energy of the same number of ZrS$_2$ formula units in the pristine host, $n_{\mathrm{Li}}$ the number of inserted Li atoms, $N_{\mathrm{ZrS}_2}$ the number of ZrS$_2$ formula units in the cell, and $\mu_{\mathrm{Li}}$ the chemical potential of bulk Li metal.

Chemical stability is assessed by constructing the lower convex hull of $E_{\mathrm{form}}(a)$ against Li content $a$. The equilibrium open-circuit voltage between two adjacent hull states $(a_1, a_2)$ follows from the Gibbs free-energy change of the reaction
\begin{equation}
\mathrm{Li}_{a_1}\mathrm{ZrS}_2 + (a_2 - a_1)\,\mathrm{Li}^+ + (a_2 - a_1)\,e^- \rightarrow \mathrm{Li}_{a_2}\mathrm{ZrS}_2.
\end{equation}
At 0~K, neglecting entropic and volumetric contributions and taking energies in eV so that the elementary charge is absorbed, the average intercalation voltage is
\begin{equation}
V(a_1 \rightarrow a_2) = -\frac{E_{\mathrm{form}}(a_2) - E_{\mathrm{form}}(a_1)}{a_2 - a_1},
\label{eq:voltage}
\end{equation}
which yields a step-wise profile in which each plateau is a two-phase coexistence region between adjacent hull states.

The theoretical specific (gravimetric) capacities are obtained from
\begin{equation}
Q_{\mathrm{grav}} = \frac{a\,F}{M_{\mathrm{ZrS}_2}},
\label{eq:specific_capacity}
\end{equation}
where $F$ is the Faraday constant, $M_{\mathrm{ZrS}_2}$ is the molar mass of the pristine host, and is converted to mAh\,g$^{-1}$ by division by 3.6.

The voltage was additionally evaluated with the uMLiP~\cite{NEQUIP}: single-point energies of uMLiP-relaxed Li$_a$ZrS$_2$ cells and of a uMLiP-relaxed body-centred-cubic bulk Li reference (FIRE relaxation to a residual force tolerance of 0.001~eV\,\AA$^{-1}$) were substituted for the VASP total energies in \eqref{eq:formation_energy} and \eqref{eq:voltage}. These uMLiP-relaxed cells are used for the voltages and capacities only, and not for phonon calculations.

\subsection{Phonons and lattice thermal conductivity}
\label{subsec:phonon_method}

For the Li$_a$ZrS$_2$, phonon and thermal-transport calculations were performed with phonopy (2.21.2) and phono3py (4.5.0)~\cite{Phonopy,Phono3py} by the finite-displacement method, using the PBE settings of \secref{subsec:dft_method} for the DFT forces. Every composition is described in a cell containing eight ZrS$_2$ formula units ($2\times2\times2$ supercell). Second- and third-order force constants were obtained from a further $2\times2\times1$ repetition of that cell, with the DFT forces evaluated on a $\Gamma$-centred $3\times3\times3$ Monkhorst--Pack mesh. All structures are the PBE-relaxed geometries.

Thermal conductivity and heat capacity were obtained in the relaxation time approximation (RTA) on a $25\times25\times25$ $q$-mesh over 0--1000~K in 10~K steps, with results quoted at 100--1000~K, and with a boundary mean free path of 1~$\mu$m imposed on every mode as an additional boundary relaxation rate. Three-phonon and boundary scattering are the only relaxation mechanisms considered. Mesh sensitivity was assessed against the same scheme (RTA) on a $10\times10\times10$ mesh (\supptabref{supptab:DLBTEkappa}): refining the mesh raises the pristine orientation-averaged conductivity from 7.30 to 9.67~W~m$^{-1}$~K$^{-1}$ but lowers that of Li$_{0.125}$ZrS$_2$ from 1.31 to 0.88~W~m$^{-1}$~K$^{-1}$.

To account for LO-TO splitting, the non-analytic correction (NAC) was applied to the pristine ZrS$_2$ only. Lithiated compositions are either semi-metallic or metallic, so the macroscopic field is screened by the carriers and the non-analytic term does not apply.

Direct solutions of the linearised Boltzmann transport equation (DLBTE) were obtained for Li$_a$ZrS$_2$. The direct solve stores the full scattering matrix, which exceeds the available memory above a $10\times10\times10$ $q$-mesh, so all DLBTE results are reported on a $10\times10\times10$ mesh; for the smaller $a=0$ (3-atom primitive cell) and $a=1$ (4-atom primitive cell) cells, $15\times15\times15$ results were additionally obtained as a mesh-convergence check. The DLBTE results are consequently $q$-mesh-limited, are reported in \suppsecref{suppsec:dlbte} for completeness, and are not used for any switching metric in the main text.

The fifty-six other systems of \secref{subsec:screening} were treated with the identical uMLiP workflow: 1T and 2H structures from Ref.~\cite{Price2023}, eight formula units per cell with one Li or Mg, force constants from a $2\times2\times1$ repetition of that cell, and RTA conductivities on a $25\times25\times25$ $q$-mesh with the same 1~$\mu$m boundary mean free path.

\subsection{Waste-heat-recovery device model}
\label{subsec:WHRmodel}

To translate the switching contrast into a device-level metric, we model a one-dimensional waste-heat-recovery cell. Per unit cross-sectional area, it consists of four regions in series on a uniform grid, inset of \subfigref{subfig:WHR_eta}.
The source generates rectangular heat pulses, each consisting of energy $E_{\mathrm{pulse}}$, and is anchored to a fixed-temperature reservoir that removes heat between pulses. Heat transport is classical Fourier conduction with temperature-independent thermal conductivity $\kappa$ and volumetric heat capacity $C_V$,
\begin{equation}
C_V\,\frac{\partial T}{\partial t} = \frac{\partial}{\partial x}\left(\kappa\,\frac{\partial T}{\partial x}\right) + Q(x,t),
\label{eq:WHRFourier}
\end{equation}
where $Q(x,t)$ is the pulsed volumetric heat input to the source region. The switch layer conducts with $\kappa^{\mathrm{on}}$ (high conductivity) or $\kappa^{\mathrm{off}}$ (low conductivity) depending on the direction of the temperature gradient across it, so the switch opens only while the gradient across it is large enough to drive heat into the heat store. The gate is directional and is only on when the power source region is hotter than the store temperature. So all backflow out of the storage occurs through the off-state conductivity.

\eqref{eq:WHRFourier} is solved with a conservative cell-centred finite-volume discretisation on a uniform grid of spacing $\Delta x$, using the harmonic mean of the two neighbouring cells at every interface~\cite{Mclean2026},
\begin{equation}
\frac{C_V\,\Delta x}{\Delta t}\left(T_i^{n+1} - T_i^{n}\right) = \sum_{j} g_{ij}\left(T_j^{n+1} - T_i^{n+1}\right) + Q_i^{n+1}\Delta x, \qquad g_{ij} = \frac{2\kappa_i\kappa_j}{(\kappa_i+\kappa_j)\Delta x},
\label{eq:WHRscheme}
\end{equation}
with backward-Euler (implicit) time stepping, a Dirichlet face at the reservoir ($g_{\mathrm{b}} = 2\kappa_{\mathrm{source}}/\Delta x$ against $T_{\mathrm{res}}$) and an adiabatic face at the far end of the sink. All three regions share the same volumetric heat capacity, $C_V = 1.69\times10^{6}$~J~m$^{-3}$~K$^{-1}$, the average of the calculated ZrS$_2$ and Li$_{0.125}$ZrS$_2$ values at 300~K. Because all properties are constant, the tridiagonal systems for the two switch states are fixed and are LU-factorized once per run, so every time step is a single back- and forward-substitution. The scheme is unconditionally stable and conserves energy exactly.

The energy audit per unit area tracks the total injected pulse energy $E_{\mathrm{pulse}}^{\mathrm{tot}}$, the heat drained through the fixed-temperature reservoir $E_{\mathrm{out}}$, the internal-energy change of each region, and the signed heat flux $q_{\mathrm{str}}$ at the store-side face of the switch layer, accumulated separately for the on and off states. Consistent with the backflow-only loss convention, the leakage is the heat returned from the storage through the off-state conductivity,
\begin{equation}
E_{\mathrm{backflow}} = \int \max\left(0, -q_{\mathrm{str}}\right)\mathrm{d}t ,
\label{eq:WHRbackflow}
\end{equation}
and heat seeping into the store through the off state counts as recovered. The recovery metrics are the capture efficiency and its ratio to the ideal switch ($\kappa^{\mathrm{off}} = 0$),
\begin{equation}
\eta = \frac{\Delta U_{\mathrm{str}}}{E_{\mathrm{pulse}}^{\mathrm{tot}}}, \qquad \eta_{\mathrm{rel}} = \frac{\eta}{\eta_{\mathrm{ideal}}},
\label{eq:WHReta}
\end{equation}
where $\Delta U_{\mathrm{str}}$ is the change in internal energy of the heat storage. The ideal switch takes the on-state conductivity of the cross-plane DFT switch of this work ($\kappa^{\mathrm{on}} = 3.73$~W~m$^{-1}$~K$^{-1}$), so every system is normalised against one common ceiling. To check simulation accuracy, two independent closure identities, $E_{\mathrm{pulse}}^{\mathrm{tot}} - E_{\mathrm{out}} = \Delta U_{\mathrm{source}} + \Delta U_{\mathrm{switch}} + \Delta U_{\mathrm{str}}$ and $\int q_{\mathrm{str}}\,\mathrm{d}t = \Delta U_{\mathrm{str}}$, are tracked, and show values smaller than $1.2\times10^{-10}$ in relative terms for every system.

\tabref{tab:WHRparams} lists the parameters, which are identical for every material; only $\kappa^{\mathrm{on}}$ and $\kappa^{\mathrm{off}}$ of the switch layer change between systems.

\begin{table}[htbp]
\centering
\caption{Waste-heat-recovery device parameters.}
Values used for every material simulated in \secref{subsec:WasteHeat}; only the on- and off-state conductivities of the switch layer vary between systems.
\begin{tabular}{ll}
\hline
Quantity & Value \\
\hline
Region thicknesses (source, switch, store) & 100, 5, 200~$\mu$m \\
Grid spacing $\Delta x$ (610 nodes) & 0.5~$\mu$m \\
Host conductor $\kappa_{\mathrm{source}}=\kappa_{\mathrm{store}}$ & 50~W~m$^{-1}$~K$^{-1}$ \\
Volumetric heat capacity $C$ (all regions) & $1.69\times10^{6}$~J~m$^{-3}$~K$^{-1}$ \\
Initial and reservoir temperature & 300~K \\
Pulse energy per unit area & $2\times10^{4}$~J~m$^{-2}$ ($\approx$118~K bulk rise of the source) \\
Pulse period, duration, number & 1~ms, 50~$\mu$s, 20 (20~ms total) \\
Time step $\Delta t$ & $10^{-7}$~s \\
Gate threshold $\Delta T_{\mathrm{crit}}$ & 10~K ($2\times10^{6}$~K~m$^{-1}$ across the switch layer) \\
Hysteresis band $\Delta T_{\mathrm{hyst}}$ & 0 \\
Total injected energy per unit area & $4.00\times10^{5}$~J~m$^{-2}$ \\
\hline
\end{tabular}
\label{tab:WHRparams}
\end{table}
\clearpage

\section{Data availability}
The data supporting the findings of this study, including the relaxed structures of all 57 candidate structures, can be found in the figshare repository~\cite{Figshare}.

\section{Code availability}
The code implementing the waste-heat-recovery device model is publicly available under the GNU General Public License version 3 (GPLv3) in a figshare repository~\cite{Figshare}.

\begin{acknowledgments}
This study was funded by the Engineering and Physical Sciences Research Council (EPSRC): H.M.\ under studentship EPSRC-690010152 and S.P.H.\ under grant EP/X013375/1. The authors acknowledge the use of resources provided by the Isambard 3 Tier-2 High Performance Computing Facility, hosted by the University of Bristol, operated by the GW4 Alliance (https://gw4.ac.uk) and funded by UK Research and Innovation and the EPSRC under grant EP/X039137/1.
\end{acknowledgments}

\section*{Author contributions}
H.M.\ conceived and performed the calculations, analysed the data, prepared the figures, wrote the code, and drafted \& revised the manuscript. S.P.H.\ supervised the project, obtained resources, contributed to the analysis, and revised the manuscript. Both authors read and approved the final manuscript.

\section*{Competing interests}
The authors declare no financial or non-financial competing interests.



\clearpage
\section*{Supplementary Information}
\setcounter{section}{0}
\renewcommand{\thesection}{S\arabic{section}}
\renewcommand{\thesubsection}{\thesection.\arabic{subsection}}
\setcounter{figure}{0}
\renewcommand{\thefigure}{S\arabic{figure}}
\setcounter{table}{0}
\renewcommand{\thetable}{S\arabic{table}}

\section{Screening of the dichalcogenide hosts}
\label{suppsec:dtmdc}

This section presents the results of our screening of the 57 candidates summarized in the main text. Twenty-six transition-metal dichalcogenide (TMDC) hosts, in the octahedrally coordinated (1T) and trigonal-prismatic (2H) phases, were each examined in the pristine state (the ``on'' state) and at 12.5\% intercalant filling (the ``off'' state).  Each was considered with  either Li or Mg, giving twenty-six pristine--intercalated pairs (43 distinct structures, since the pristine cells are shared between the Li and Mg pairs of a host). Li$_{0.125}$ZrS$_2$, the subject of the main study, is the twenty-seventh pair and is included in \tabref{supptab:DTMDCscreen} and in the Li panel of the main-text. Every calculation uses the universal machine-learned interatomic potential (uMLiP, NequIP-OAM-L~\cite{NEQUIP}) on the DFT-relaxed geometries, and solves the Boltzmann transport equation in the relaxation time approximation (BTE-RTA) on a $25\times25\times25$ $q$-mesh from a $2\times2\times1$ supercell with a 1~$\mu$m boundary mean free path, over 100--1000~K in 50~K steps; \tabref{supptab:DTMDCscreen} collects the 300~K results. In this table the pristine and intercalated members of each pair are labelled ``on'' and ``off''; in the deposited data and figure file names they are labelled D0 and D1 respectively.

All 43 structures are dynamically stable at the harmonic level: the most negative frequency anywhere in the set is $\approx-$0.01~THz which we consider within the numerical accuracy of the approach. The $25\times25\times25$ anharmonic runs of the two pristine CrSe$_2$ and VSe$_2$ hosts (shared by their Li and Mg pairs) do develop small imaginary frequencies at isolated $q$-points, down to $\approx-0.09$~THz, so the conductivities of the four pairs built on these six distinct structures carry additional uncertainty; all intercalated runs remain stable. In \tabref{supptab:DTMDCscreen}, the $^{\dagger}$ row is Li$_{0.125}$ZrS$_2$, the material studied with DFT in the main text, whose cross-plane conductivities there are 3.73 and 0.046~W~m$^{-1}$~K$^{-1}$ ($S=0.99$); the entry in the table is the uMLiP value used in the screen.

In \tabref{supptab:DTMDCscreen}, $\kappa_{zz}$ and $\kappa_{xx}$ are the cross-plane and in-plane lattice thermal conductivities of the pristine (on) and 12.5\%-intercalated (off) states; $\Delta\kappa_{\perp}/\kappa_{\perp}^{\mathrm{on}}=(\kappa_{\perp}^{\mathrm{on}}-\kappa_{\perp}^{\mathrm{off}})/\kappa_{\perp}^{\mathrm{on}}$ is the signed fractional change in cross-plane conductivity (a negative value means that intercalation raises the cross-plane conductivity; its absolute value is the switch score $S$ of the main text); $C_v$ is the mean volumetric heat capacity of the two states; and $\alpha_{zz}=(\kappa_{\perp}^{\mathrm{on}}/C_v^{\mathrm{on}}+\kappa_{\perp}^{\mathrm{off}}/C_v^{\mathrm{off}})/2$ is the state-averaged cross-plane thermal diffusivity, the mean of the on- and off-state diffusivities. All values are from the uMLiP (BTE-RTA, $25\times25\times25$ $q$-mesh, 1~$\mu$m boundary mean free path); conductivities are in W~m$^{-1}$~K$^{-1}$, heat capacities in J~K$^{-1}$~cm$^{-3}$ and diffusivities in mm$^{2}$~s$^{-1}$. Rows are grouped by host phase and, within each phase, ordered by decreasing $|\Delta\kappa_{\perp}/\kappa_{\perp}^{\mathrm{on}}|$.

\begin{table*}[h!]
\centering
\scriptsize
\renewcommand{\arraystretch}{1.0}
\caption{Screening of intercalated transition-metal dichalcogenides at 300~K.
Cross-plane and in-plane lattice thermal conductivities (on: pristine; off: 12.5\% intercalated), the signed fractional change in cross-plane conductivity (negative values, shown as $-$, mark hosts whose cross-plane conductivity rises on intercalation), the mean volumetric heat capacity $C_v$ of the two states and the state-averaged cross-plane thermal diffusivity $\alpha_{\perp}$, all from the uMLiP (BTE-RTA); definitions and ordering as described in the text.}
\begin{tabular}{llcccccccc}
\hline
Phase & Host & Ion & $\kappa_{\perp}^{\mathrm{on}}$ & $\kappa_{\perp}^{\mathrm{off}}$ & $\Delta\kappa_{\perp}/\kappa_{\perp}^{\mathrm{on}}$ & $\kappa_{\parallel}^{\mathrm{on}}$ & $\kappa_{\parallel}^{\mathrm{off}}$ & $C_v$ & $\alpha_{\perp}$ \\
\hline
1T & ZrS$_2^{\dagger}$ & Li & 3.95 & 0.165 & +0.958 & 11.9 & 3.2 & 1.688 & 1.24 \\
1T & TiS$_{2}$ & Li & 1.96 & 0.187 & +0.905 & 6.1 & 3.0 & 1.968 & 0.55 \\
1T & VS$_{2}$ & Li & 0.69 & 0.069 & +0.899 & 3.2 & 0.6 & 2.195 & 0.17 \\
1T & ZrSe$_{2}$ & Li & 2.20 & 0.323 & +0.853 & 7.4 & 3.8 & 1.581 & 0.81 \\
1T & HfS$_{2}$ & Mg & 3.11 & 0.534 & +0.828 & 11.8 & 4.2 & 1.732 & 1.07 \\
1T & NiSe$_{2}$ & Mg & 4.66 & 0.875 & +0.812 & 13.7 & 3.4 & 2.200 & 1.24 \\
1T & CoS$_{2}$ & Li & 1.87 & 0.377 & +0.798 & 9.6 & 4.3 & 2.539 & 0.44 \\
1T & NiS$_{2}$ & Li & 4.68 & 0.963 & +0.794 & 25.6 & 13.3 & 2.459 & 1.13 \\
1T & CoS$_{2}$ & Mg & 1.87 & 0.393 & +0.789 & 9.6 & 4.3 & 2.490 & 0.45 \\
1T & NiS$_{2}$ & Mg & 4.68 & 1.002 & +0.786 & 25.6 & 8.6 & 2.431 & 1.14 \\
1T & ZrSe$_{2}$ & Mg & 2.19 & 0.538 & +0.755 & 7.4 & 3.1 & 1.582 & 0.87 \\
1T & CoSe$_{2}$ & Mg & 1.89 & 0.503 & +0.734 & 5.5 & 3.6 & 2.222 & 0.53 \\
1T & HfSe$_{2}$ & Mg & 2.20 & 0.625 & +0.715 & 7.0 & 2.9 & 1.621 & 0.88 \\
1T & TiS$_{2}$ & Mg & 1.97 & 0.626 & +0.682 & 6.1 & 3.0 & 1.960 & 0.66 \\
1T & TiSe$_{2}$ & Mg & 0.98 & 0.512 & +0.475 & 4.5 & 2.2 & 1.815 & 0.41 \\
\hline
2H & VSe$_{2}$ & Mg & 0.21 & 0.651 & $-$2.149 & 13.6 & 8.5 & 1.920 & 0.22 \\
2H & WSe$_{2}$ & Mg & 1.65 & 0.481 & +0.708 & 36.9 & 14.1 & 1.893 & 0.56 \\
2H & TiSe$_{2}$ & Li & 0.94 & 0.292 & +0.688 & 9.6 & 5.7 & 1.783 & 0.35 \\
2H & WSe$_{2}$ & Li & 1.65 & 0.552 & +0.665 & 36.9 & 16.1 & 1.922 & 0.58 \\
2H & CrS$_{2}$ & Li & 0.88 & 0.381 & +0.569 & 40.1 & 15.4 & 2.158 & 0.29 \\
2H & VS$_{2}$ & Mg & 1.93 & 0.847 & +0.561 & 30.3 & 13.0 & 2.090 & 0.66 \\
2H & WS$_{2}$ & Li & 0.89 & 0.519 & +0.415 & 74.2 & 31.9 & 2.000 & 0.35 \\
2H & VSe$_{2}$ & Li & 0.21 & 0.277 & $-$0.342 & 13.6 & 9.6 & 1.931 & 0.12 \\
2H & WS$_{2}$ & Mg & 0.88 & 1.114 & $-$0.259 & 74.2 & 30.8 & 1.976 & 0.51 \\
2H & CrSe$_{2}$ & Li & 0.49 & 0.439 & +0.109 & 20.3 & 9.9 & 2.021 & 0.23 \\
2H & CrS$_{2}$ & Mg & 0.88 & 0.973 & $-$0.100 & 40.1 & 15.4 & 2.139 & 0.43 \\
2H & CrSe$_{2}$ & Mg & 0.49 & 0.523 & $-$0.062 & 20.3 & 7.4 & 2.009 & 0.25 \\
\hline
\end{tabular}
\label{supptab:DTMDCscreen}
\end{table*}

\subsection{Heat capacities and cross-plane switching fractions}
\label{suppsec:dtmdc_cv}

\begin{figure*}[h!]
    \centering
    \includegraphics[width=0.9\linewidth]{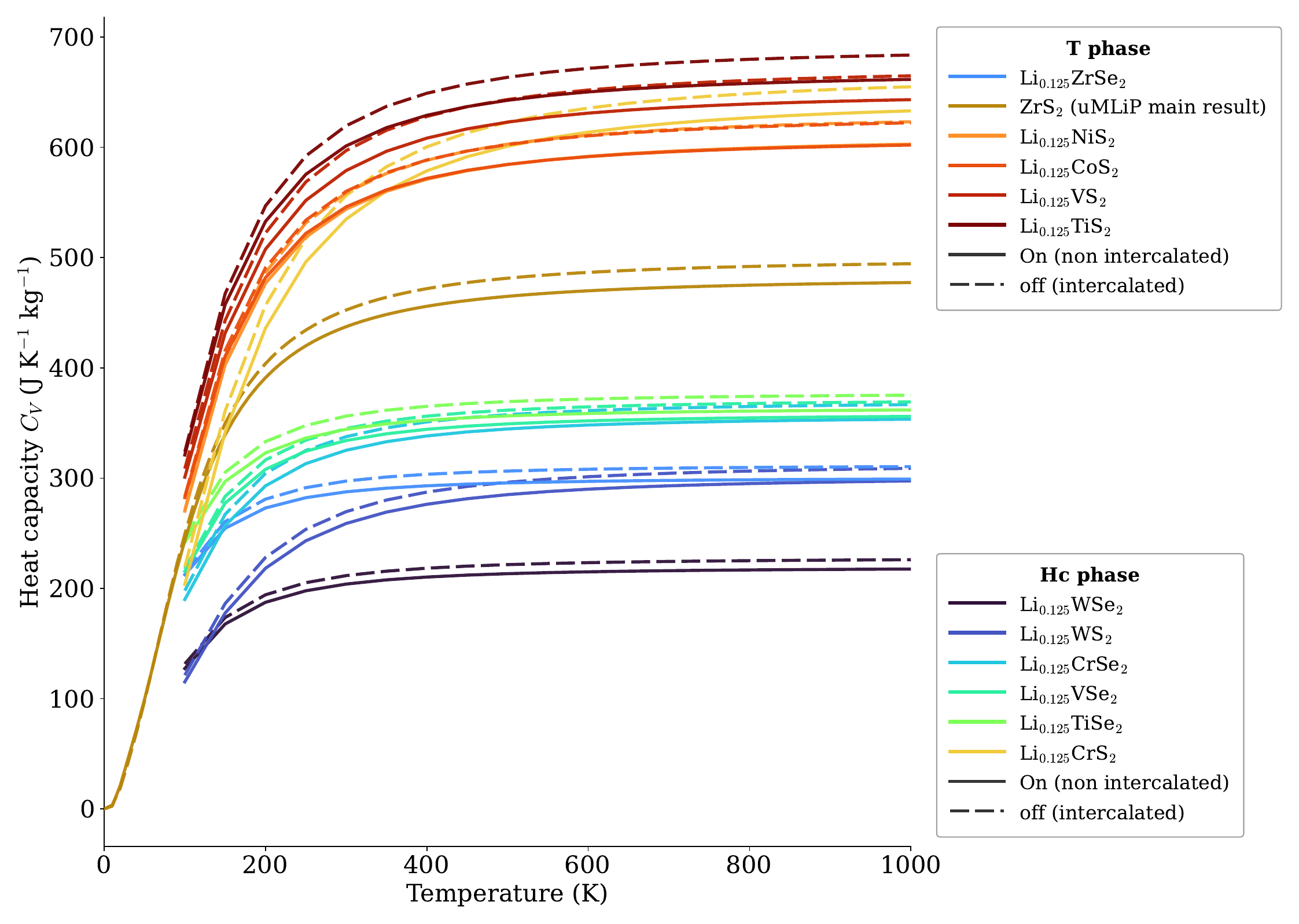}
    \caption{Specific heat capacity of the Li-intercalated screen. Specific heat capacity $C_V$ per unit mass versus temperature for the screened transition-metal dichalcogenides intercalated with Li to 12.5\% (dashed curves) and for their pristine hosts (solid curves). Each material is drawn in a different colour, and the legend is grouped into 1T- and 2H-phase blocks ordered by mean heat capacity.}
    \label{suppfig:DTMDCCvLi}
\end{figure*}

\begin{figure*}[h!]
    \centering
    \includegraphics[width=0.9\linewidth]{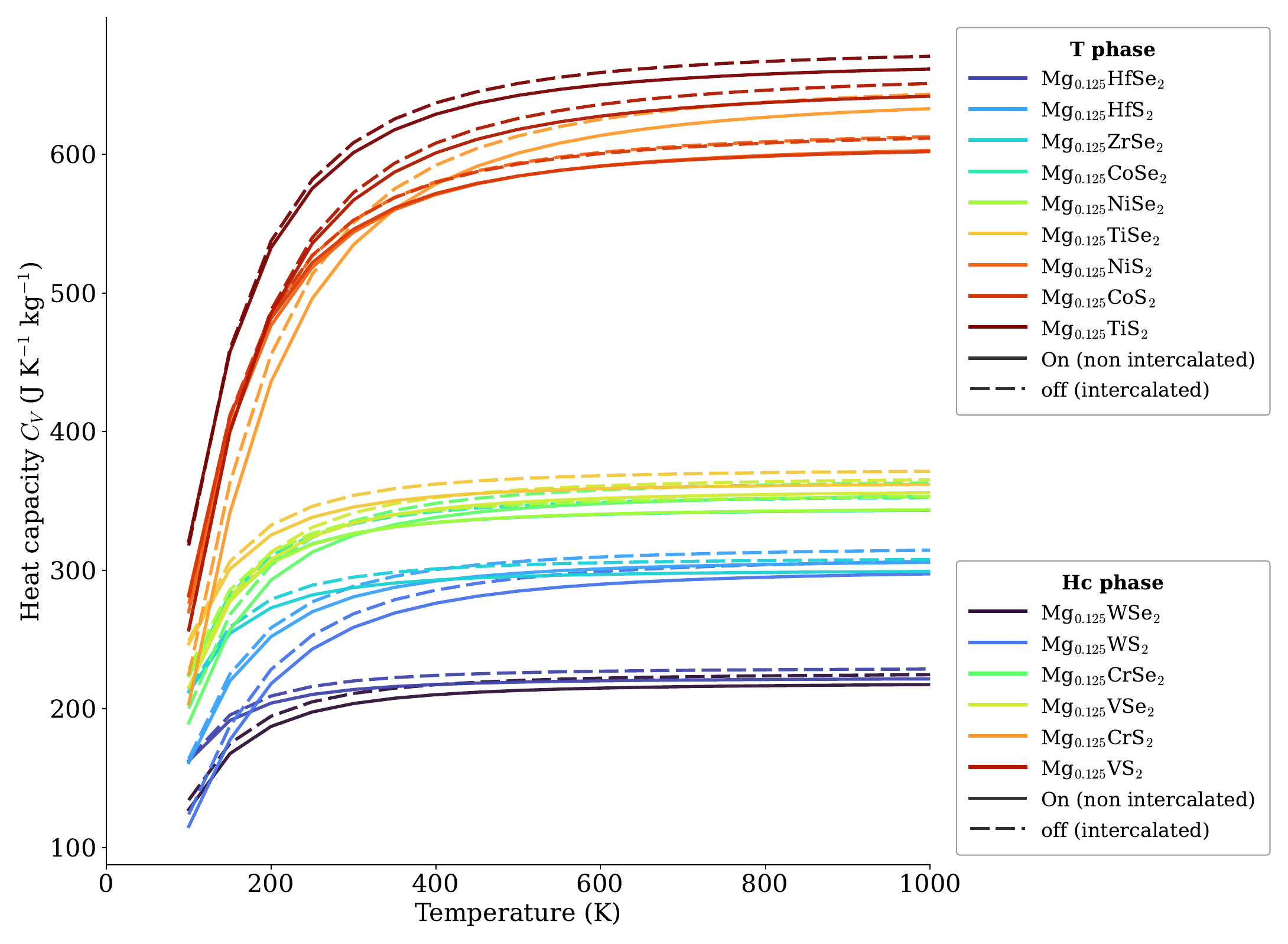}
    \caption{Specific heat capacity of the Mg-intercalated screen. As \figref{suppfig:DTMDCCvLi}, but for the transition-metal dichalcogenides intercalated with Mg to 12.5\%.}
    \label{suppfig:DTMDCCvMg}
\end{figure*}

\figref{suppfig:DTMDCCvLi} and \figref{suppfig:DTMDCCvMg} show the specific heat capacity per unit mass as a function of temperature for the Li- and Mg-intercalated families respectively, obtained from the mode-resolved heat capacities of the BTE-RTA runs. Solid curves are the pristine hosts and dashed curves the intercalated structures; each material is drawn in a different colour, and the materials are ordered by mean heat capacity within the 1T and 2H blocks.

\begin{figure*}[h!]
    \centering
    \includegraphics[width=0.9\linewidth]{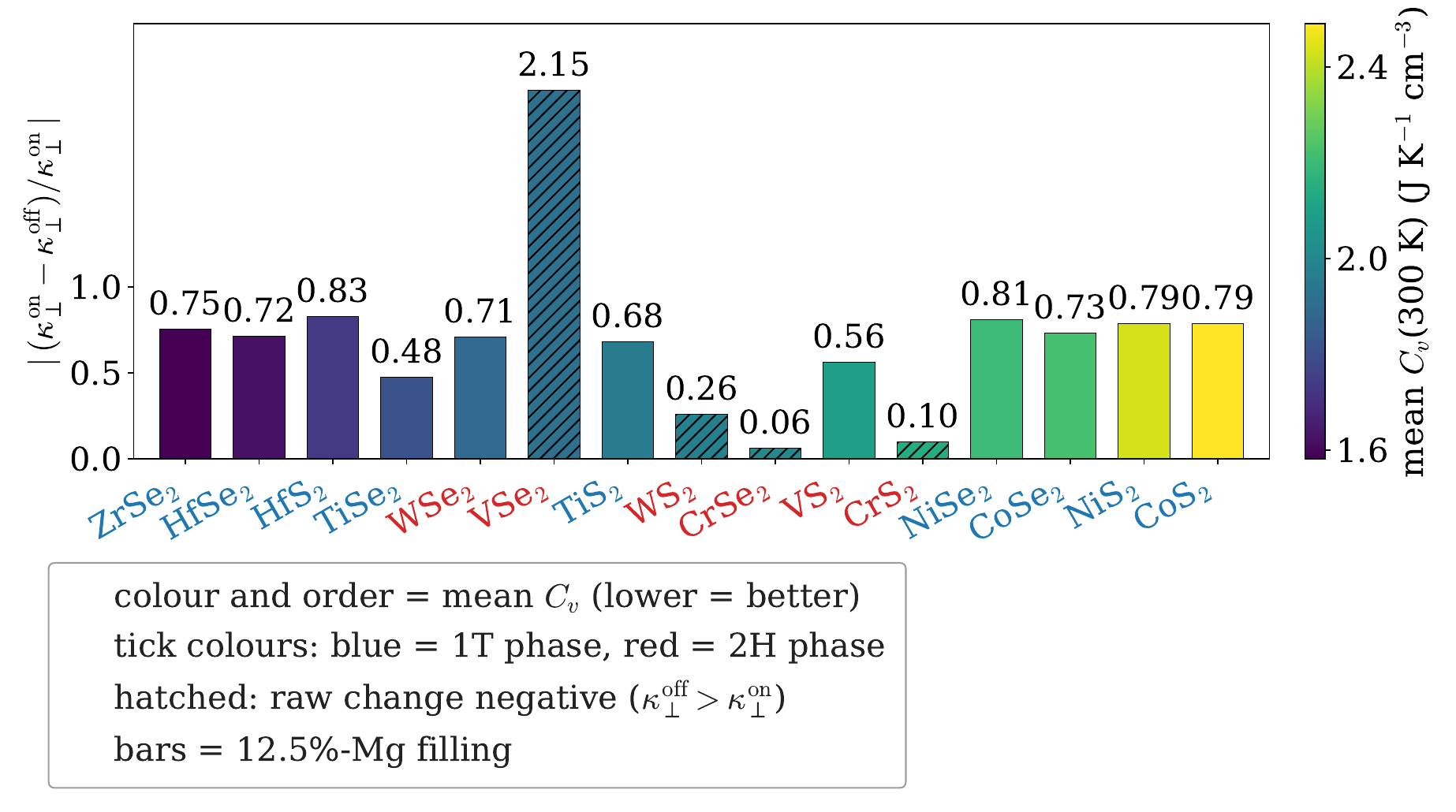}
    \caption{Cross-plane switching fractions of the Mg-intercalated screen. Absolute value of the cross-plane switch score $S$ at 300~K for the screened transition-metal dichalcogenides intercalated with Mg to 12.5\%. Bars are ordered along the horizontal axis by increasing mean volumetric heat capacity and are filled from a sequential colour scale that encodes that heat capacity, running from dark purple (low) through blue and green to yellow (high). Hatched bars mark the four pairs whose cross-plane conductivity rises on intercalation, i.e.\ $\kappa_{\perp}^{\mathrm{off}}>\kappa_{\perp}^{\mathrm{on}}$. Tick labels are coloured blue for 1T hosts and red for 2H hosts.}
    \label{suppfig:DTMDCKappaMg}
\end{figure*}

\figref{suppfig:DTMDCKappaMg} shows the corresponding cross-plane switching fractions at 12.5\% Mg filling. Four of the fifteen Mg pairs (Mg$_{0.125}$VSe$_2$, Mg$_{0.125}$CrS$_2$, Mg$_{0.125}$CrSe$_2$ and Mg$_{0.125}$WS$_2$) switch negatively, i.e.\ their intercalated cross-plane conductivity exceeds the pristine one; together with Li$_{0.125}$VSe$_2$ these are the only five negative switches in the set, and all five are 2H hosts.

\newpage
\subsection{Phonon dispersions of the screened hosts}
\label{suppsec:dtmdc_bands}

In this section we present the phonon dispersions and total densities of states  of stable phonon systems explored by our screening approach. The materials are organized into families that share a bulk host, and each figure combines the on-state panel with the off-state panels for Li and, where calculated, Mg filling.


\begin{figure*}[h!]
    \centering
    \subfloat[]{\includegraphics[width=0.5\linewidth]{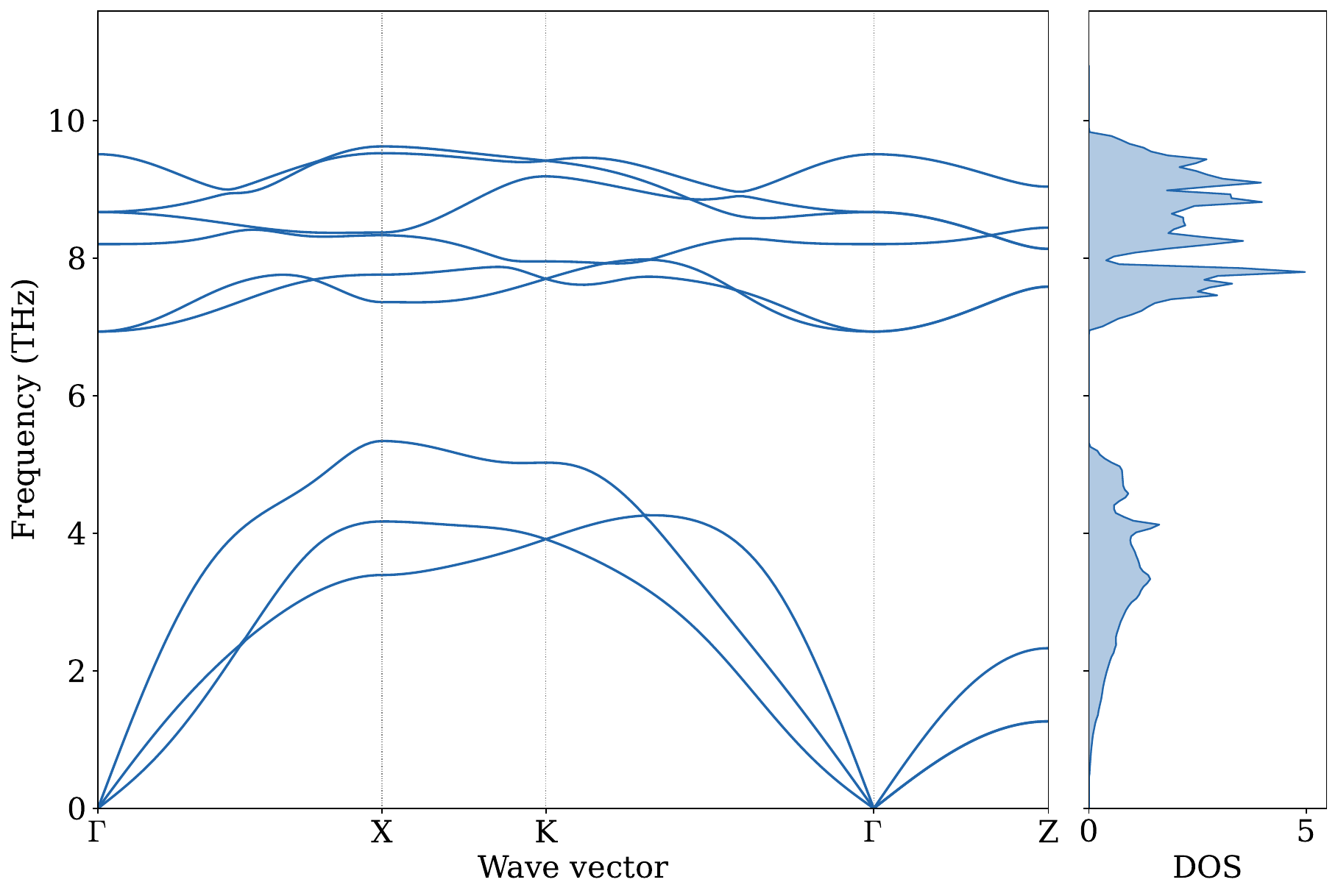}\label{subfig:DTMDC_T_DCo_S2_Li_D0}}
    \subfloat[]{\includegraphics[width=0.5\linewidth]{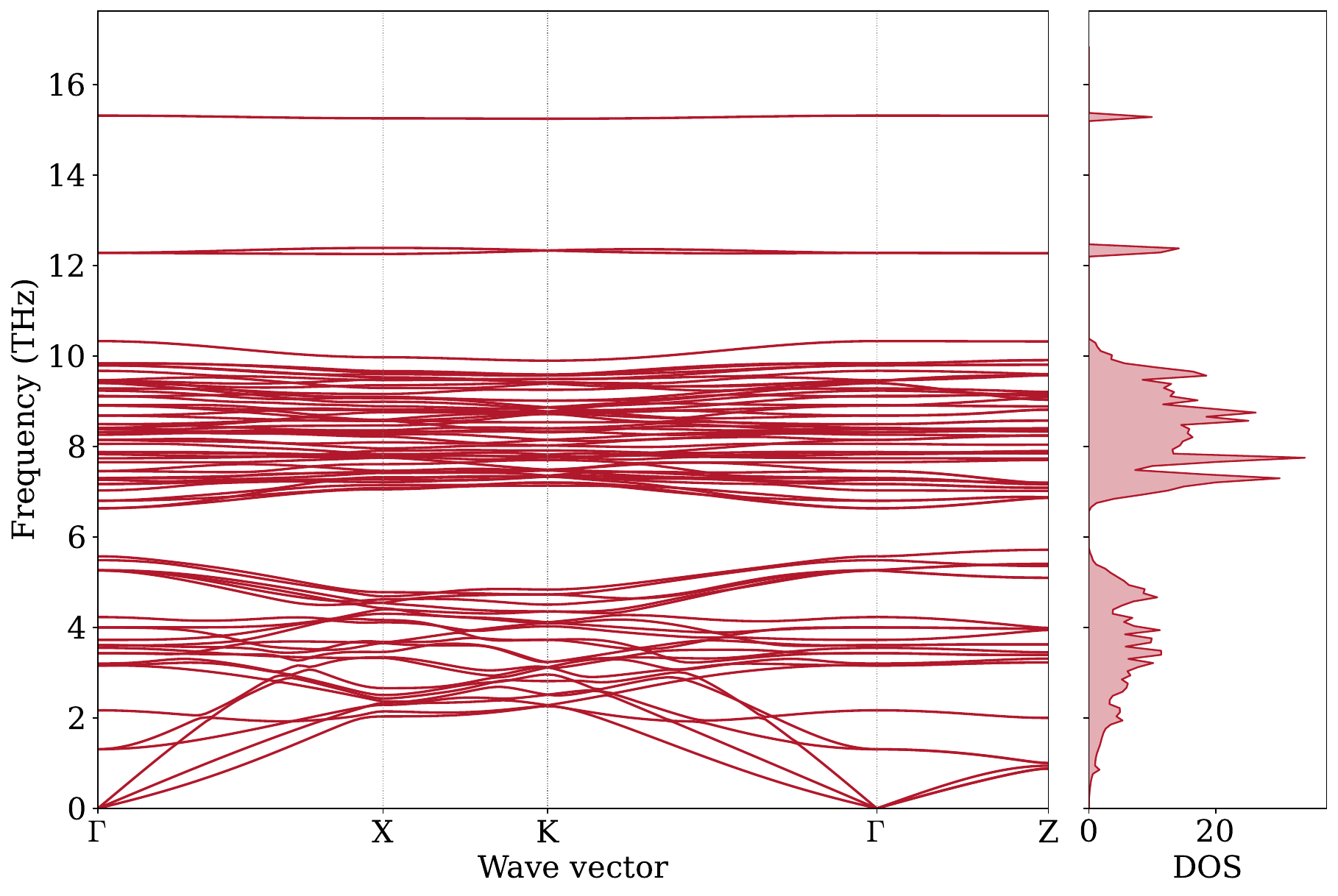}\label{subfig:DTMDC_T_DCo_S2_Li_D1}}
    \hfill
    \subfloat[]{\includegraphics[width=0.5\linewidth]{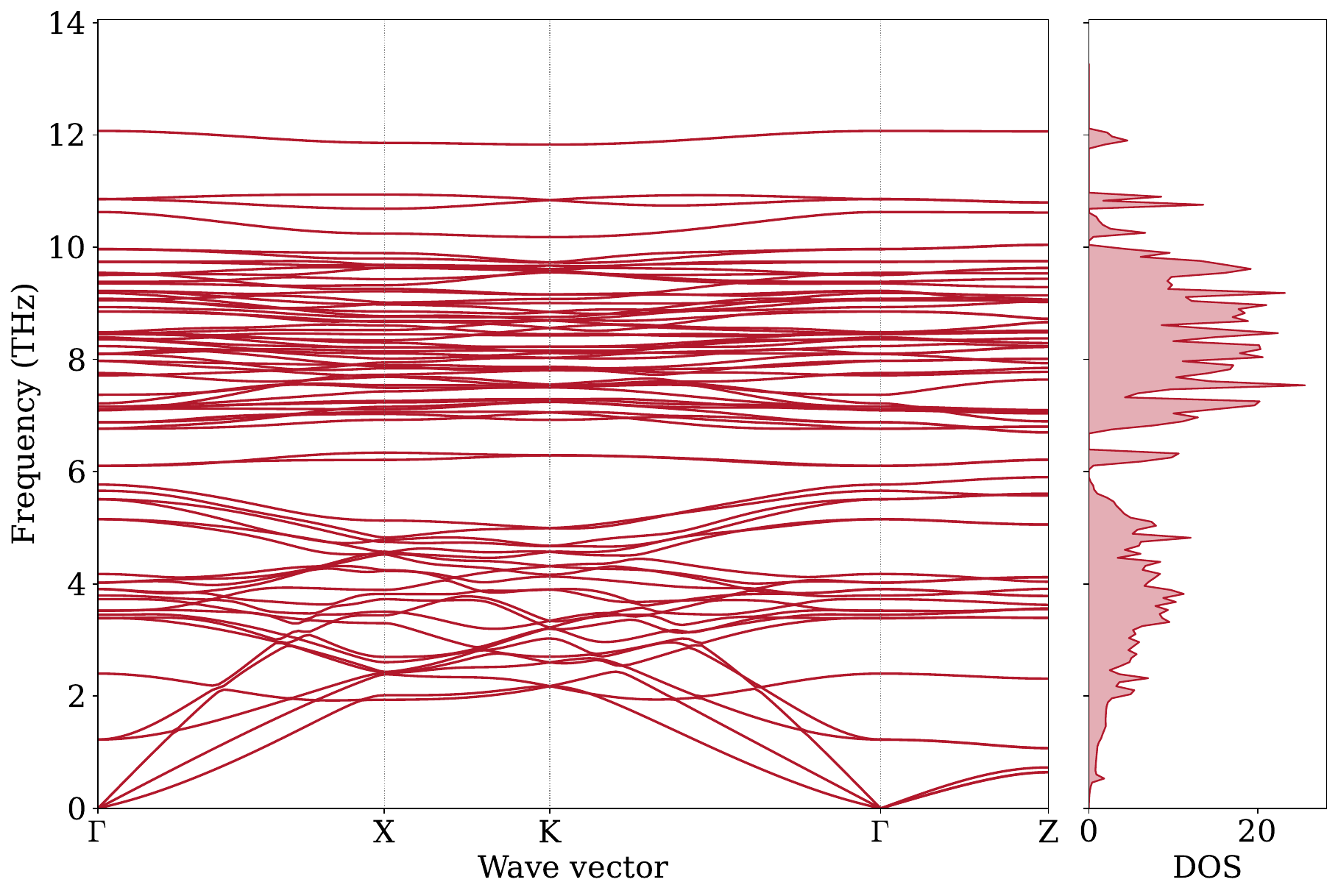}\label{subfig:DTMDC_T_DCo_S2_Mg_D1}}
    \caption{Phonon dispersion and total density of states (DOS) for the CoS$_2$ family: (a)~bulk CoS$_2$ (shared by both intercalation runs); (b)~Li$_{0.125}$CoS$_2$; (c)~Mg$_{0.125}$CoS$_2$.}
    \label{suppfig:DTMDCT_CoS}
\end{figure*}

\begin{figure*}[h!]
    \centering
    \subfloat[]{\includegraphics[width=0.5\linewidth]{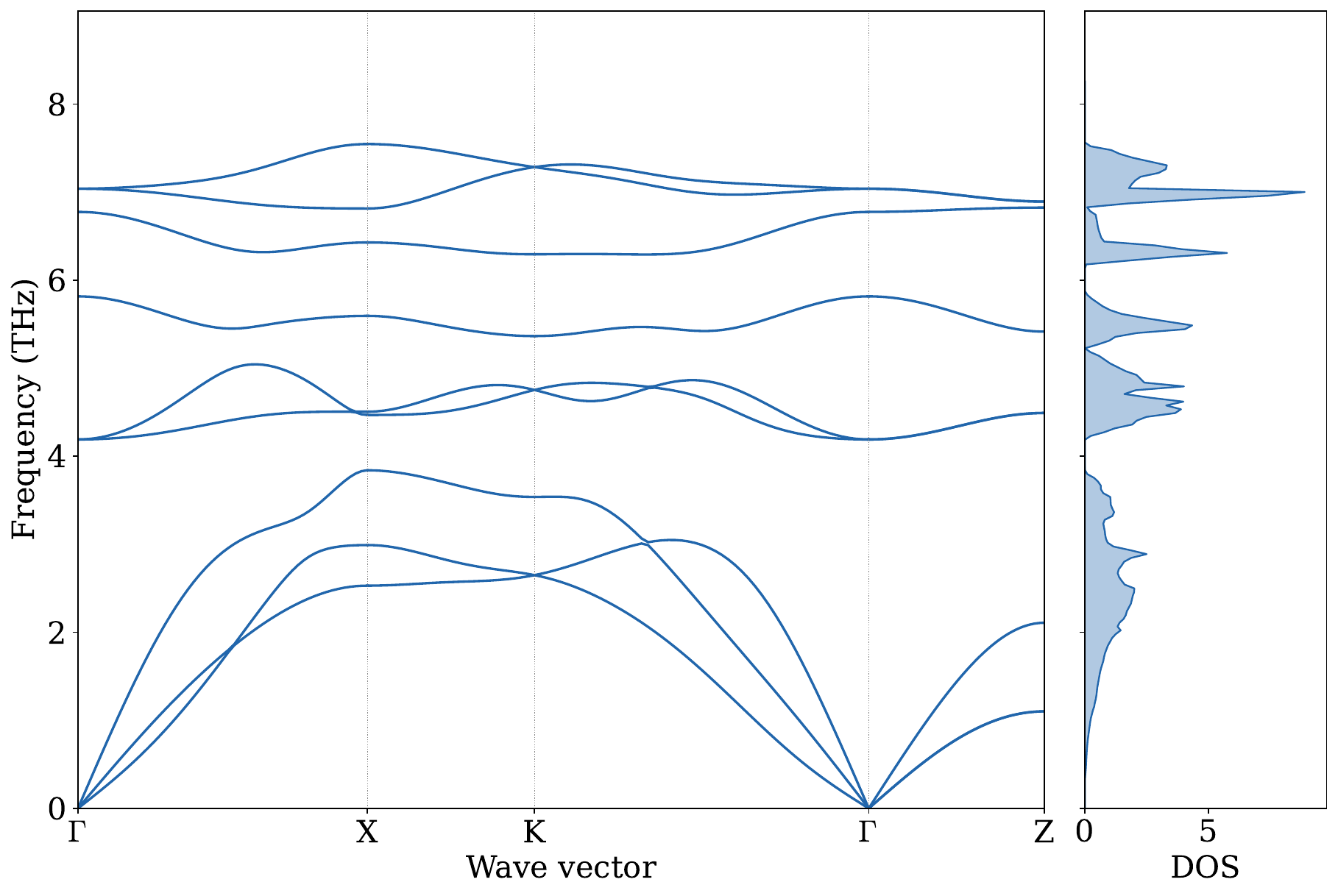}\label{subfig:DTMDC_T_DCo_Se2_Mg_D0}}
    \subfloat[]{\includegraphics[width=0.5\linewidth]{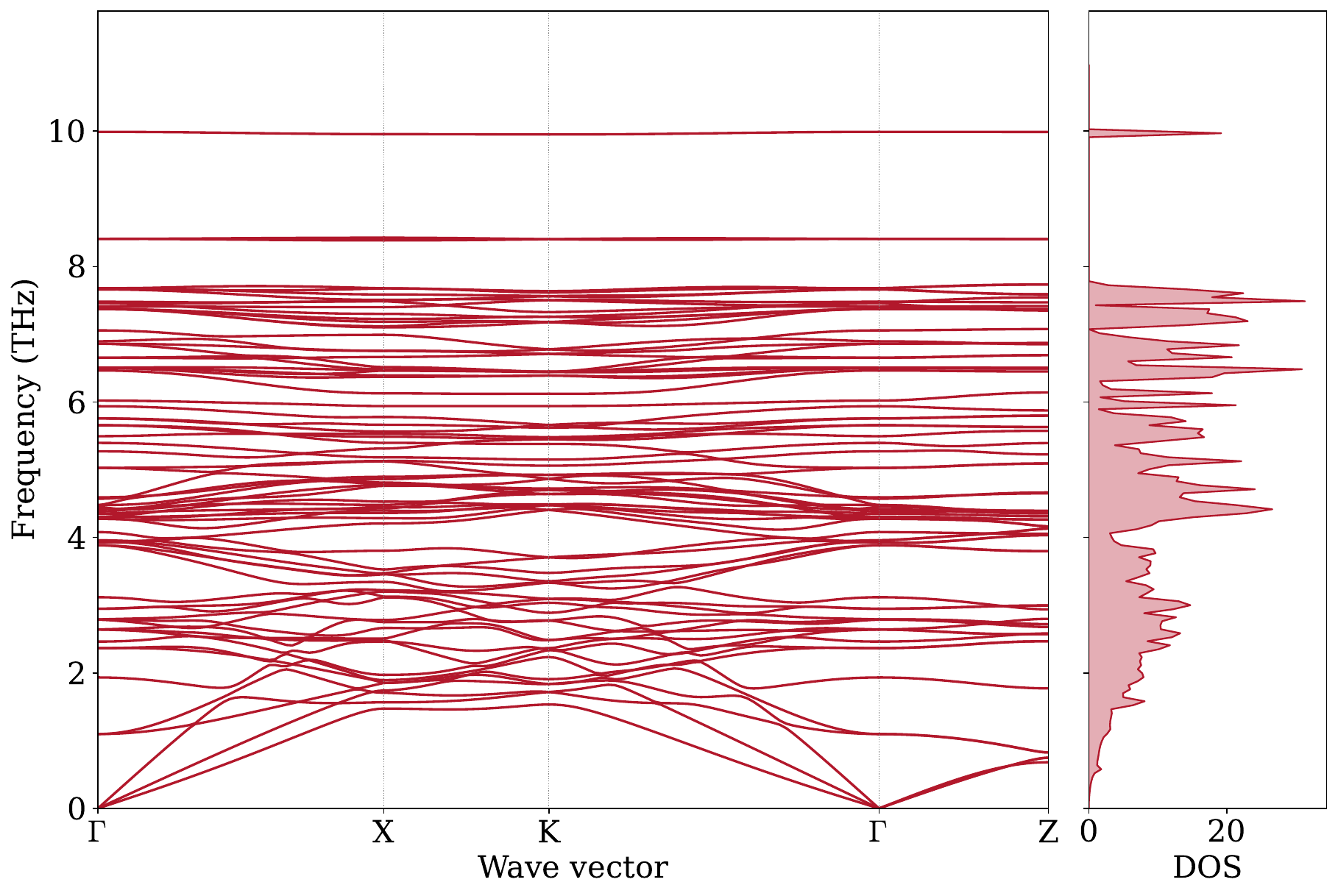}\label{subfig:DTMDC_T_DCo_Se2_Mg_D1}}
    \caption{Phonon dispersion and total DOS for the CoSe$_2$ family: (a)~bulk CoSe$_2$; (b)~Mg$_{0.125}$CoSe$_2$.}
    \label{suppfig:DTMDCT_CoSe}
\end{figure*}

\begin{figure*}[h!]
    \centering
    \subfloat[]{\includegraphics[width=0.5\linewidth]{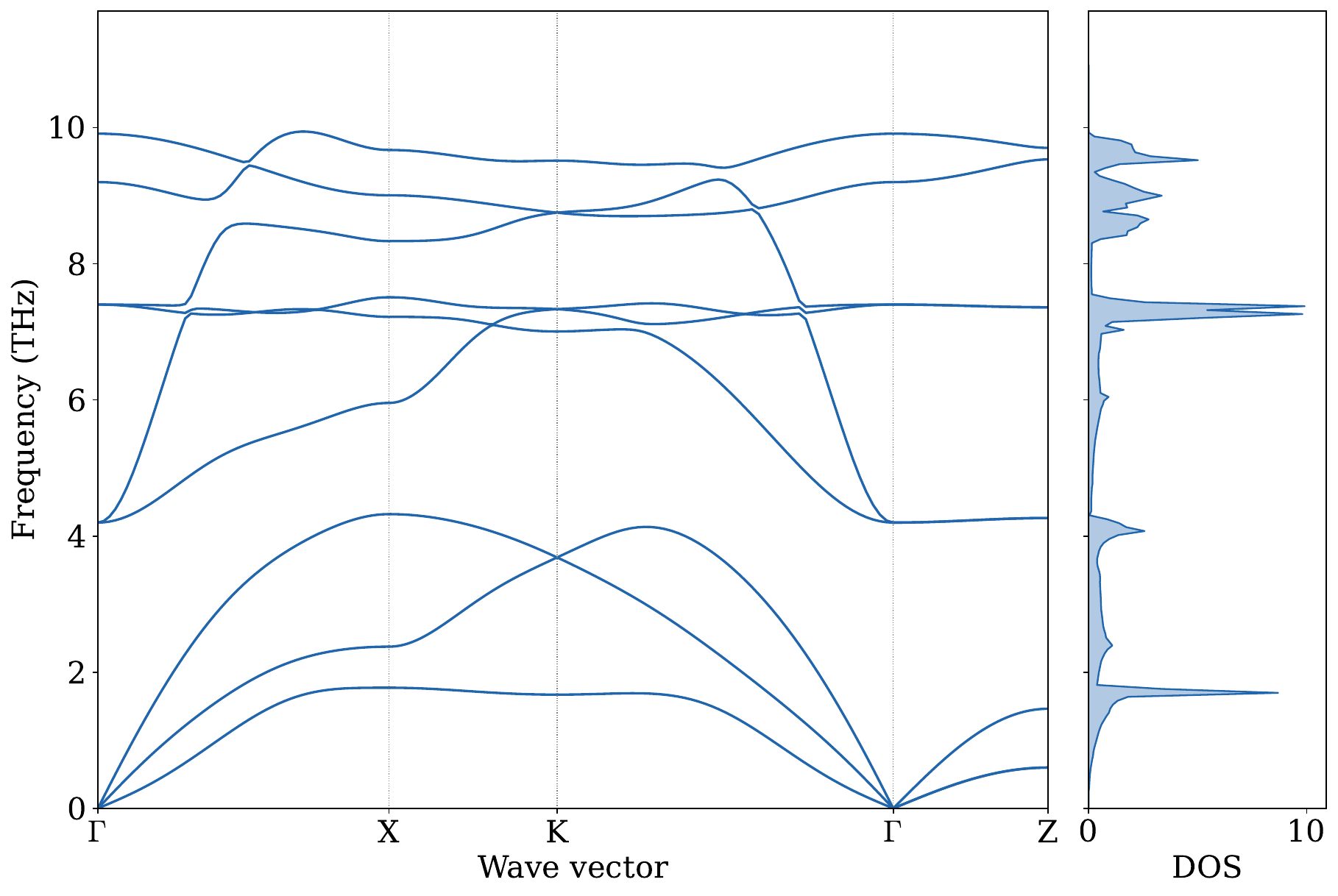}\label{subfig:DTMDC_T_DHf_S2_Mg_D0}}
    \subfloat[]{\includegraphics[width=0.5\linewidth]{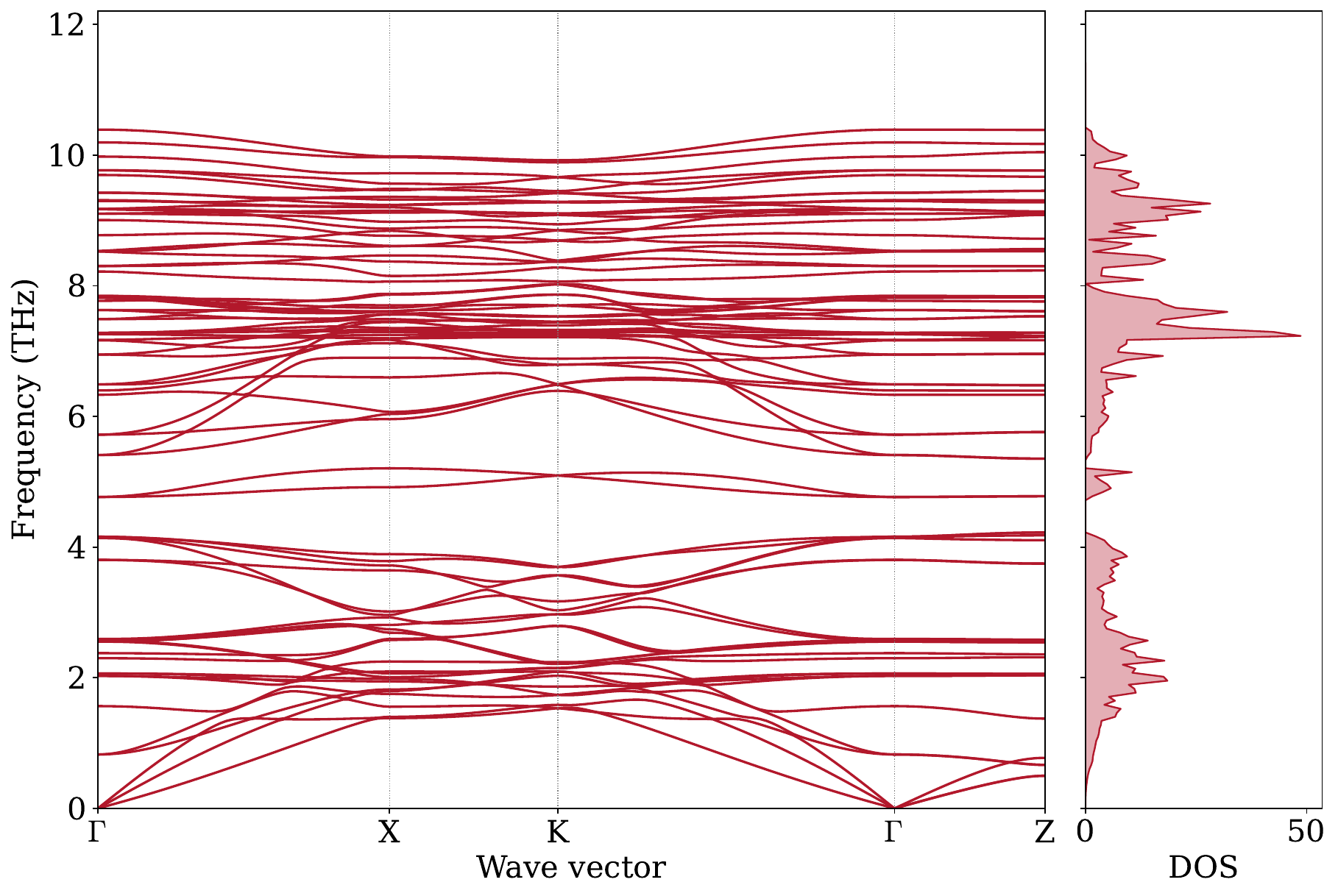}\label{subfig:DTMDC_T_DHf_S2_Mg_D1}}
    \caption{Phonon dispersion and total DOS for the HfS$_2$ family: (a)~bulk HfS$_2$; (b)~Mg$_{0.125}$HfS$_2$.}
    \label{suppfig:DTMDCT_HfS}
\end{figure*}

\begin{figure*}[h!]
    \centering
    \subfloat[]{\includegraphics[width=0.5\linewidth]{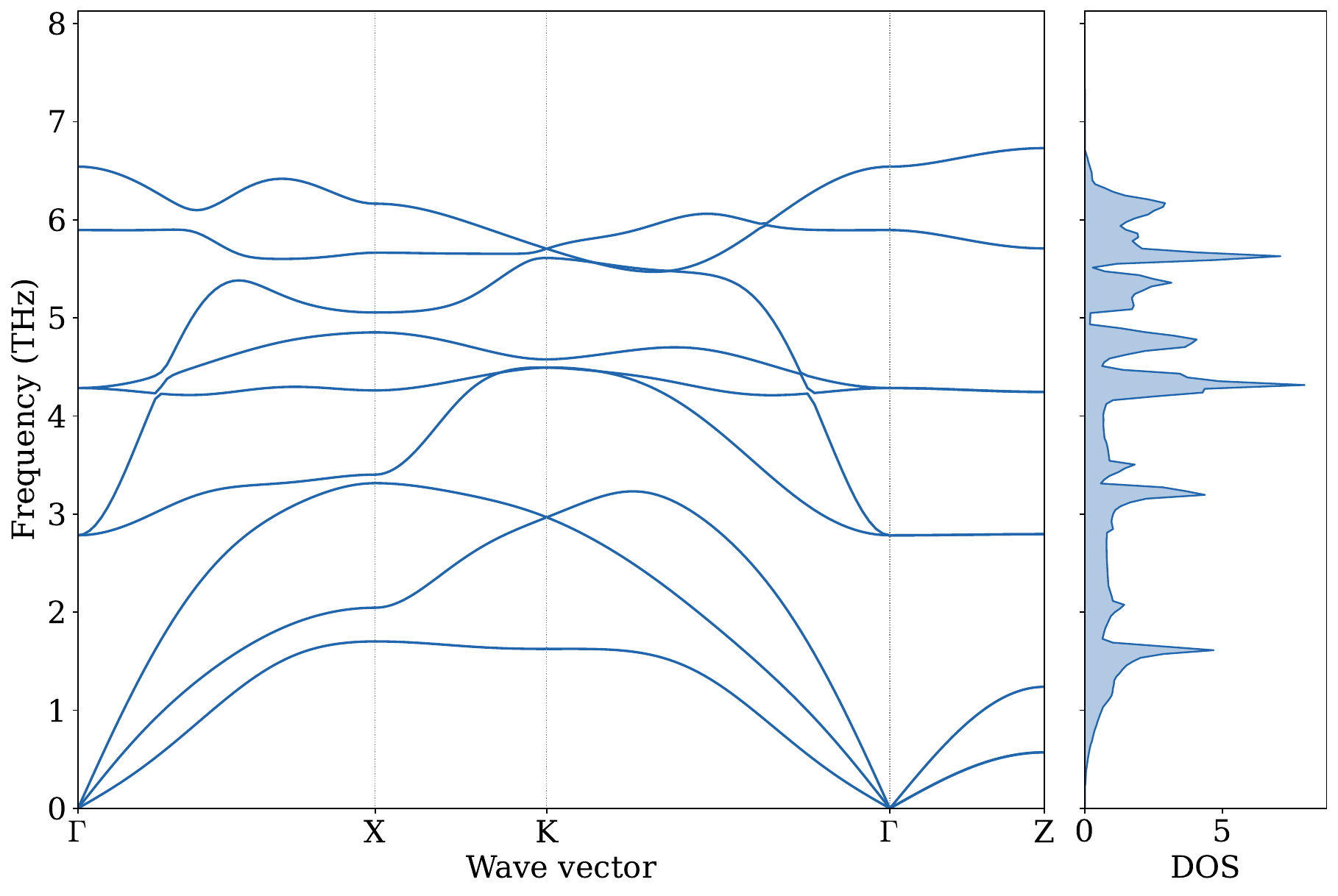}\label{subfig:DTMDC_T_DHf_Se2_Mg_D0}}
    \subfloat[]{\includegraphics[width=0.5\linewidth]{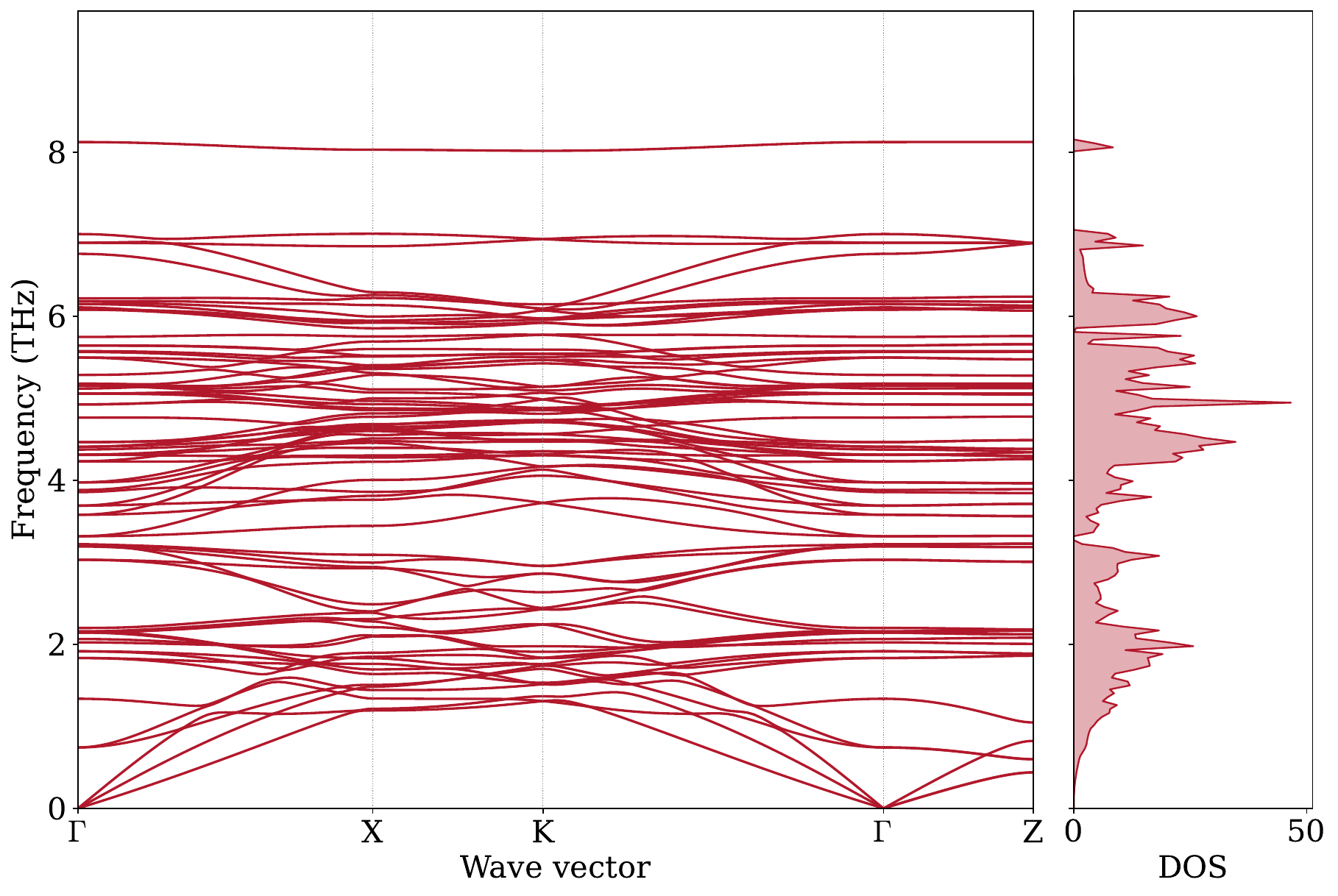}\label{subfig:DTMDC_T_DHf_Se2_Mg_D1}}
    \caption{Phonon dispersion and total DOS for the HfSe$_2$ family: (a)~bulk HfSe$_2$; (b)~Mg$_{0.125}$HfSe$_2$.}
    \label{suppfig:DTMDCT_HfSe}
\end{figure*}

\begin{figure*}[h!]
    \centering
    \subfloat[]{\includegraphics[width=0.5\linewidth]{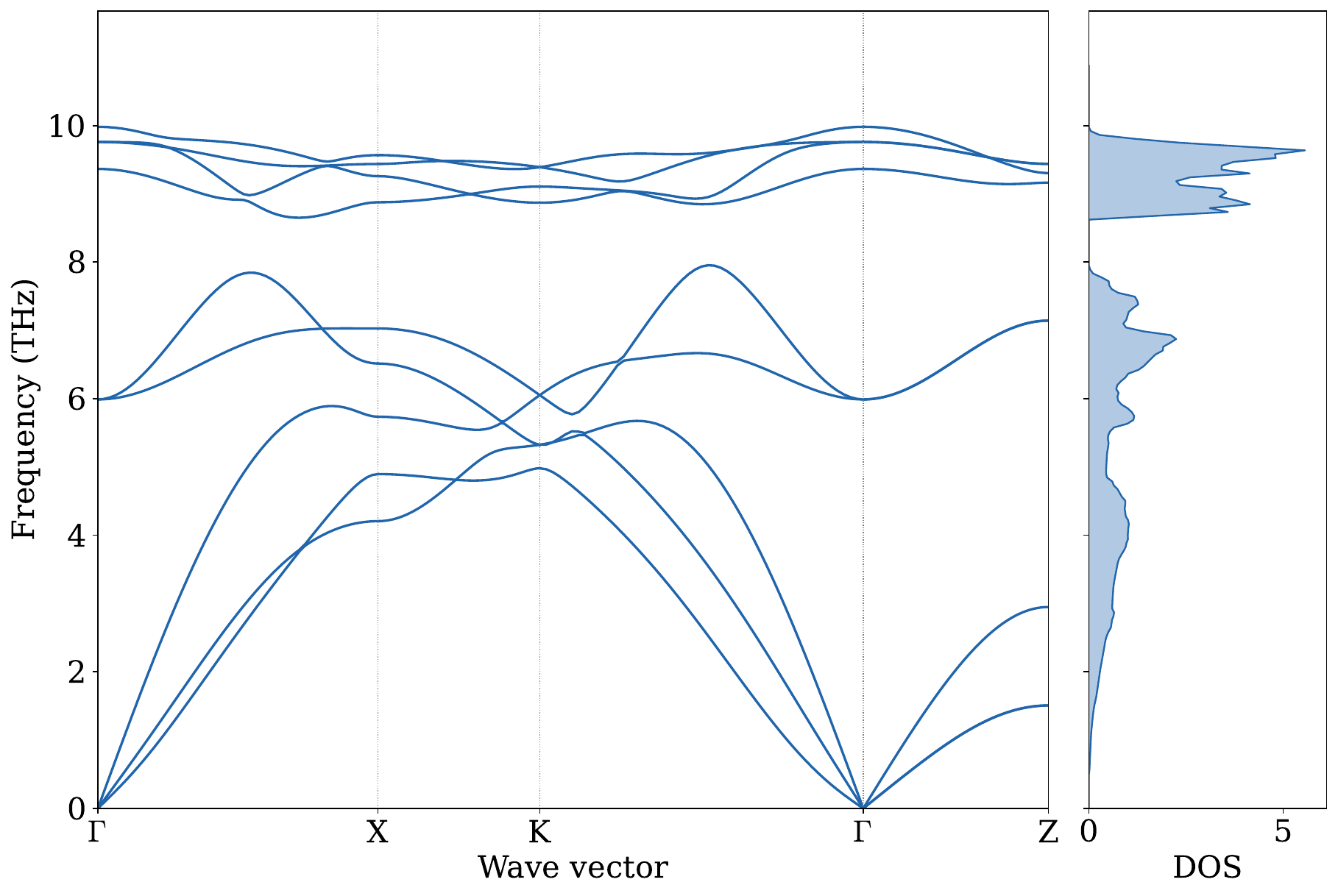}\label{subfig:DTMDC_T_DNi_S2_Li_D0}}
    \subfloat[]{\includegraphics[width=0.5\linewidth]{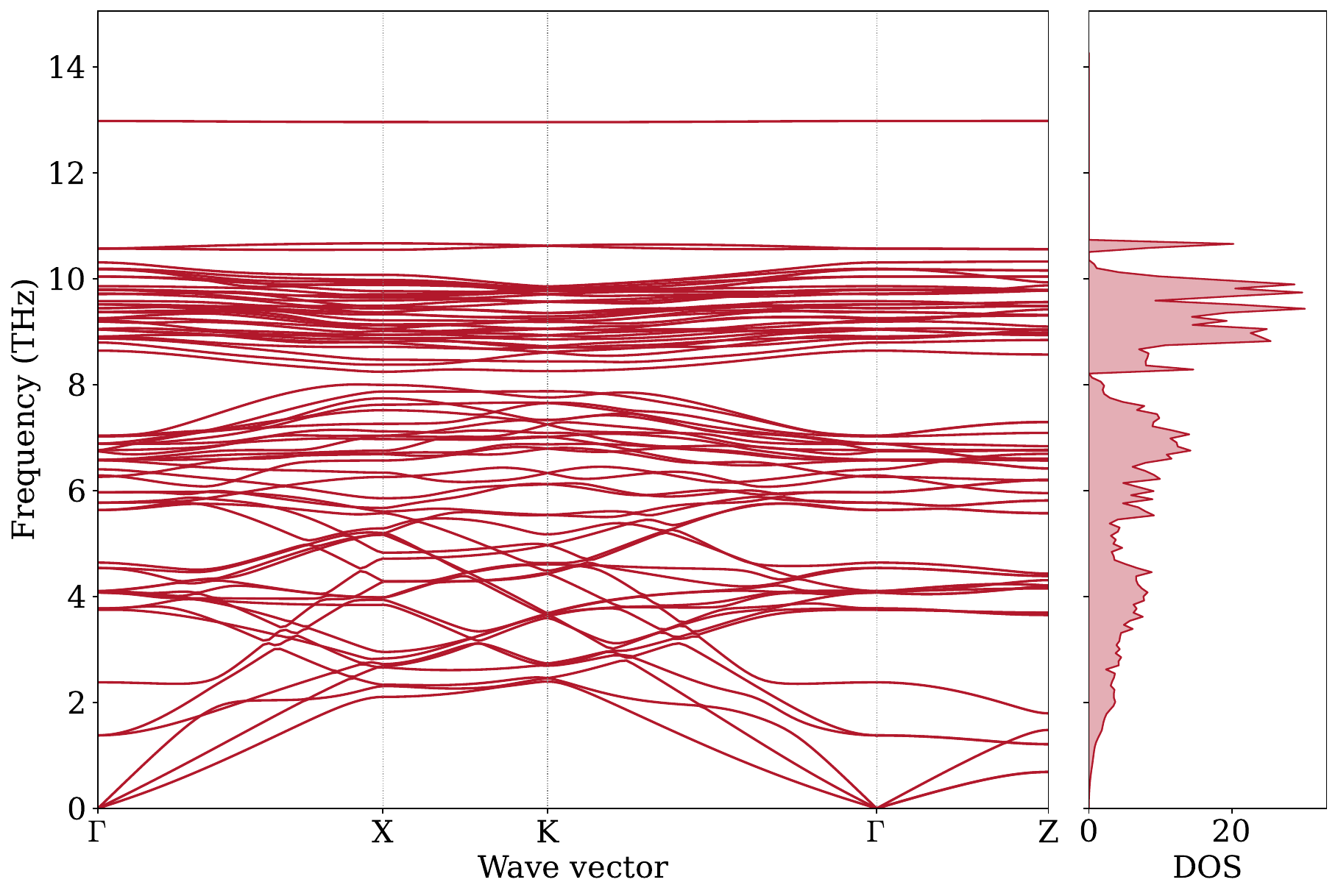}\label{subfig:DTMDC_T_DNi_S2_Li_D1}}
    \hfill
    \subfloat[]{\includegraphics[width=0.5\linewidth]{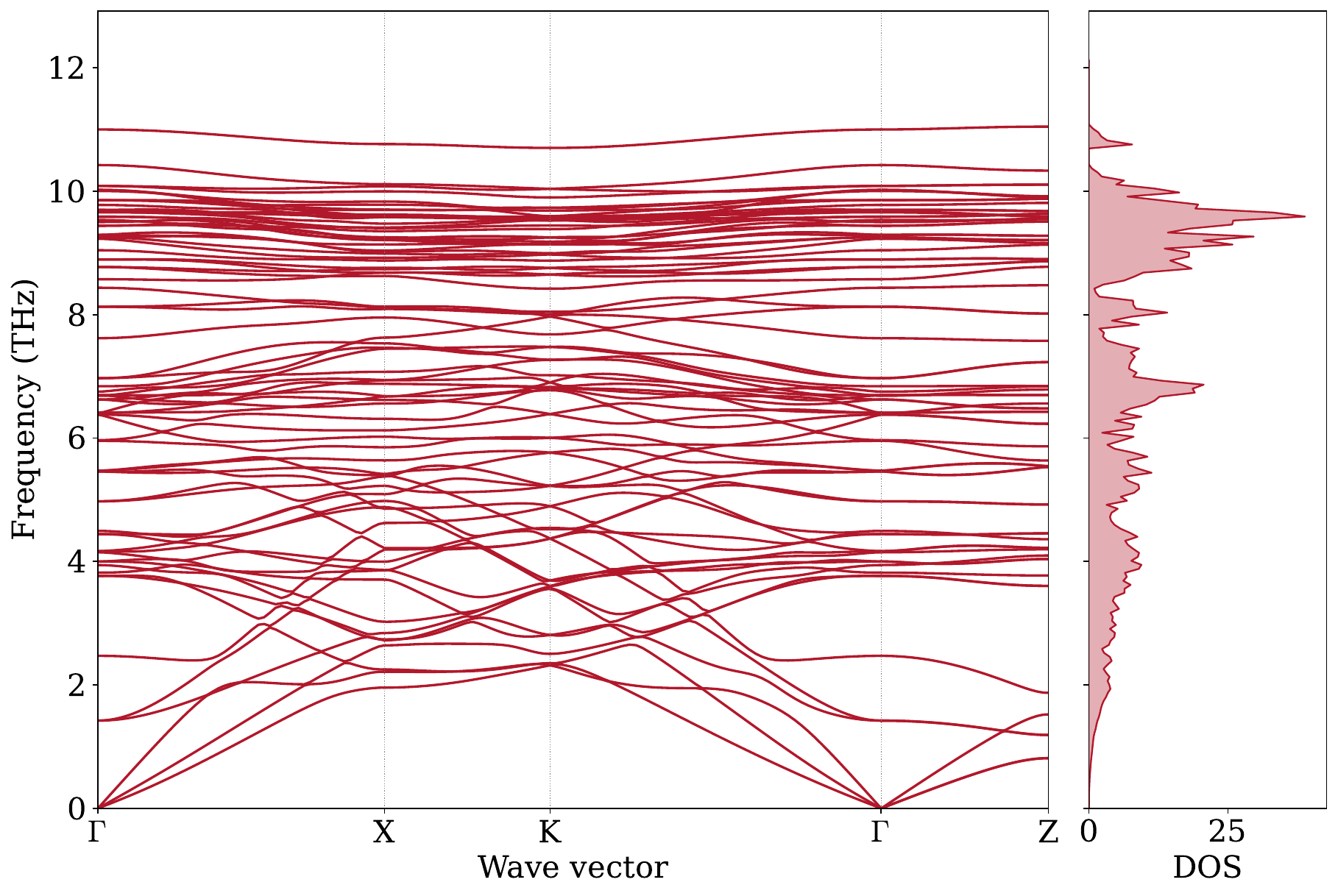}\label{subfig:DTMDC_T_DNi_S2_Mg_D1}}
    \caption{Phonon dispersion and total DOS for the NiS$_2$ family: (a)~bulk NiS$_2$; (b)~Li$_{0.125}$NiS$_2$; (c)~Mg$_{0.125}$NiS$_2$.}
    \label{suppfig:DTMDCT_NiS}
\end{figure*}

\begin{figure*}[h!]
    \centering
    \subfloat[]{\includegraphics[width=0.5\linewidth]{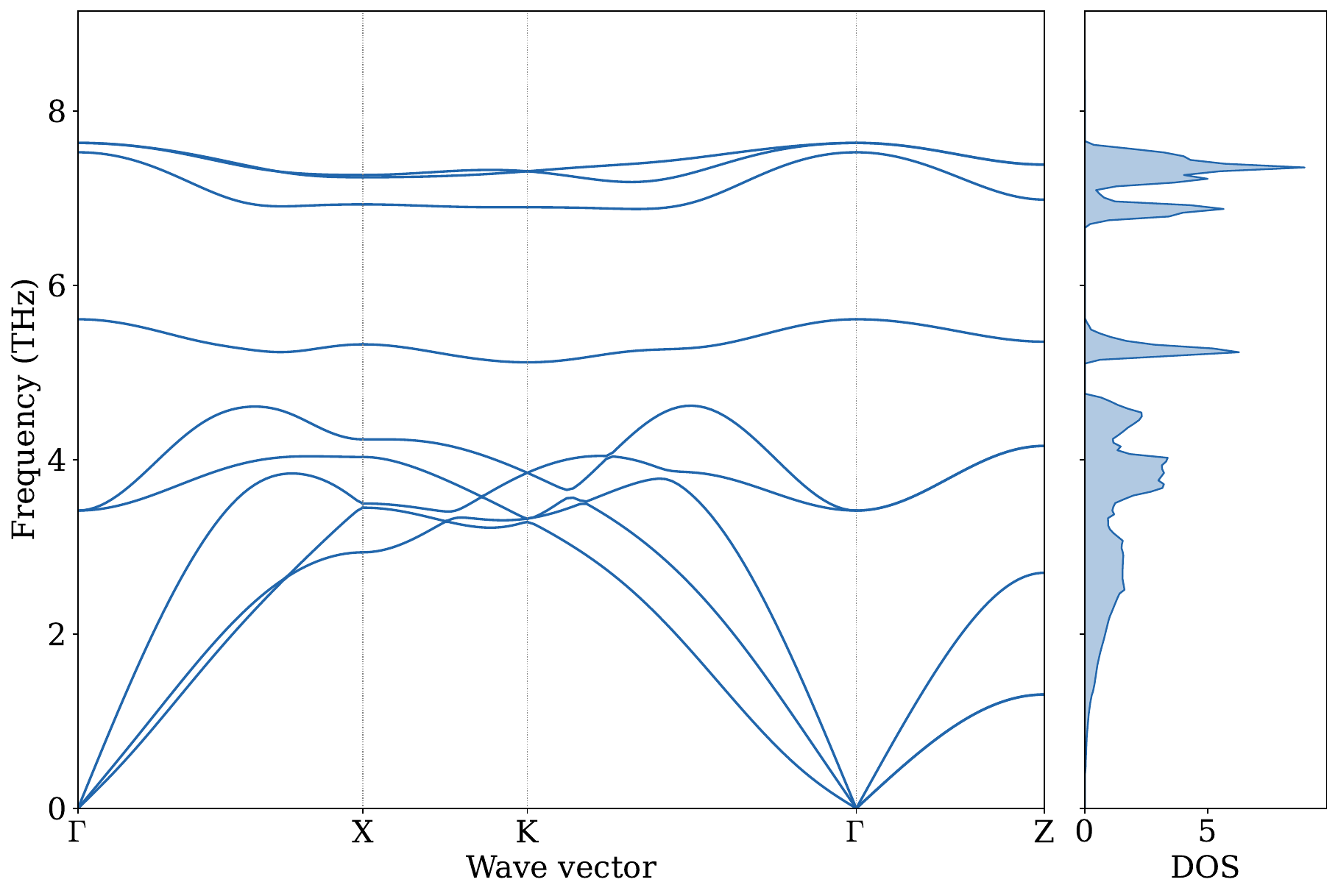}\label{subfig:DTMDC_T_DNi_Se2_Mg_D0}}
    \subfloat[]{\includegraphics[width=0.5\linewidth]{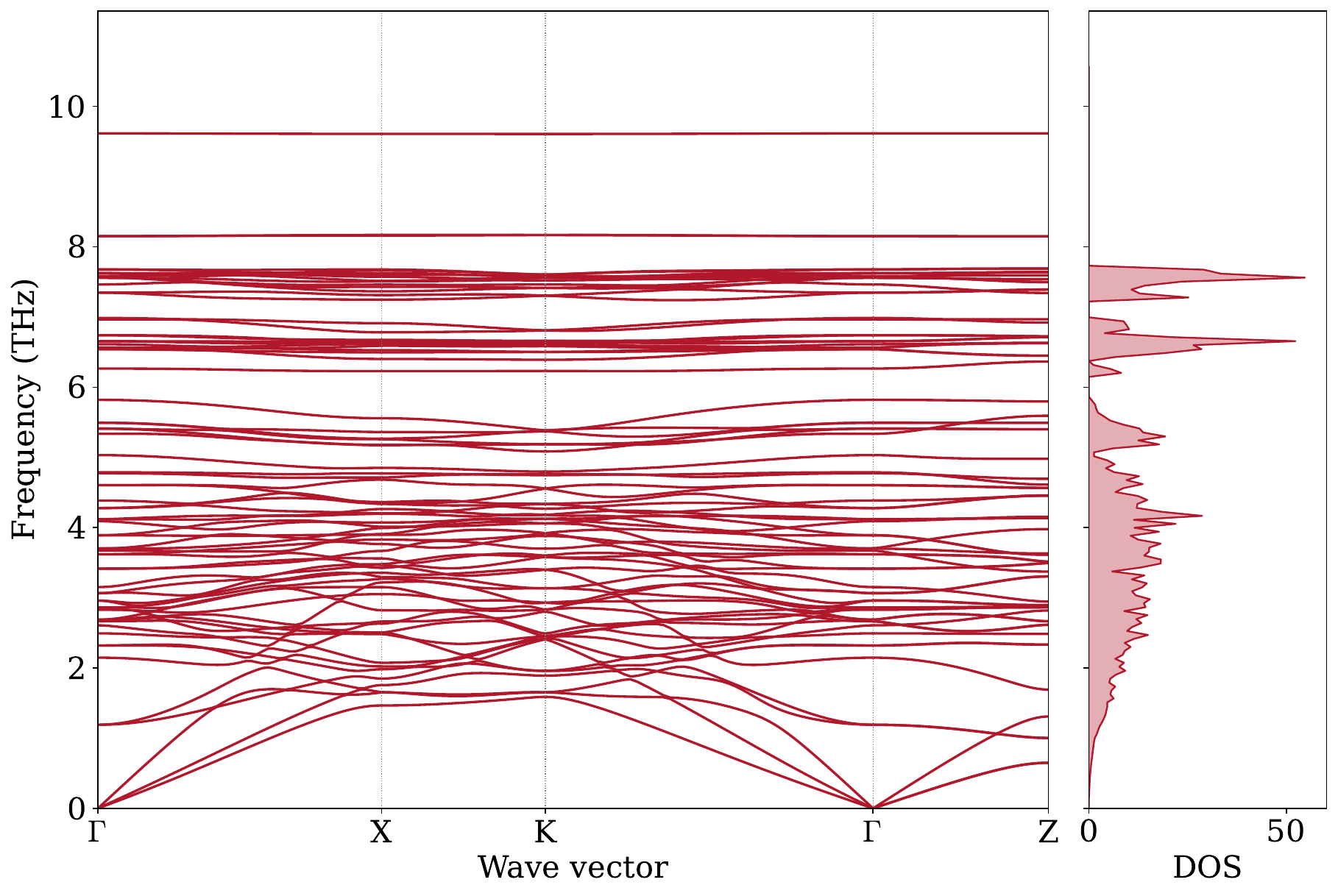}\label{subfig:DTMDC_T_DNi_Se2_Mg_D1}}
    \caption{Phonon dispersion and total DOS for the NiSe$_2$ family: (a)~bulk NiSe$_2$; (b)~Mg$_{0.125}$NiSe$_2$.}
    \label{suppfig:DTMDCT_NiSe}
\end{figure*}

\begin{figure*}[h!]
    \centering
    \subfloat[]{\includegraphics[width=0.5\linewidth]{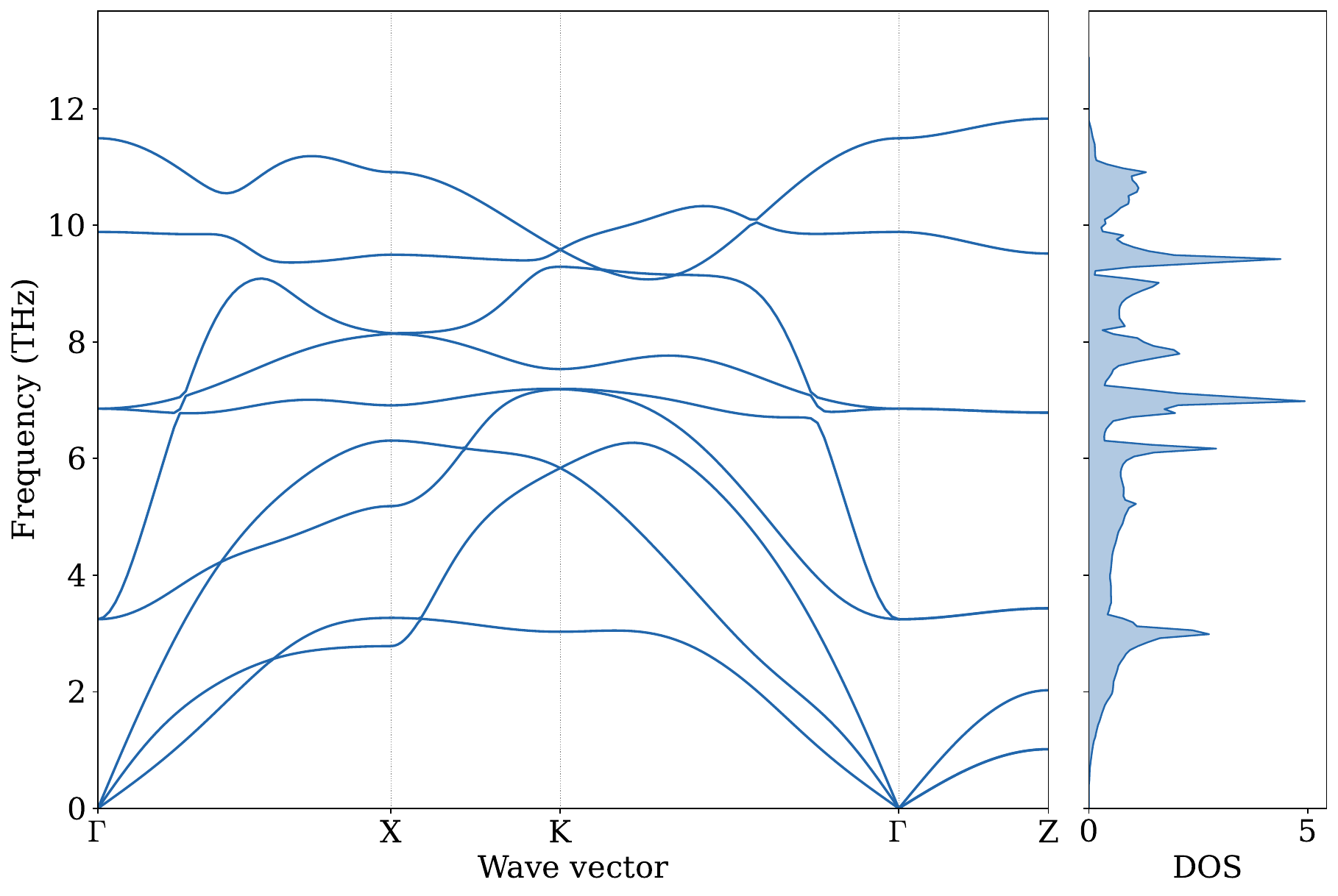}\label{subfig:DTMDC_T_DTi_S2_Li_D0}}
    \subfloat[]{\includegraphics[width=0.5\linewidth]{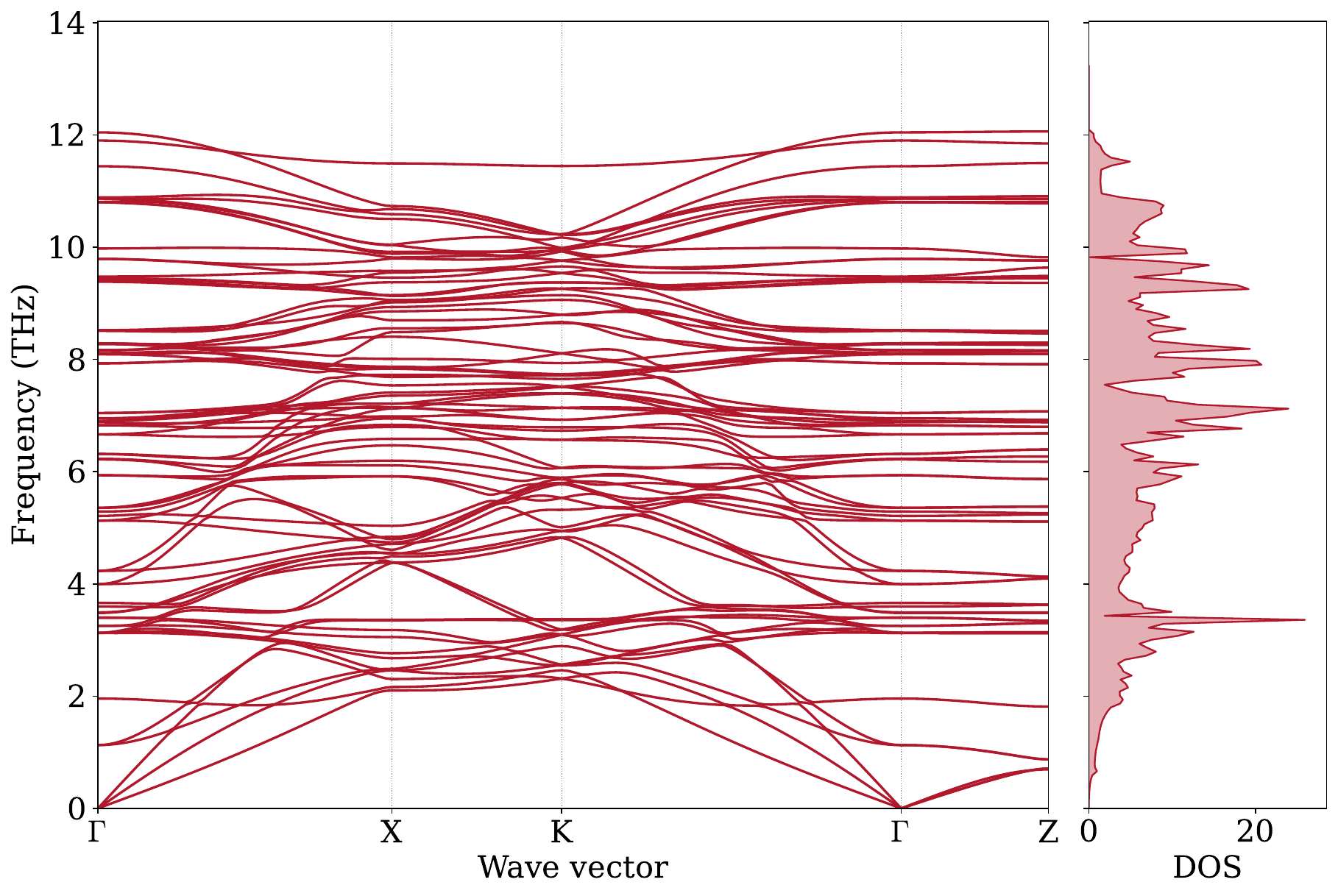}\label{subfig:DTMDC_T_DTi_S2_Li_D1}}
    \hfill
    \subfloat[]{\includegraphics[width=0.5\linewidth]{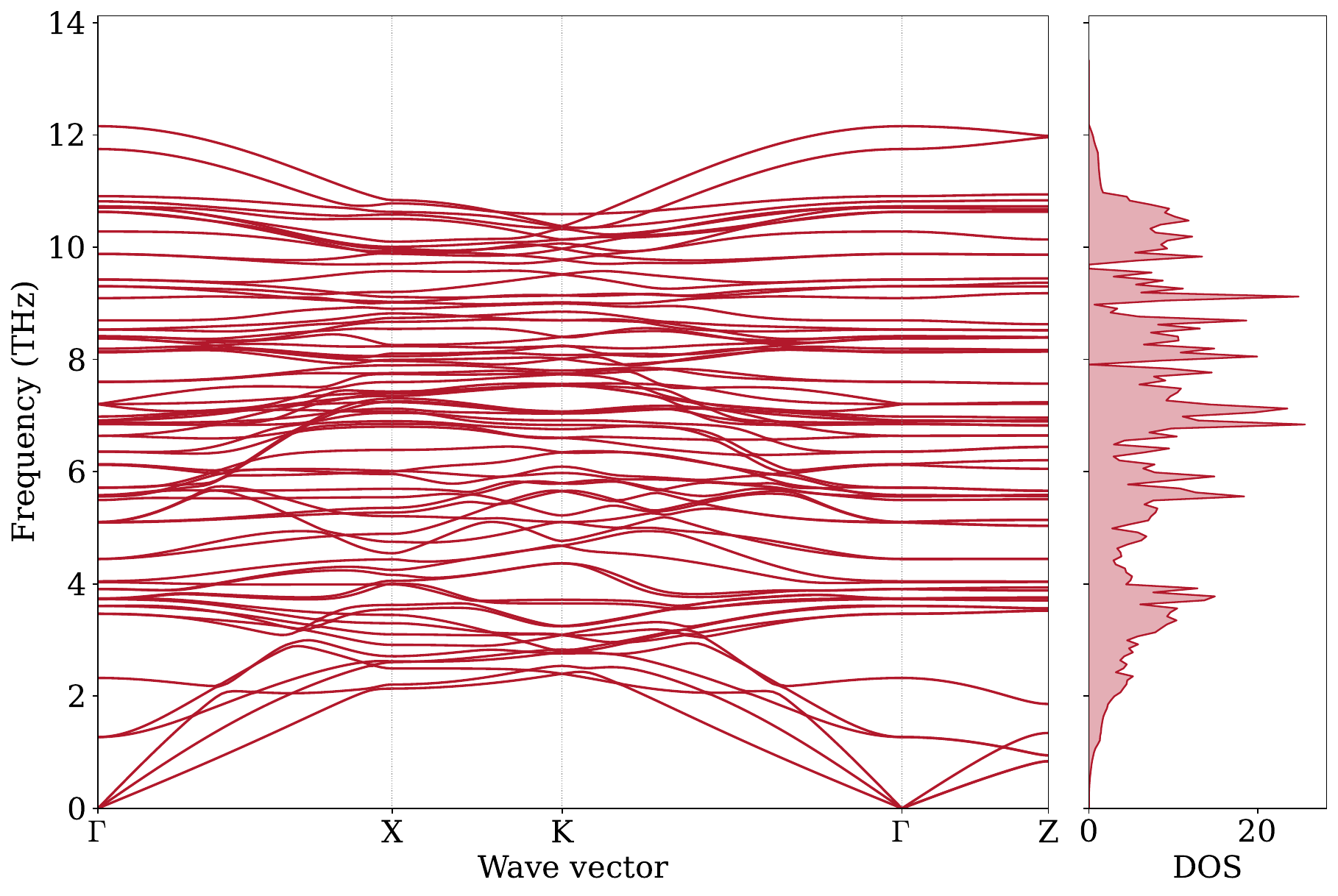}\label{subfig:DTMDC_T_DTi_S2_Mg_D1}}
    \caption{Phonon dispersion and total DOS for the TiS$_2$ family: (a)~bulk TiS$_2$; (b)~Li$_{0.125}$TiS$_2$; (c)~Mg$_{0.125}$TiS$_2$.}
    \label{suppfig:DTMDCT_TiS}
\end{figure*}

\begin{figure*}[h!]
    \centering
    \subfloat[]{\includegraphics[width=0.5\linewidth]{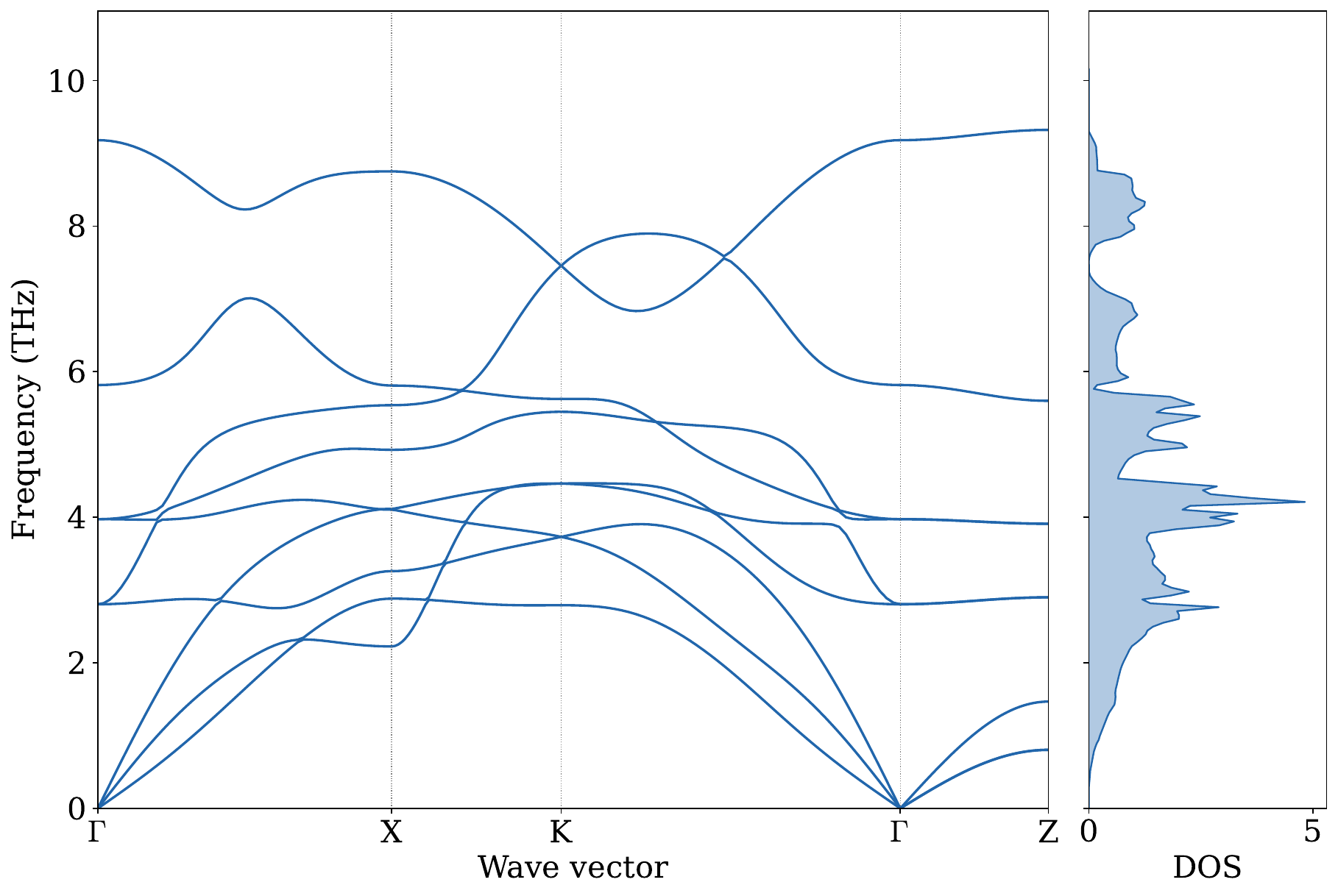}\label{subfig:DTMDC_T_DTi_Se2_Mg_D0}}
    \subfloat[]{\includegraphics[width=0.5\linewidth]{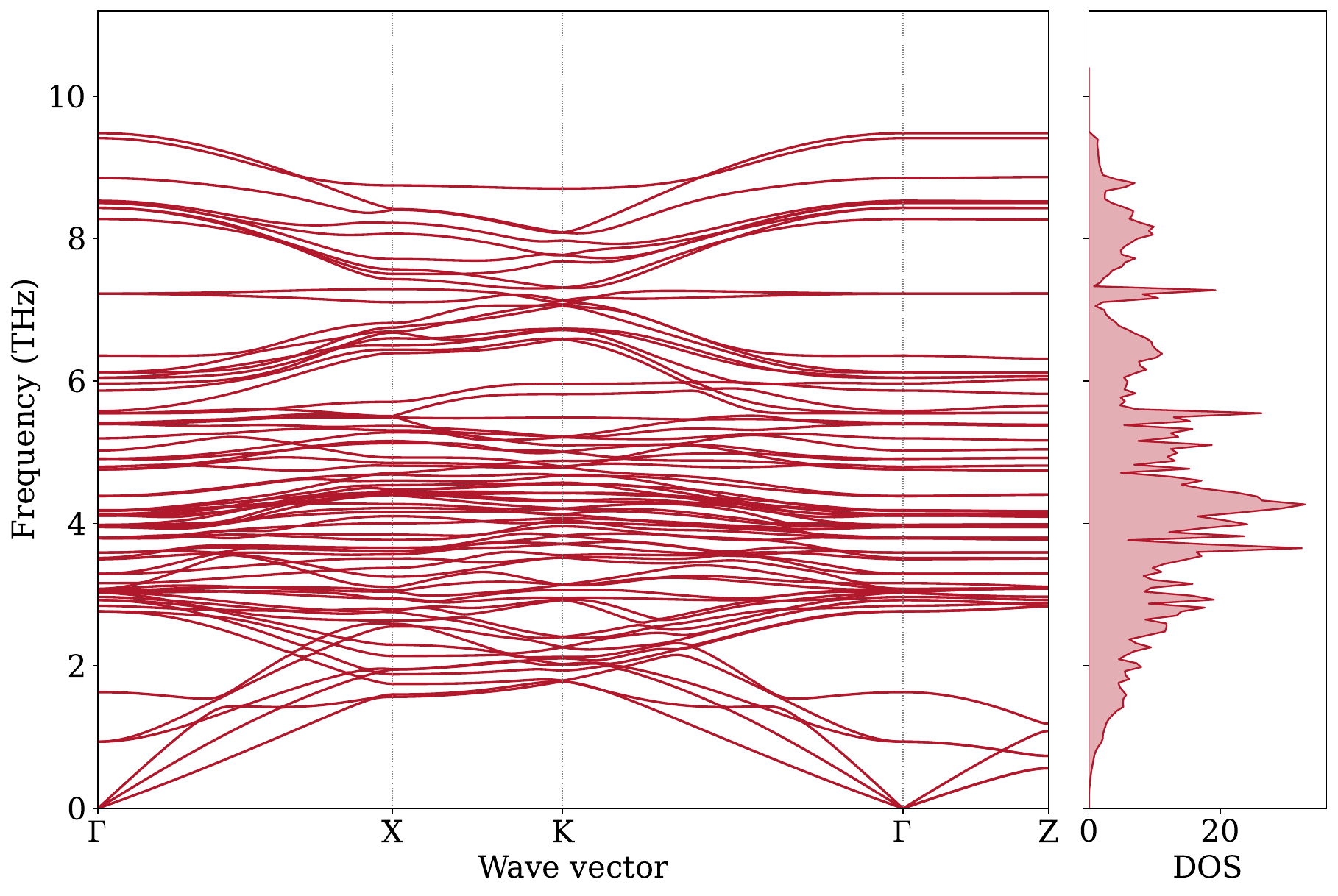}\label{subfig:DTMDC_T_DTi_Se2_Mg_D1}}
    \caption{Phonon dispersion and total DOS for the TiSe$_2$ family: (a)~bulk TiSe$_2$; (b)~Mg$_{0.125}$TiSe$_2$.}
    \label{suppfig:DTMDCT_TiSe}
\end{figure*}

\begin{figure*}[h!]
    \centering
    \subfloat[]{\includegraphics[width=0.5\linewidth]{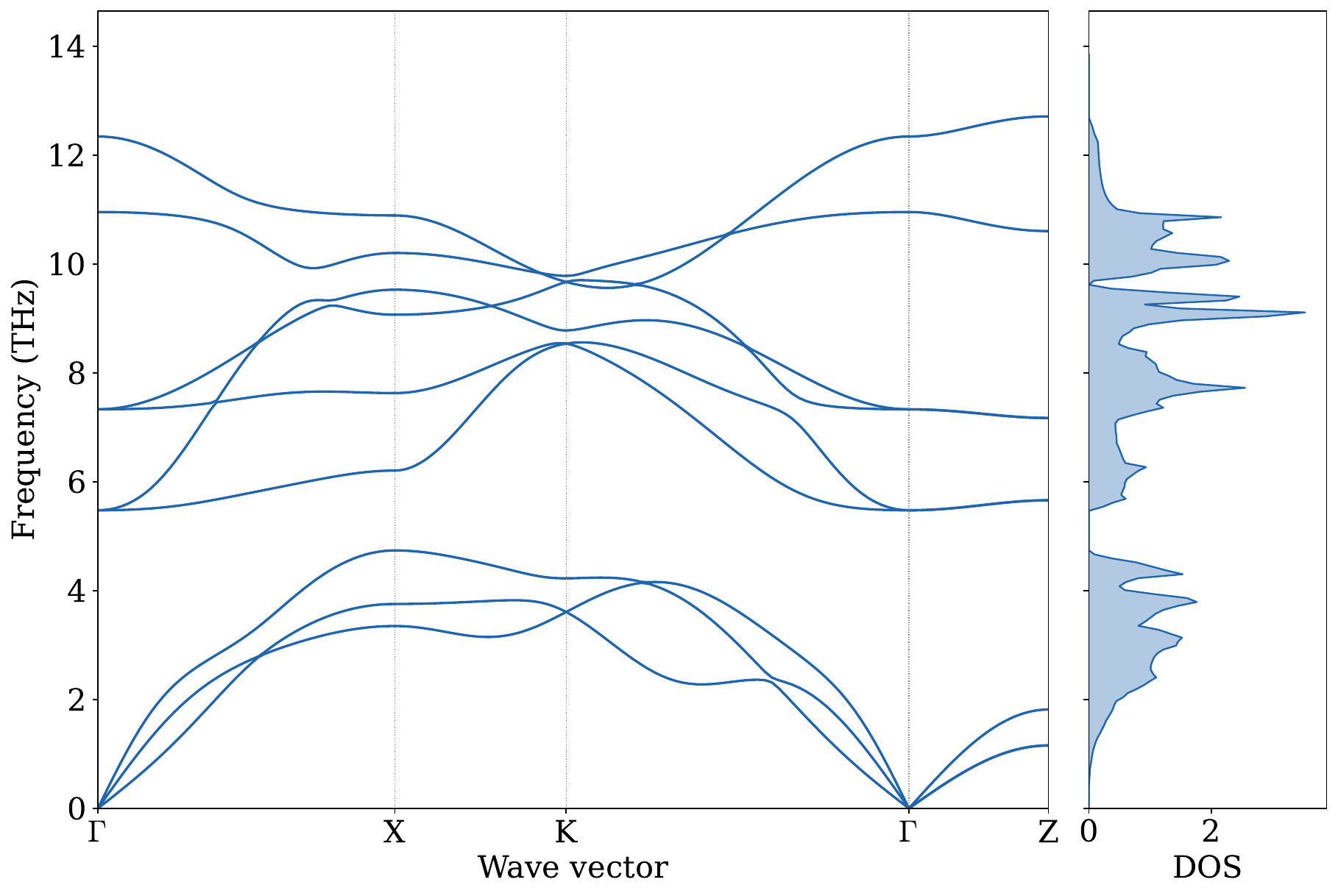}\label{subfig:DTMDC_T_DV_S2_Li_D0}}
    \subfloat[]{\includegraphics[width=0.5\linewidth]{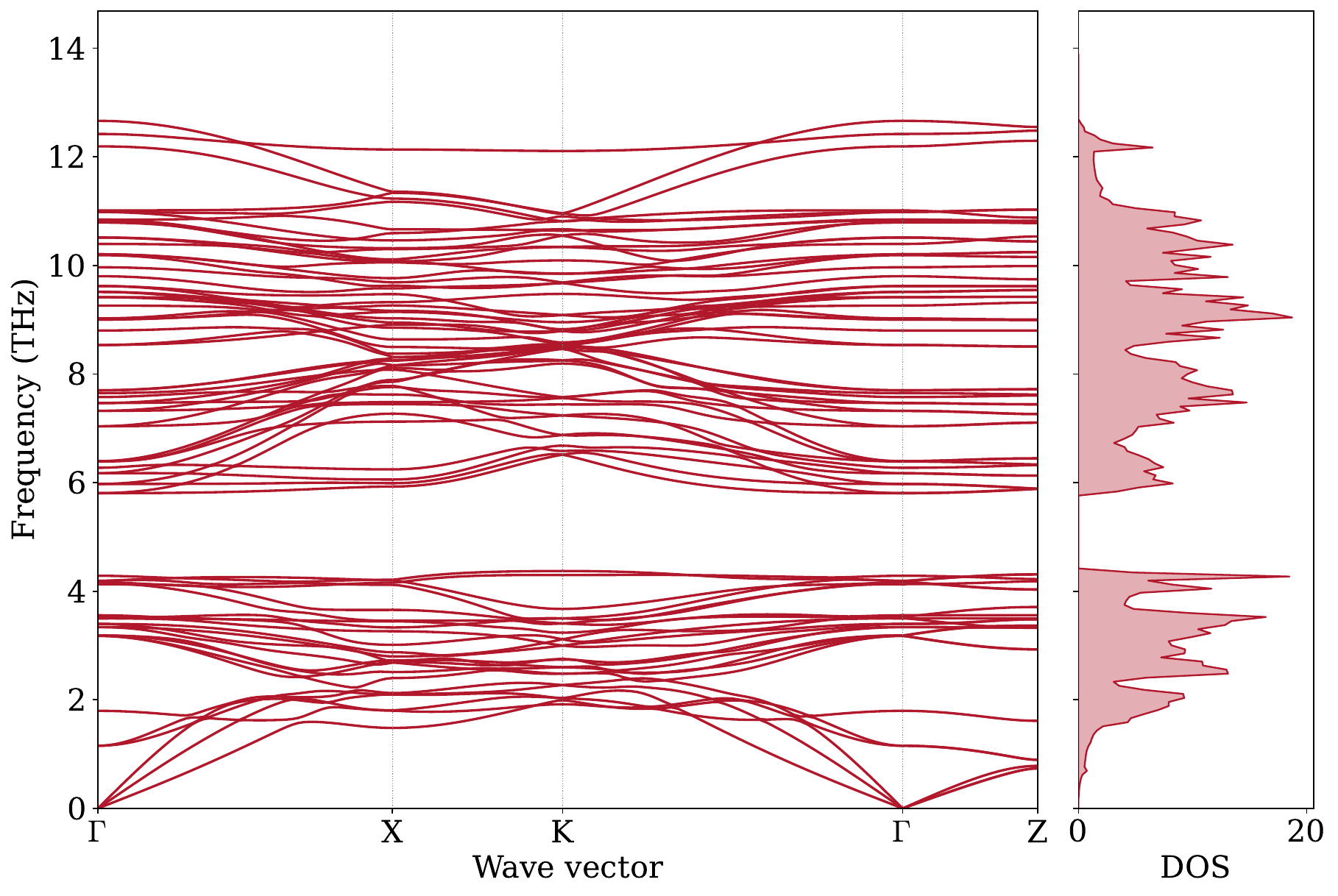}\label{subfig:DTMDC_T_DV_S2_Li_D1}}
    \caption{Phonon dispersion and total DOS for the VS$_2$ family: (a)~bulk VS$_2$; (b)~Li$_{0.125}$VS$_2$.}
    \label{suppfig:DTMDCT_VS}
\end{figure*}

\begin{figure*}[h!]
    \centering
    \subfloat[]{\includegraphics[width=0.5\linewidth]{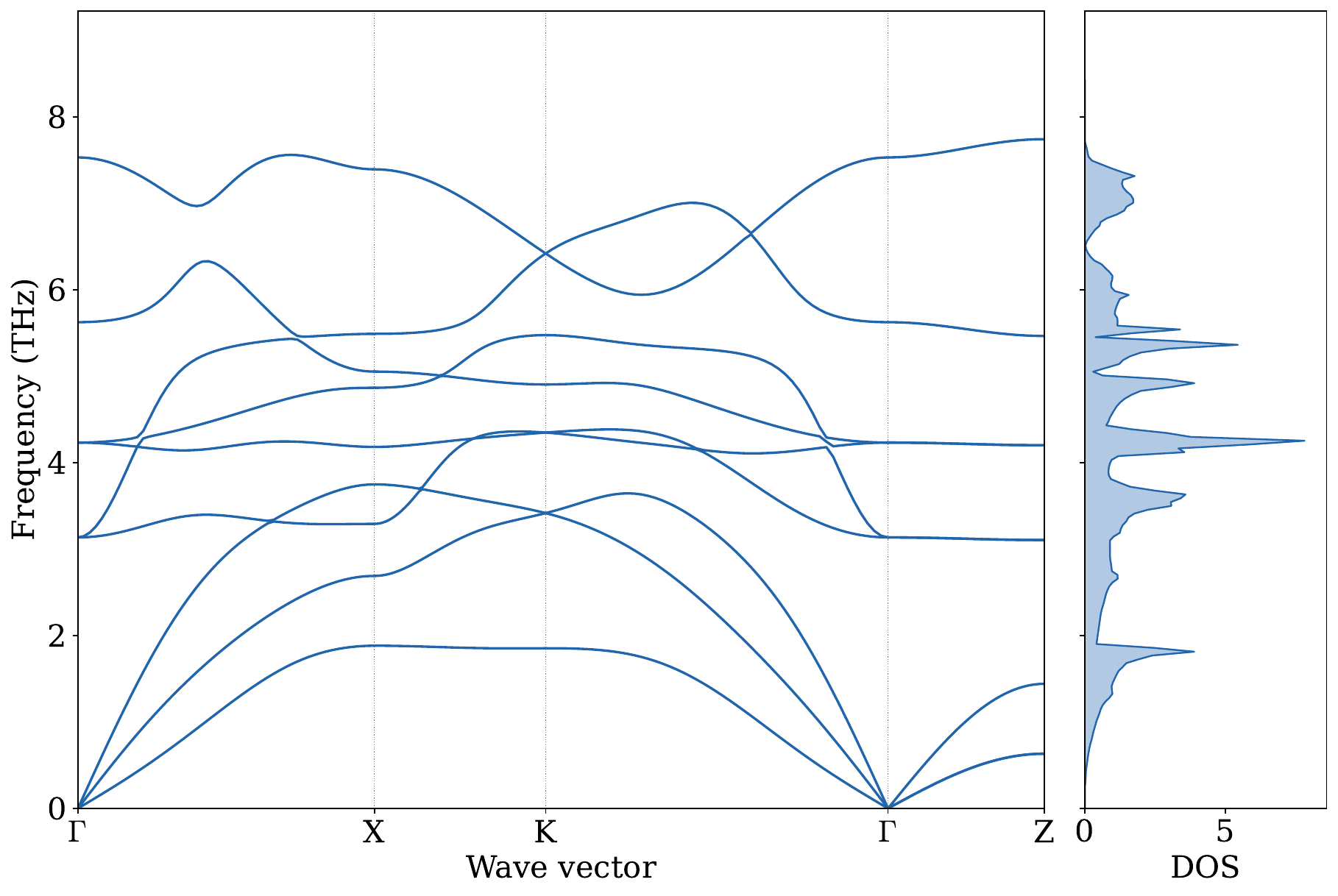}\label{subfig:DTMDC_T_DZr_Se_Li_D0}}
    \subfloat[]{\includegraphics[width=0.5\linewidth]{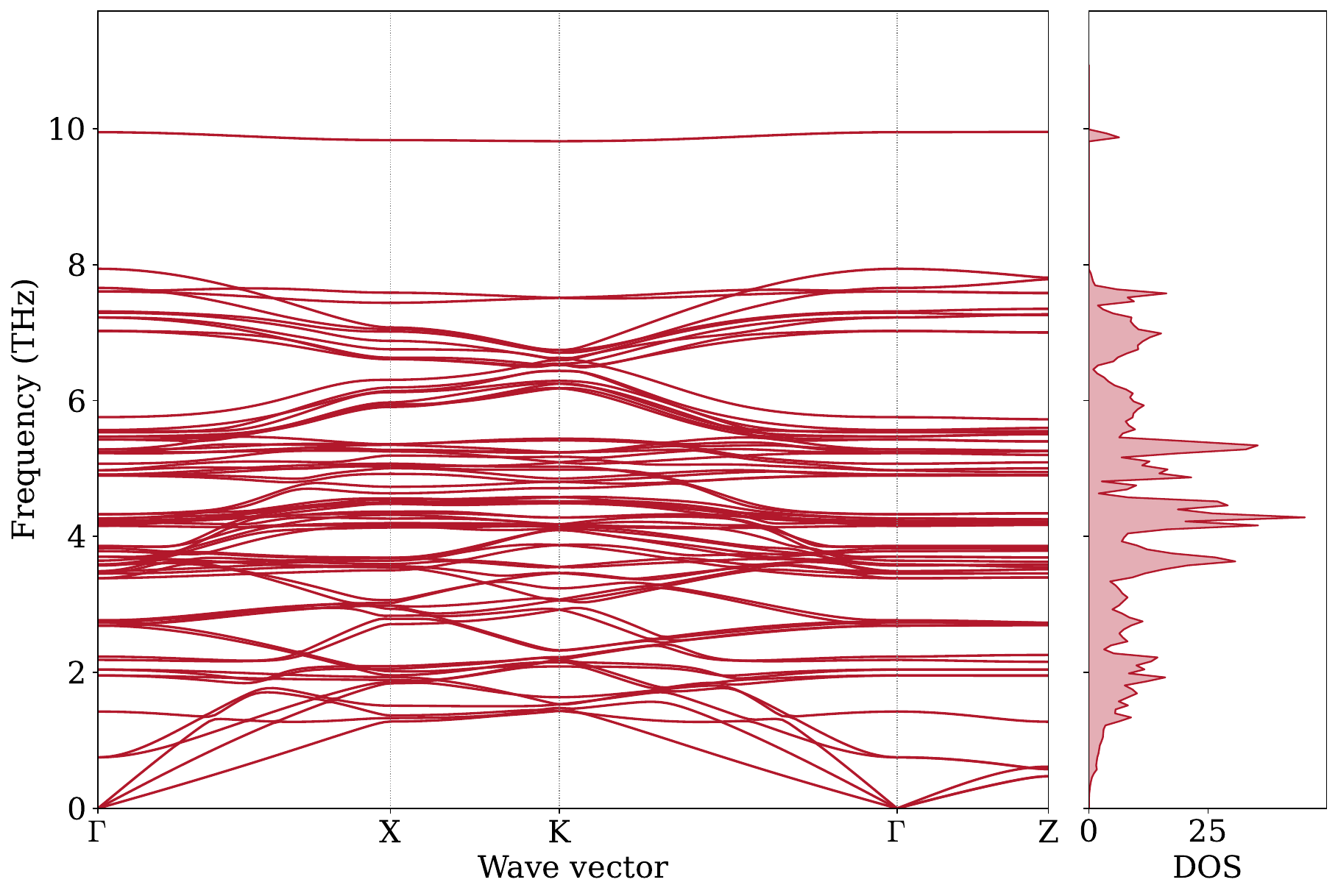}\label{subfig:DTMDC_T_DZr_Se_Li_D1}}
    \hfill
    \subfloat[]{\includegraphics[width=0.5\linewidth]{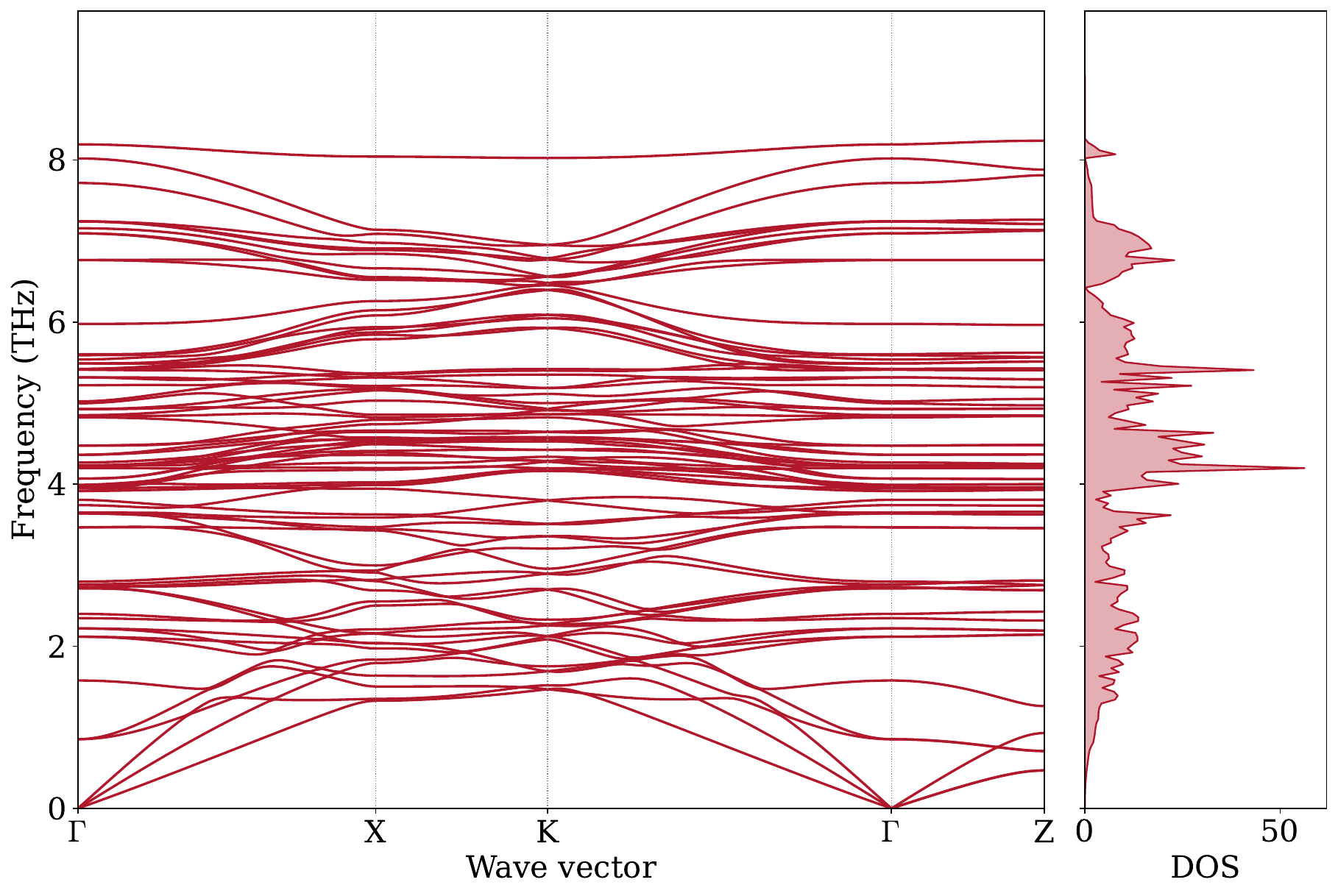}\label{subfig:DTMDC_T_DZr_Se_Mg_D1}}
    \caption{Phonon dispersion and total DOS for the ZrSe$_2$ family: (a)~bulk ZrSe$_2$; (b)~Li$_{0.125}$ZrSe$_2$; (c)~Mg$_{0.125}$ZrSe$_2$.}
    \label{suppfig:DTMDCT_ZrSe}
\end{figure*}


\begin{figure*}[h!]
    \centering
    \subfloat[]{\includegraphics[width=0.5\linewidth]{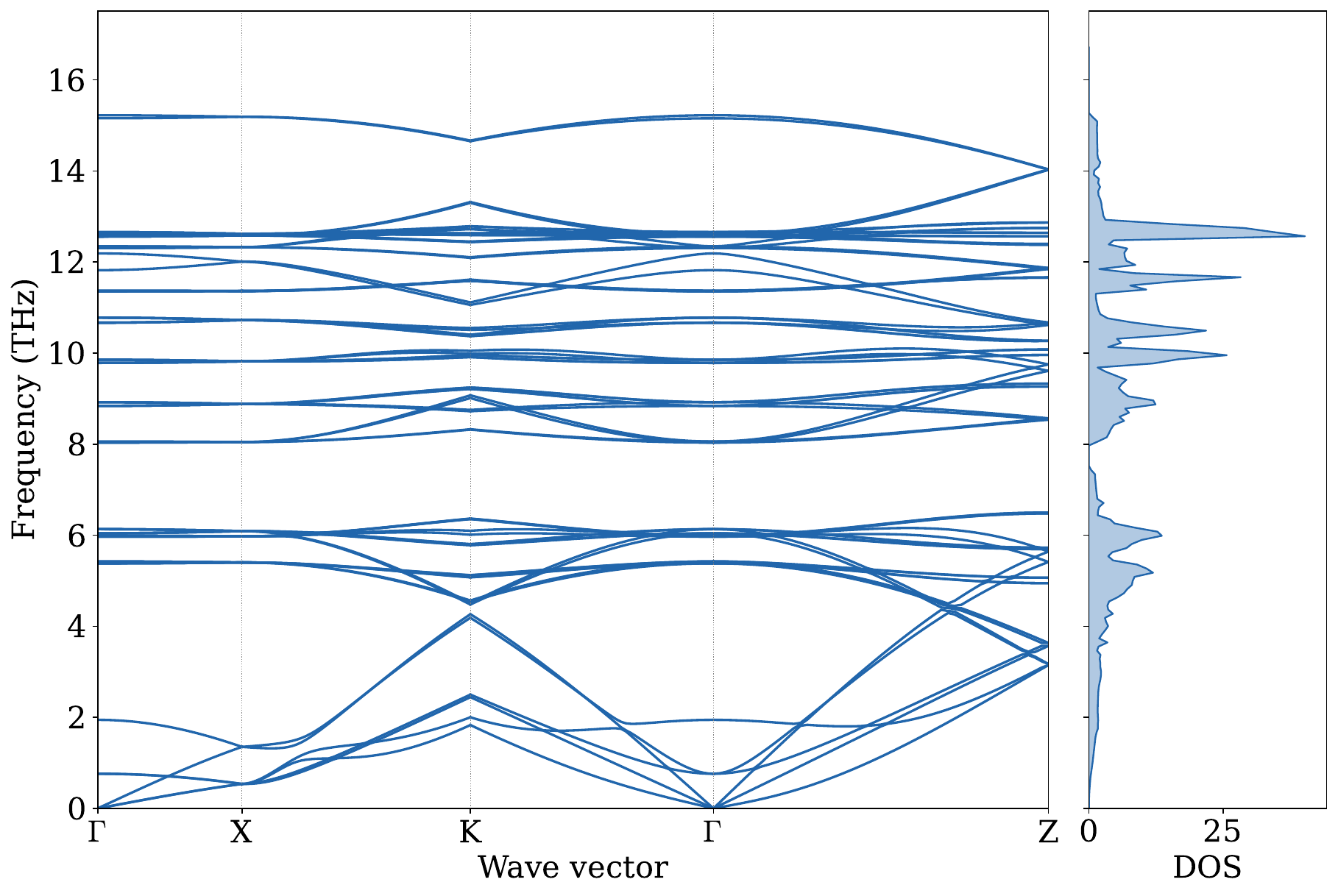}\label{subfig:DTMDC_Hc_DCr_S2_Li_D0}}
    \subfloat[]{\includegraphics[width=0.5\linewidth]{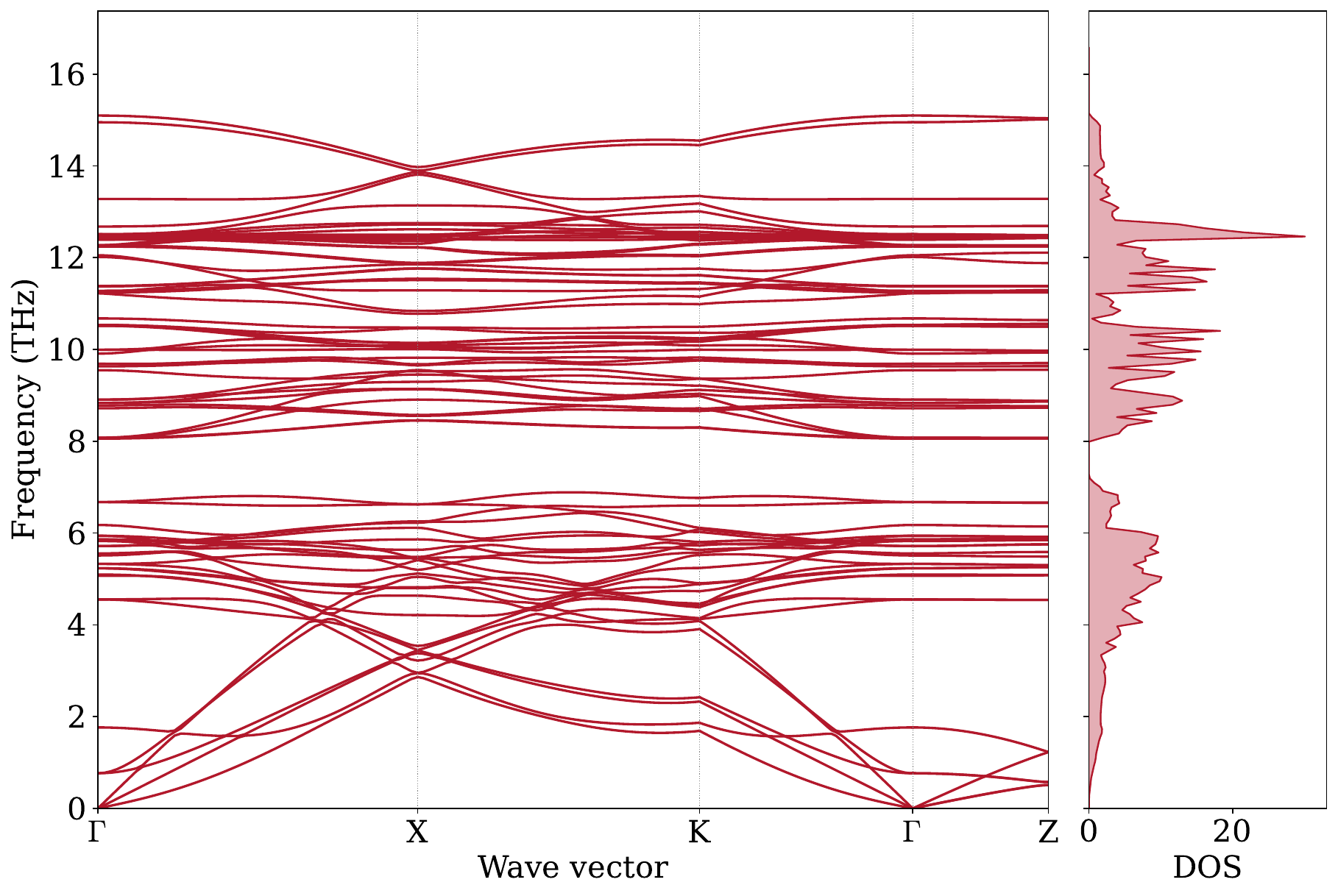}\label{subfig:DTMDC_Hc_DCr_S2_Li_D1}}
    \hfill
    \subfloat[]{\includegraphics[width=0.5\linewidth]{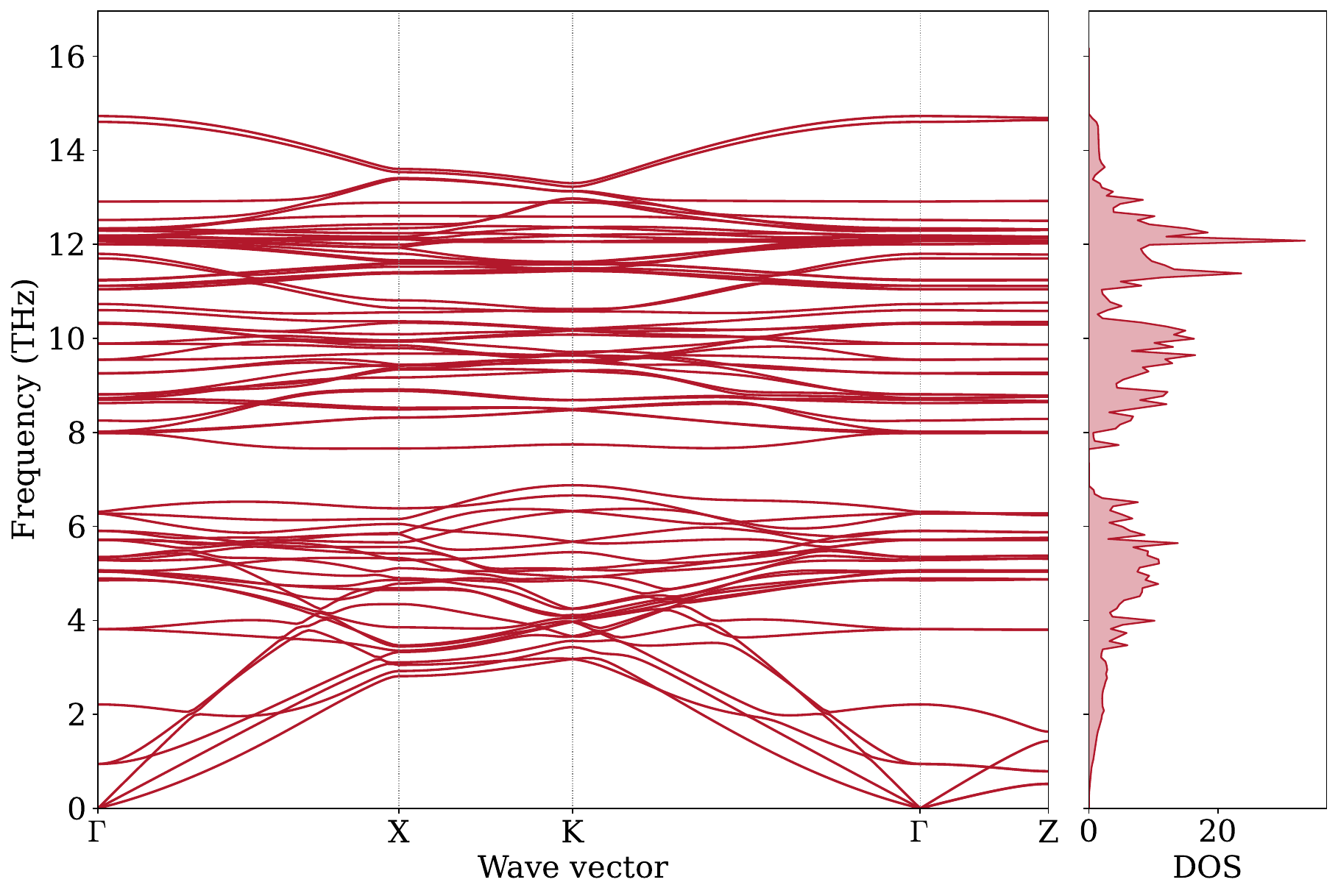}\label{subfig:DTMDC_Hc_DCr_S2_Mg_D1}}
    \caption{Phonon dispersion and total DOS for the CrS$_2$ family: (a)~bulk CrS$_2$; (b)~Li$_{0.125}$CrS$_2$; (c)~Mg$_{0.125}$CrS$_2$.}
    \label{suppfig:DTMDCHc_CrS}
\end{figure*}

\begin{figure*}[h!]
    \centering
    \subfloat[]{\includegraphics[width=0.5\linewidth]{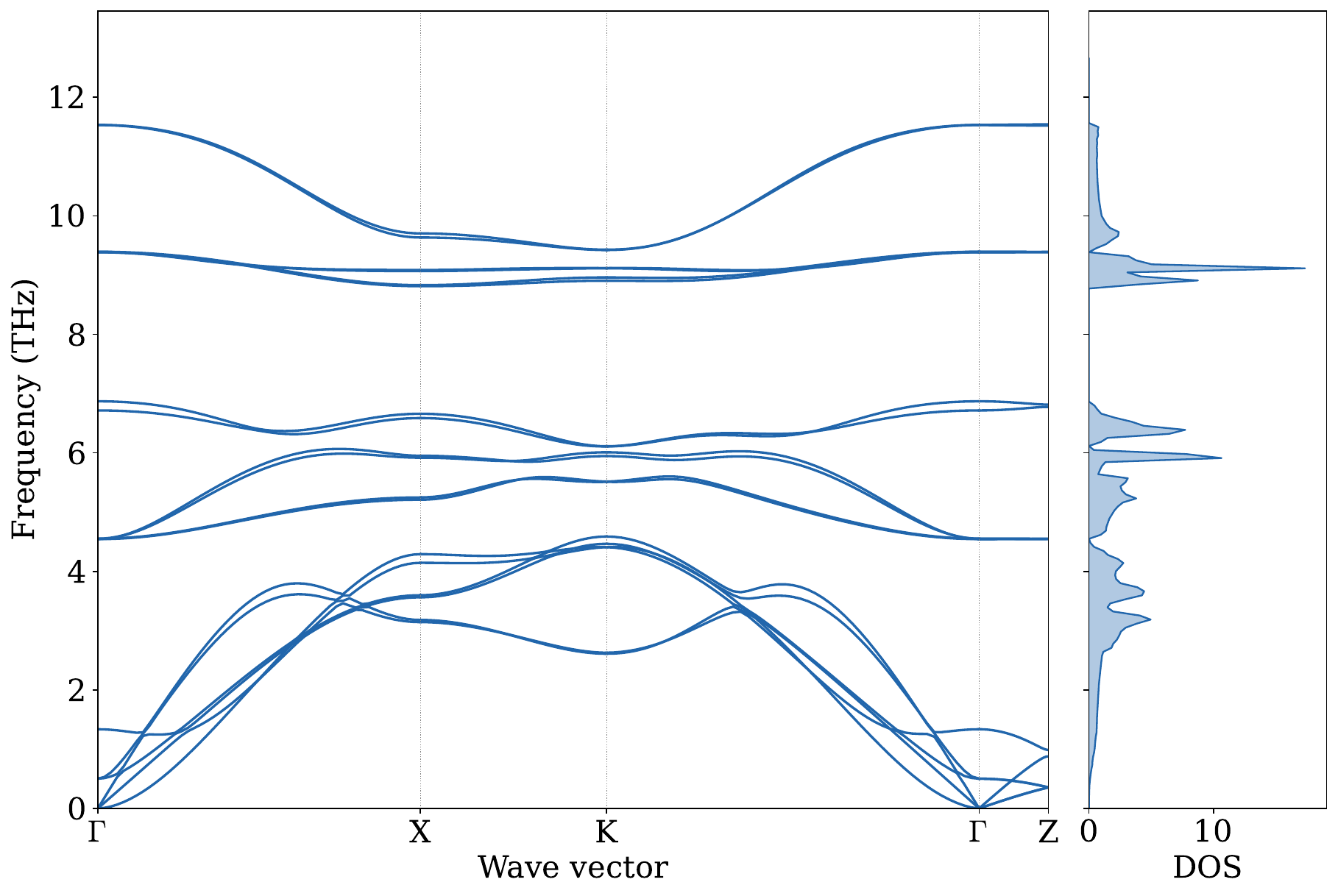}\label{subfig:DTMDC_Hc_DCr_Se2_Li_D0}}
    \subfloat[]{\includegraphics[width=0.5\linewidth]{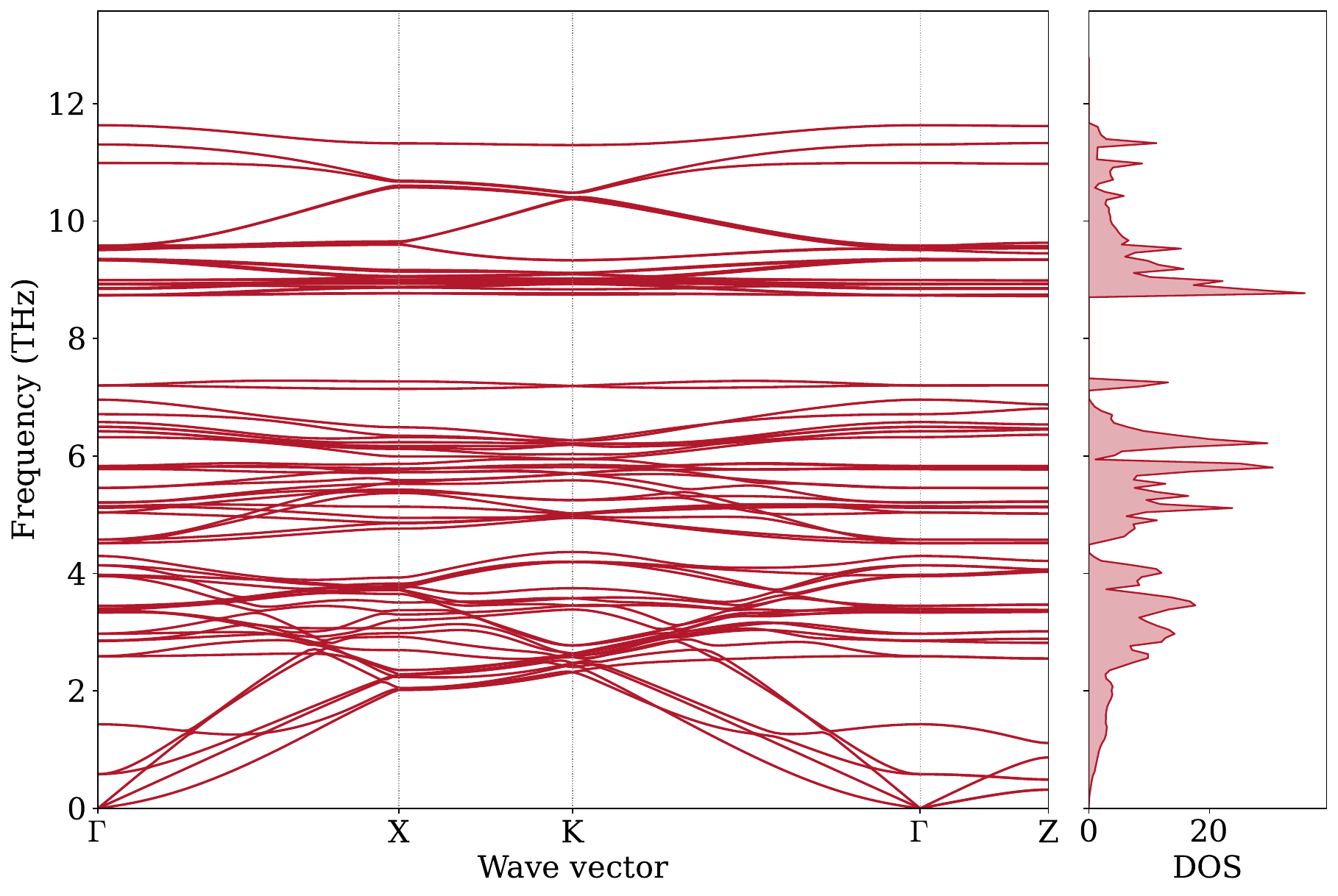}\label{subfig:DTMDC_Hc_DCr_Se2_Li_D1}}
    \hfill
    \subfloat[]{\includegraphics[width=0.5\linewidth]{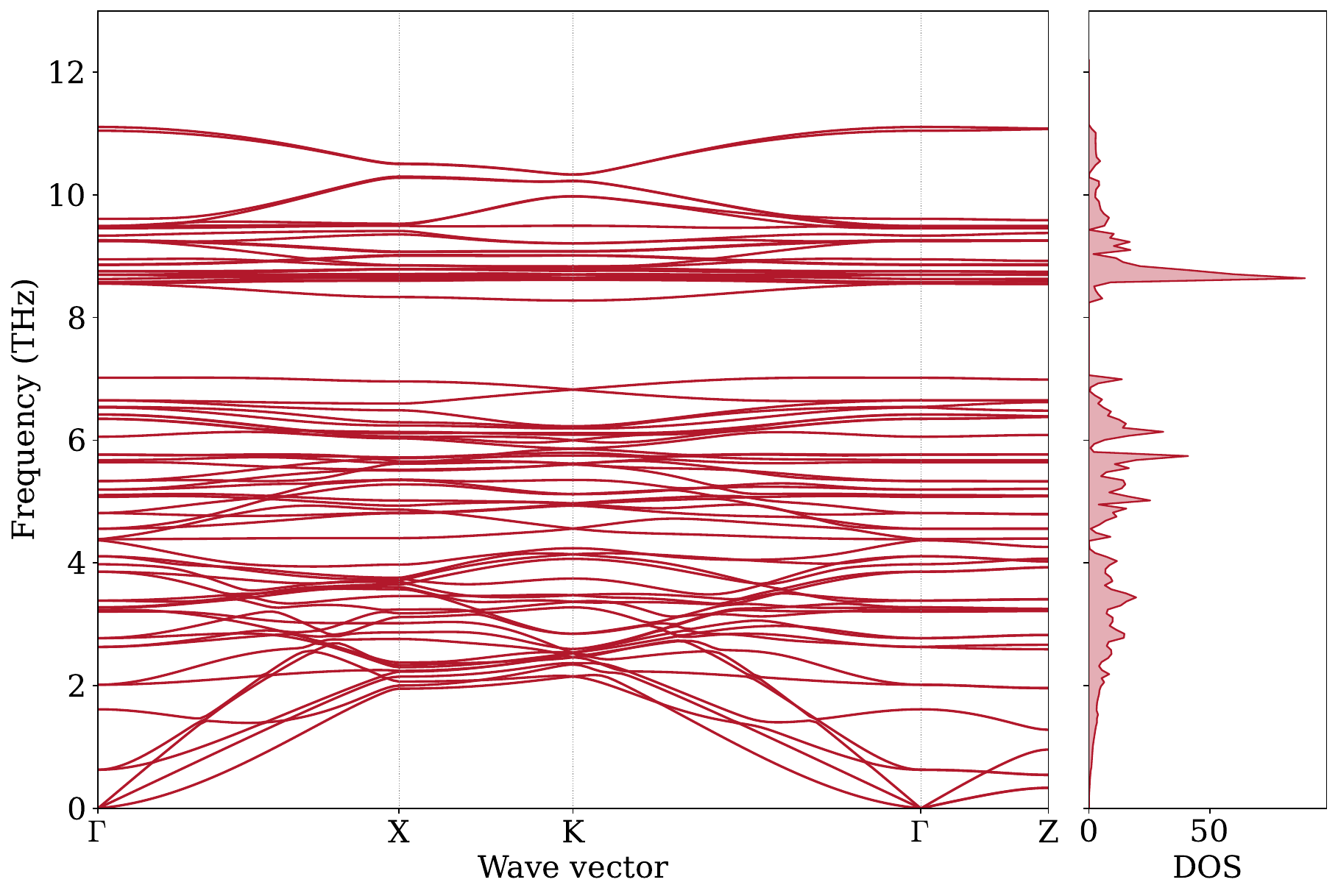}\label{subfig:DTMDC_Hc_DCr_Se2_Mg_D1}}
    \caption{Phonon dispersion and total DOS for the CrSe$_2$ family: (a)~bulk CrSe$_2$; (b)~Li$_{0.125}$CrSe$_2$; (c)~Mg$_{0.125}$CrSe$_2$.}
    \label{suppfig:DTMDCHc_CrSe}
\end{figure*}

\begin{figure*}[h!]
    \centering
    \subfloat[]{\includegraphics[width=0.5\linewidth]{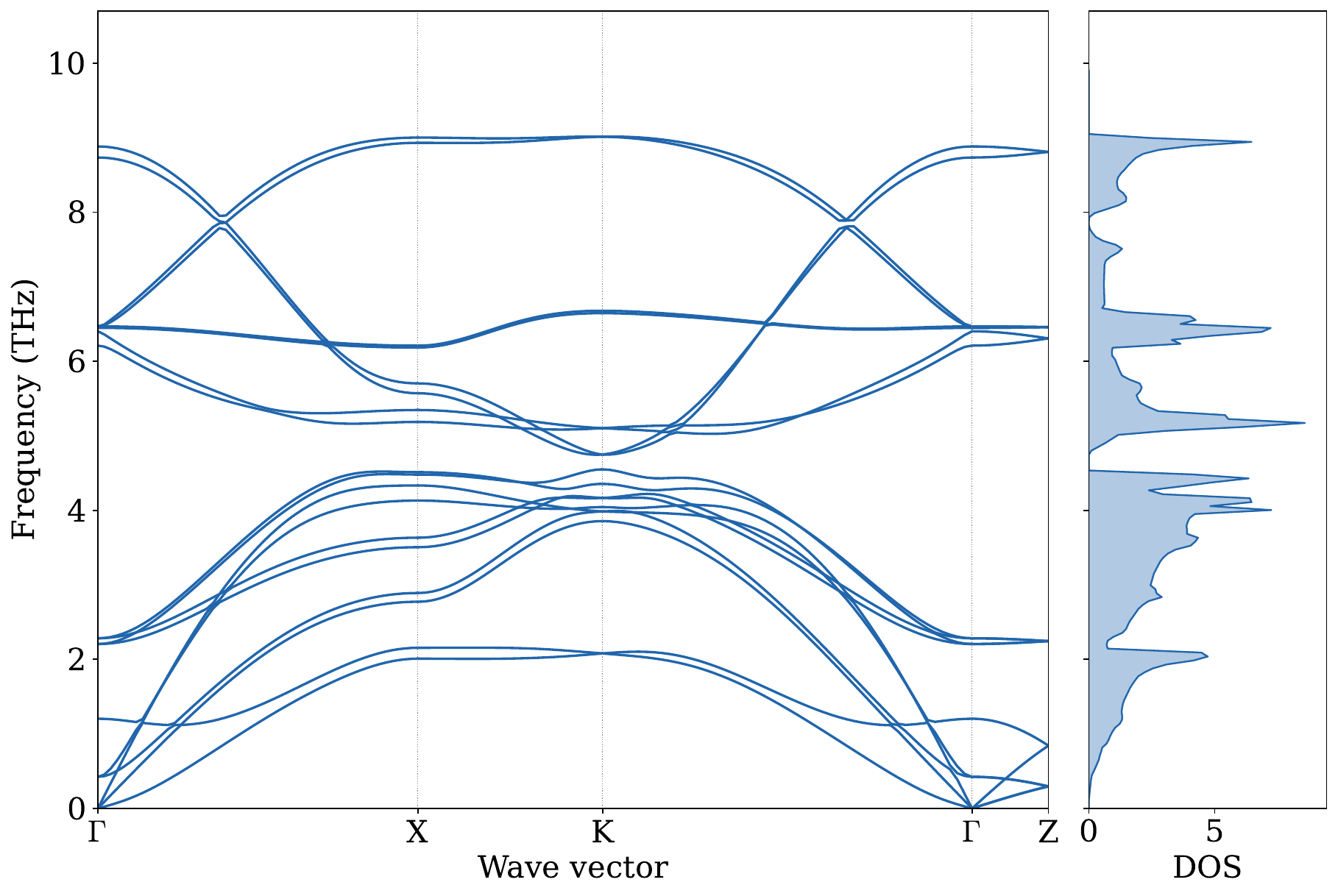}\label{subfig:DTMDC_Hc_DTi_Se2_Li_D0}}
    \subfloat[]{\includegraphics[width=0.5\linewidth]{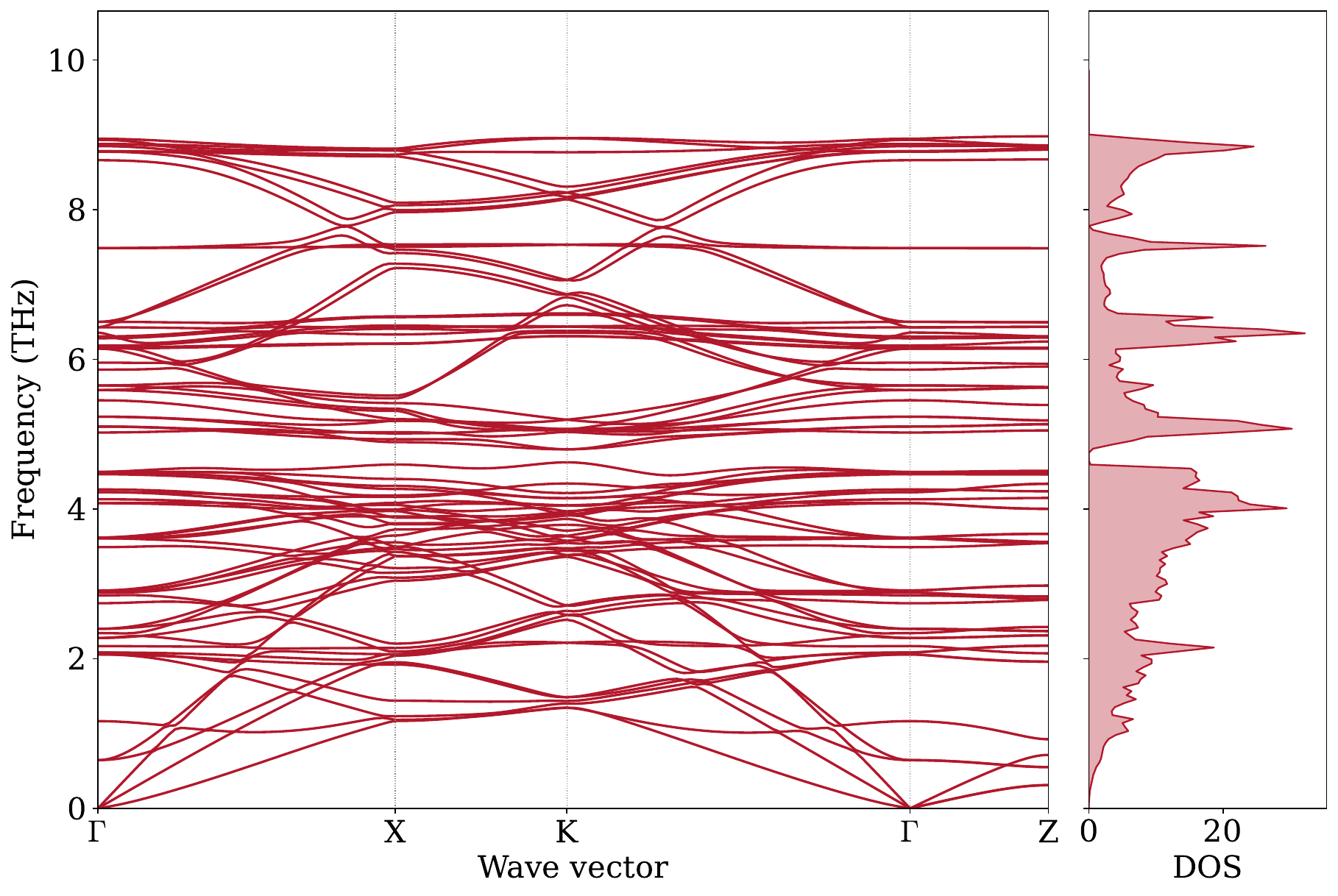}\label{subfig:DTMDC_Hc_DTi_Se2_Li_D1}}
    \caption{Phonon dispersion and total DOS for the TiSe$_2$ family: (a)~bulk TiSe$_2$; (b)~Li$_{0.125}$TiSe$_2$.}
    \label{suppfig:DTMDCHc_TiSe}
\end{figure*}

\begin{figure*}[h!]
    \centering
    \subfloat[]{\includegraphics[width=0.5\linewidth]{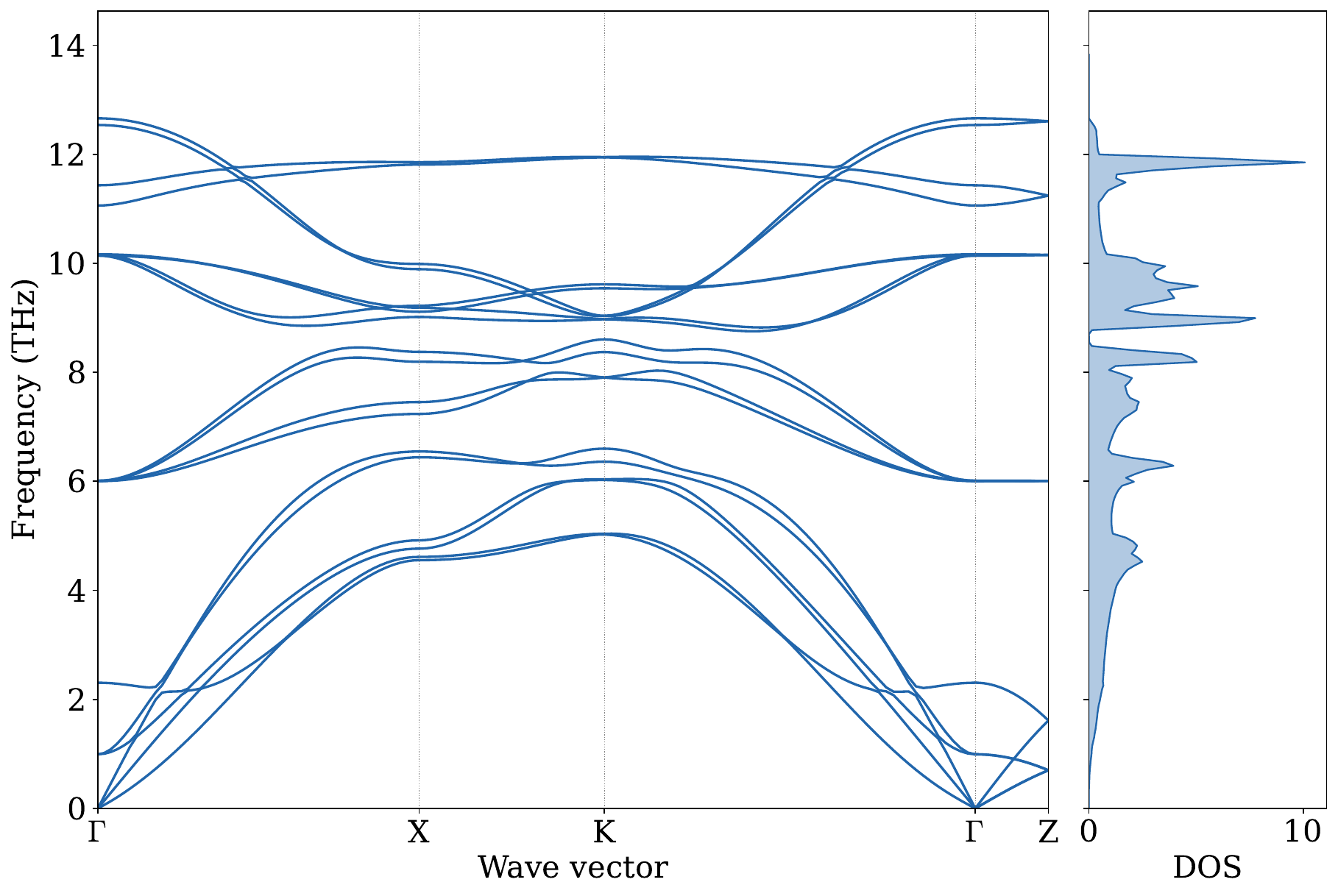}\label{subfig:DTMDC_Hc_DV_S2_Mg_D0}}
    \subfloat[]{\includegraphics[width=0.5\linewidth]{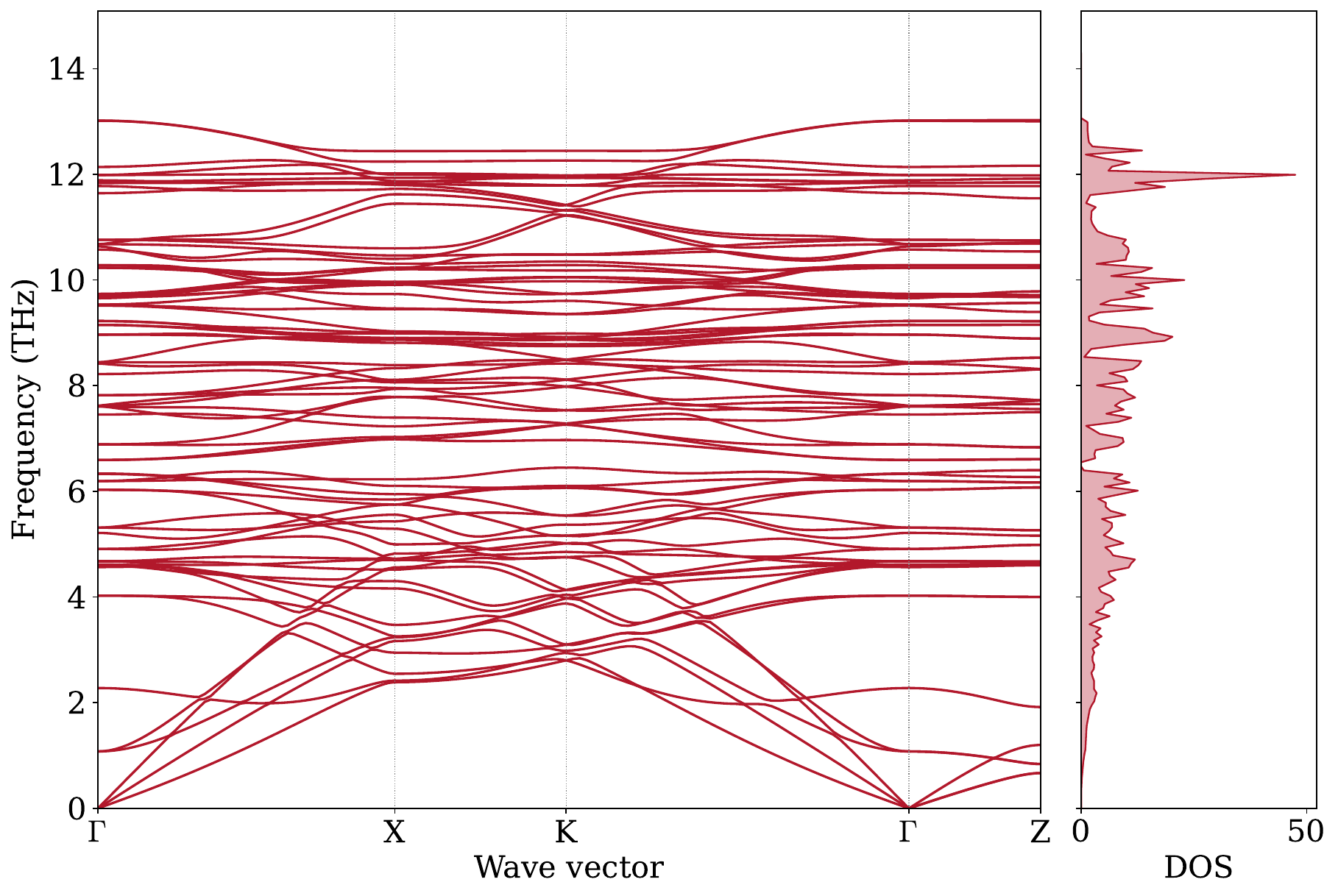}\label{subfig:DTMDC_Hc_DV_S2_Mg_D1}}
    \caption{Phonon dispersion and total DOS for the VS$_2$ family: (a)~bulk VS$_2$; (b)~Mg$_{0.125}$VS$_2$.}
    \label{suppfig:DTMDCHc_VS}
\end{figure*}

\begin{figure*}[h!]
    \centering
    \subfloat[]{\includegraphics[width=0.5\linewidth]{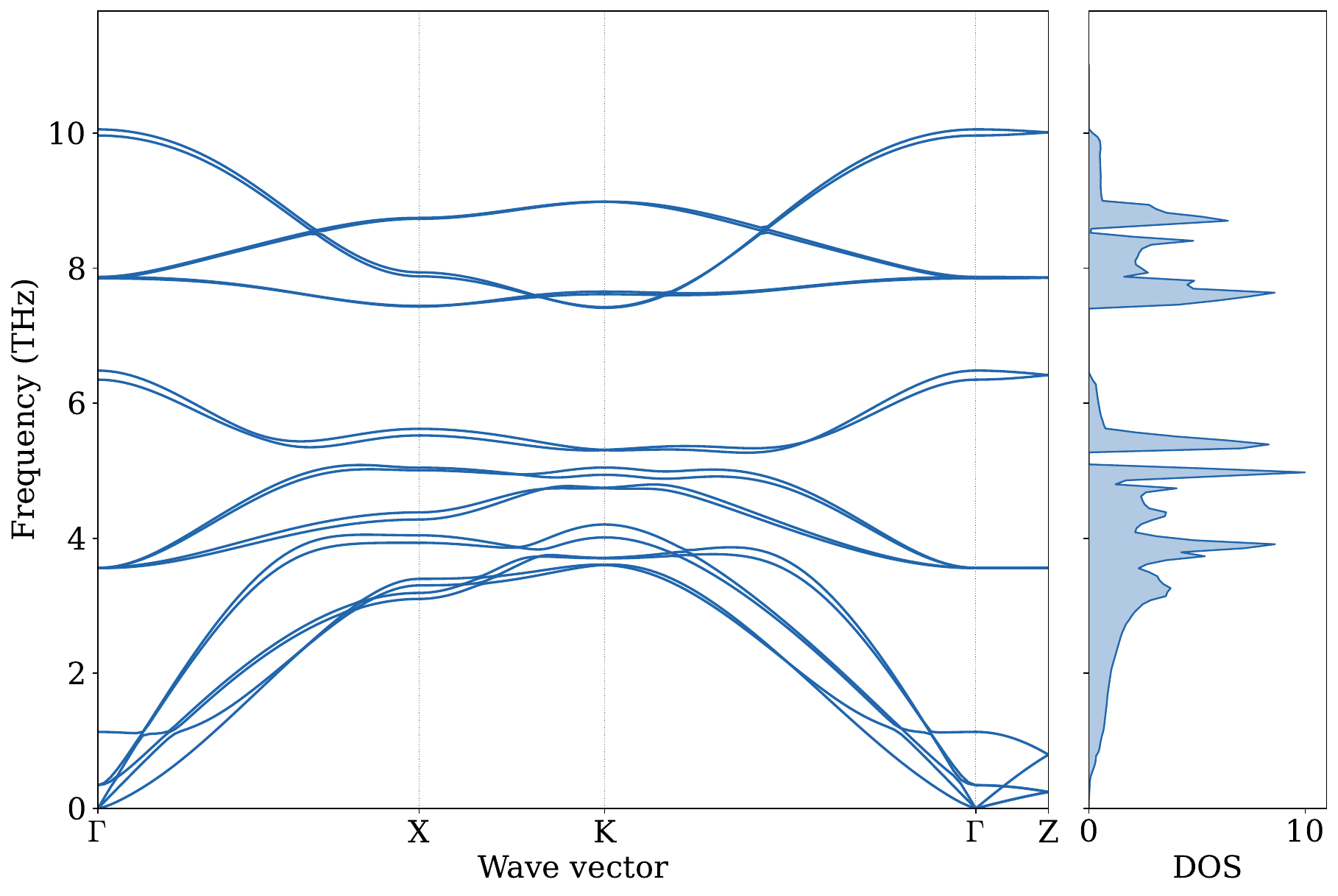}\label{subfig:DTMDC_Hc_DV_Se2_Li_D0}}
    \subfloat[]{\includegraphics[width=0.5\linewidth]{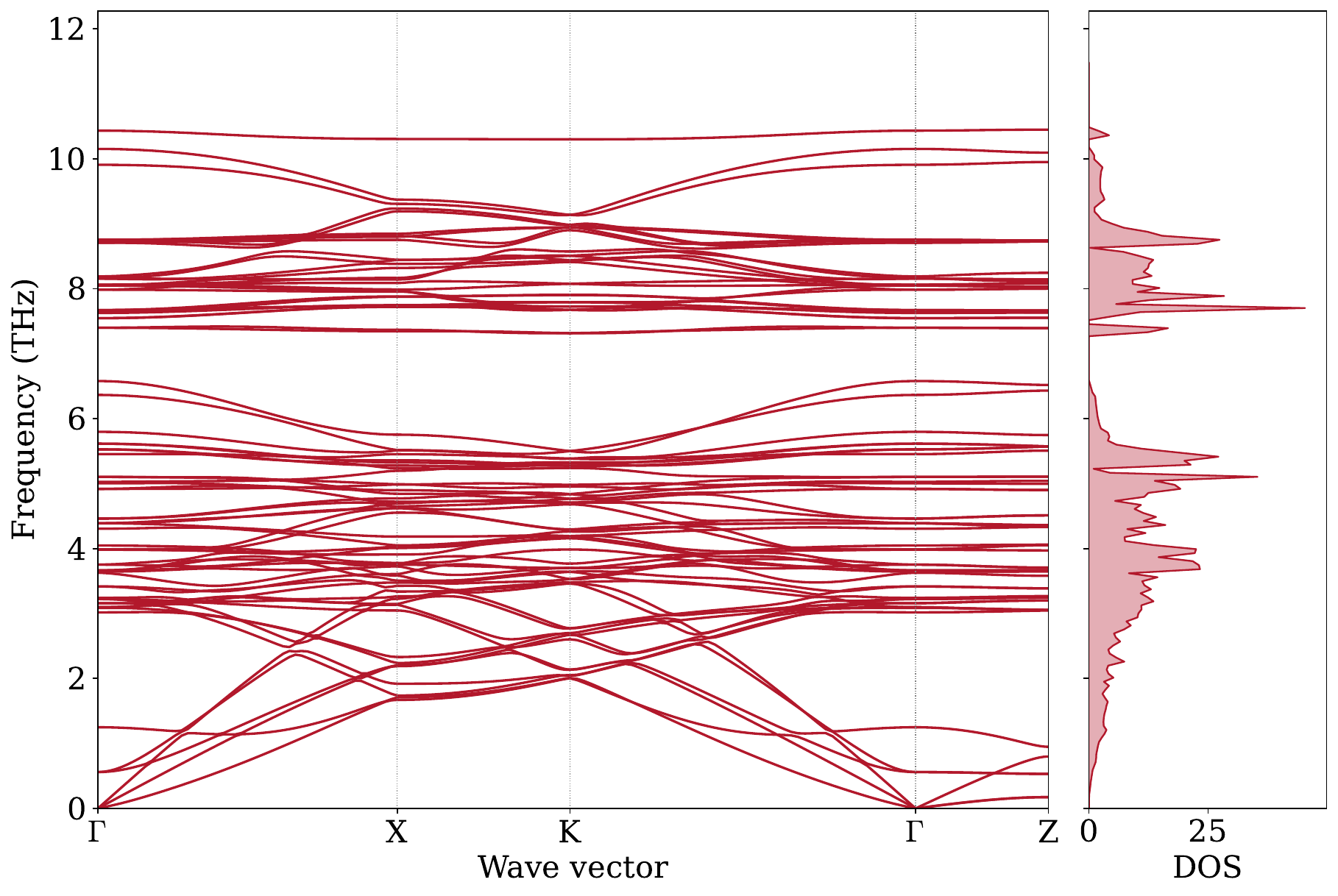}\label{subfig:DTMDC_Hc_DV_Se2_Li_D1}}
    \hfill
    \subfloat[]{\includegraphics[width=0.5\linewidth]{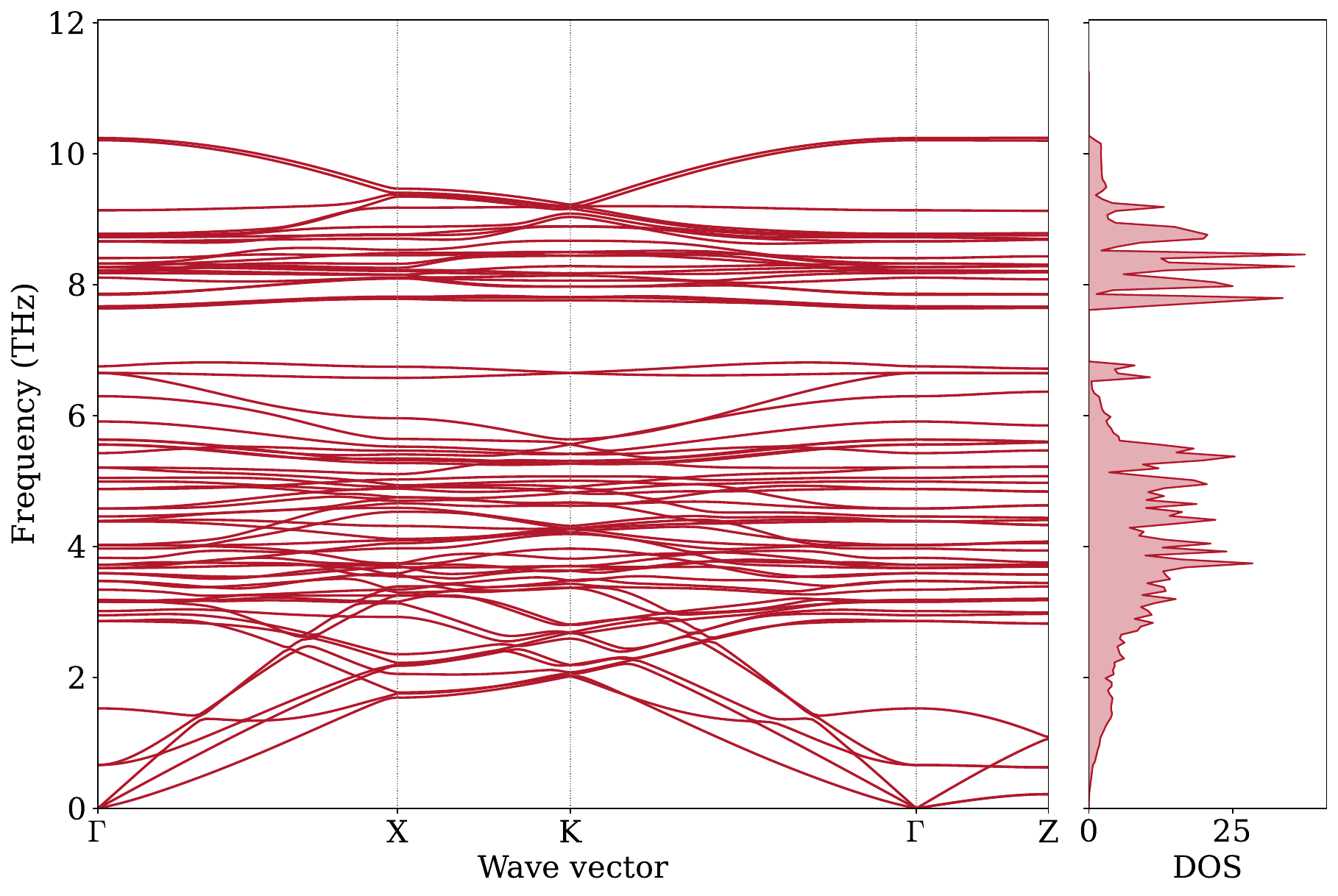}\label{subfig:DTMDC_Hc_DV_Se2_Mg_D1}}
    \caption{Phonon dispersion and total DOS for the VSe$_2$ family: (a)~bulk VSe$_2$; (b)~Li$_{0.125}$VSe$_2$; (c)~Mg$_{0.125}$VSe$_2$.}
    \label{suppfig:DTMDCHc_VSe}
\end{figure*}

\begin{figure*}[h!]
    \centering
    \subfloat[]{\includegraphics[width=0.5\linewidth]{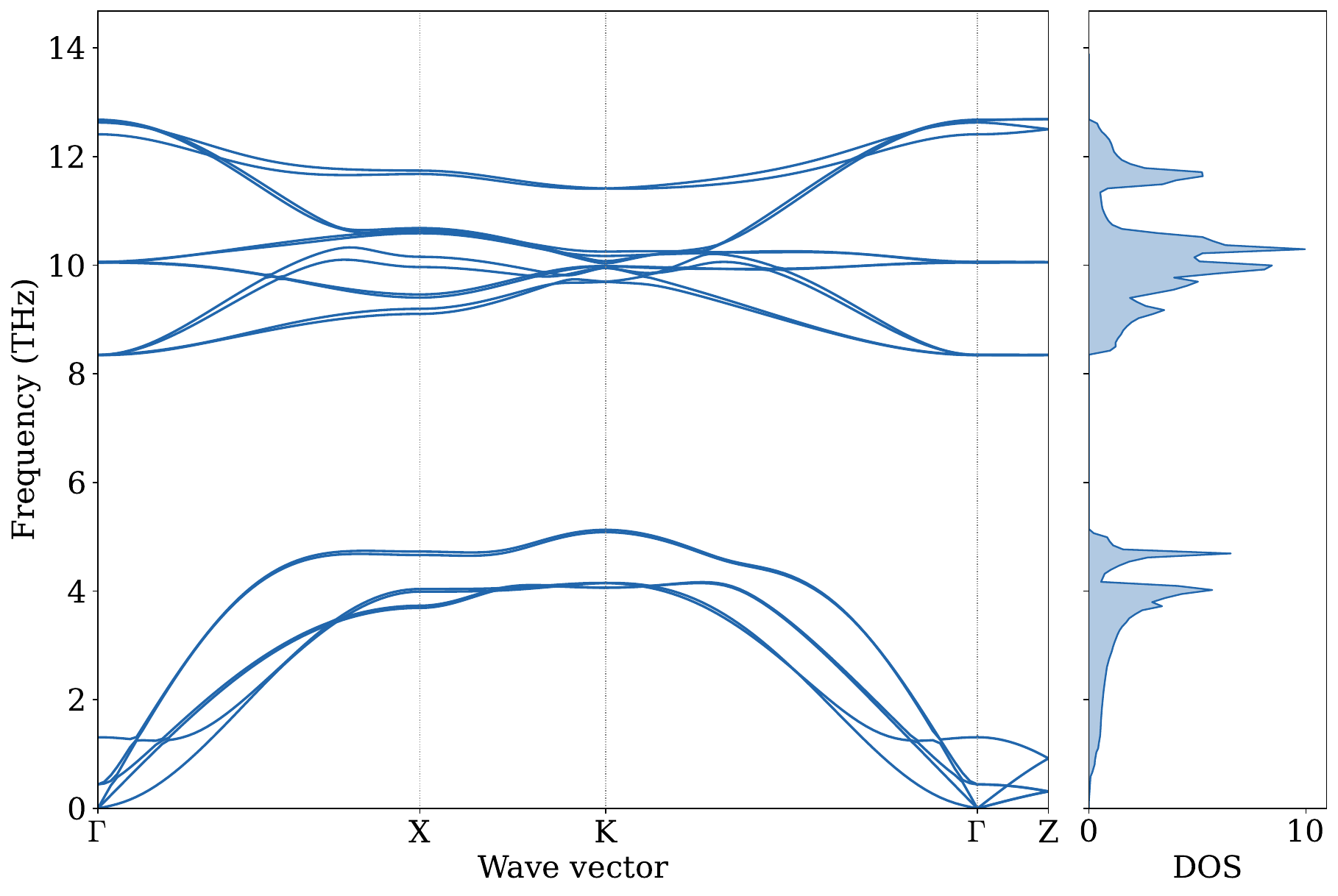}\label{subfig:DTMDC_Hc_DW_S2_Li_D0}}
    \subfloat[]{\includegraphics[width=0.5\linewidth]{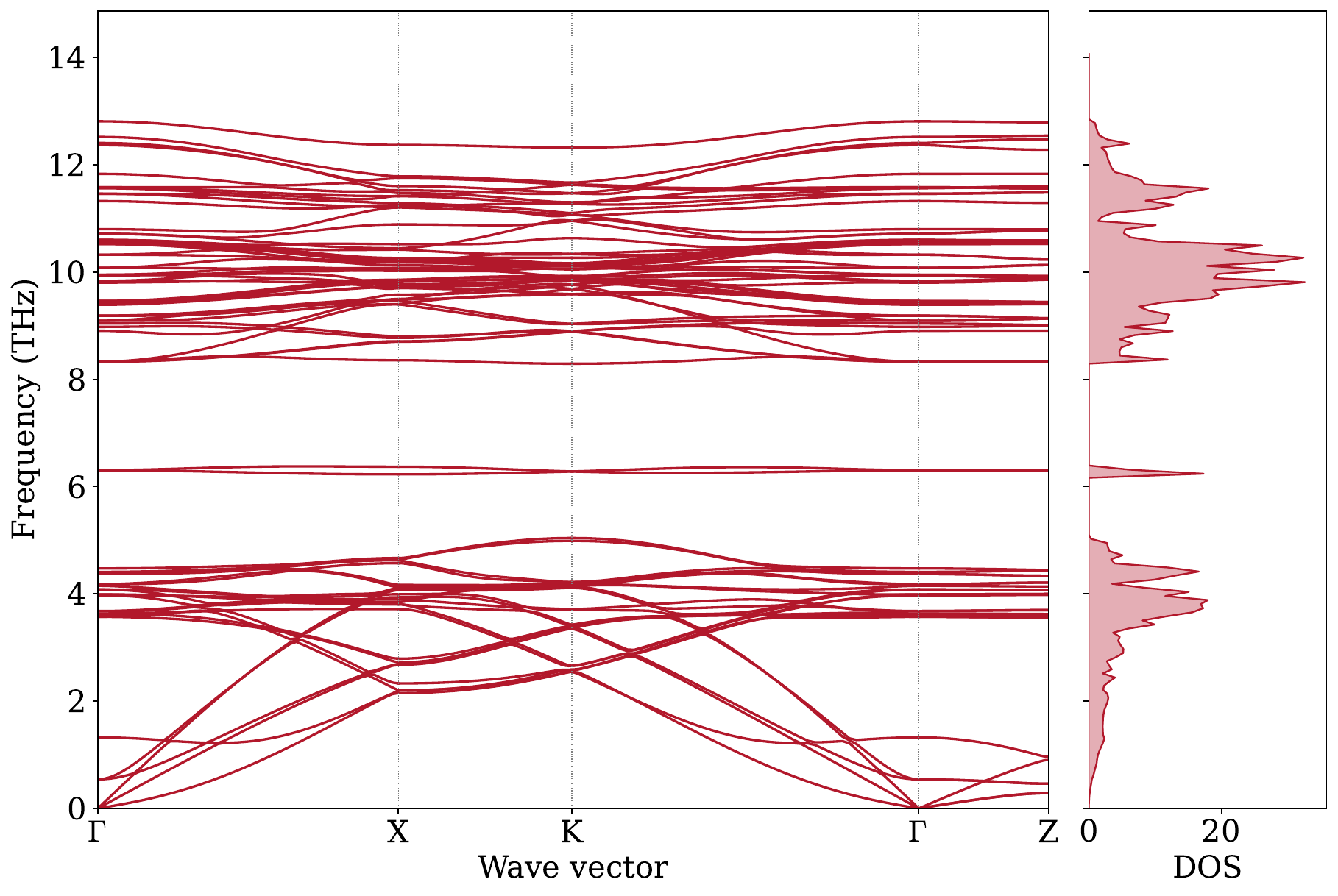}\label{subfig:DTMDC_Hc_DW_S2_Li_D1}}
    \hfill
    \subfloat[]{\includegraphics[width=0.5\linewidth]{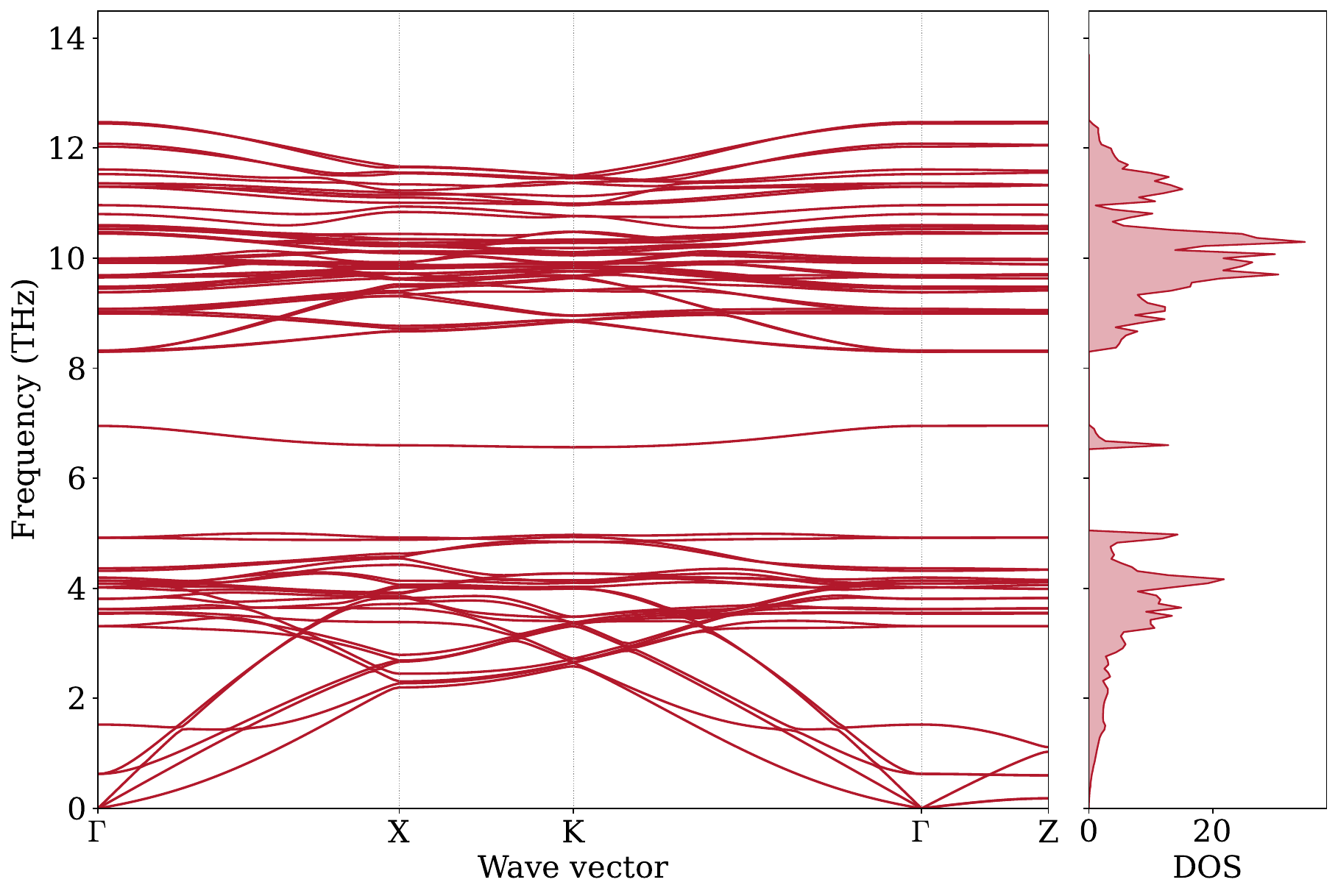}\label{subfig:DTMDC_Hc_DW_S2_Mg_D1}}
    \caption{Phonon dispersion and total DOS for the WS$_2$ family: (a)~bulk WS$_2$; (b)~Li$_{0.125}$WS$_2$; (c)~Mg$_{0.125}$WS$_2$.}
    \label{suppfig:DTMDCHc_WS}
\end{figure*}

\begin{figure*}[h!]
    \centering
    \subfloat[]{\includegraphics[width=0.5\linewidth]{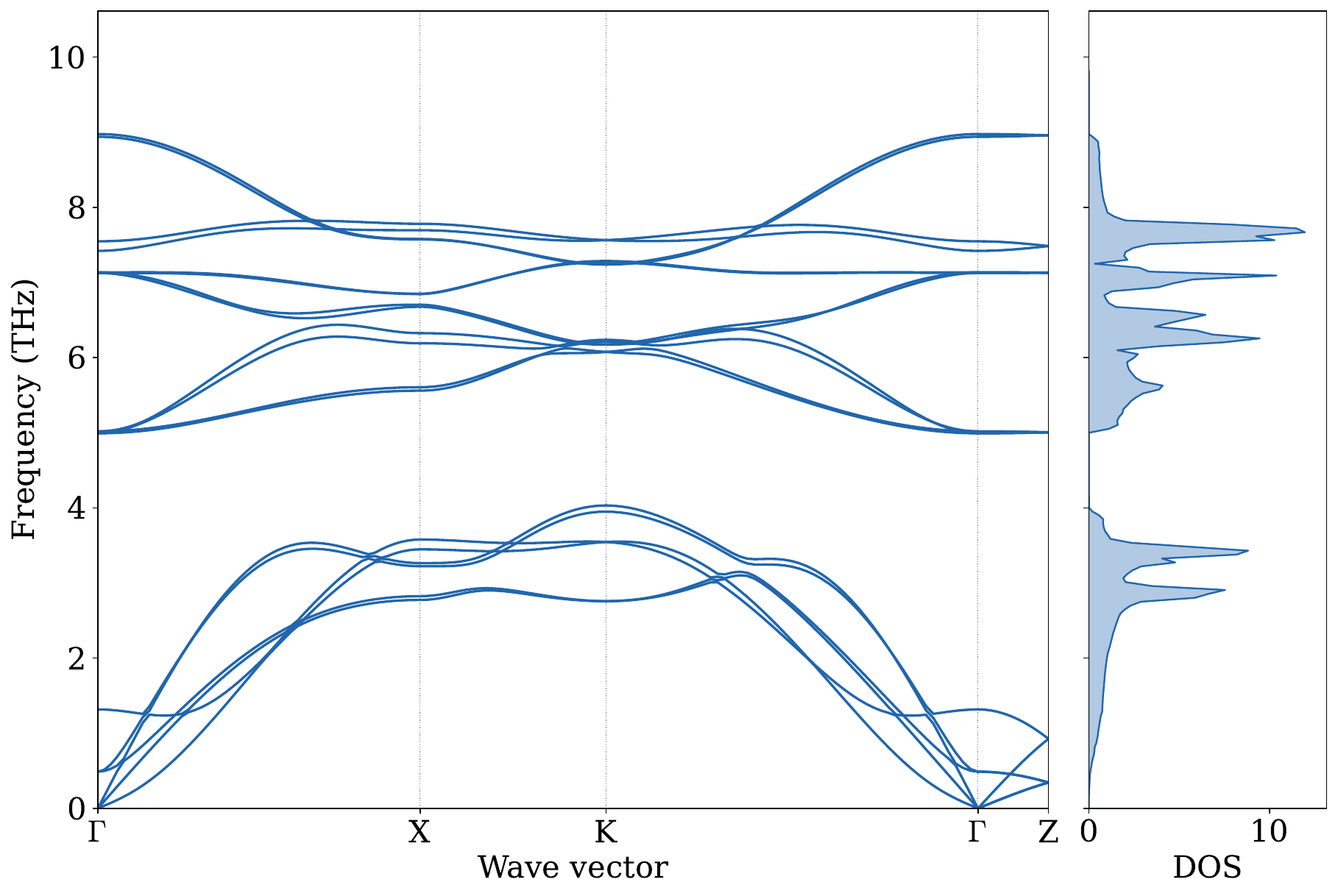}\label{subfig:DTMDC_Hc_DW_Se2_Li_D0}}
    \subfloat[]{\includegraphics[width=0.5\linewidth]{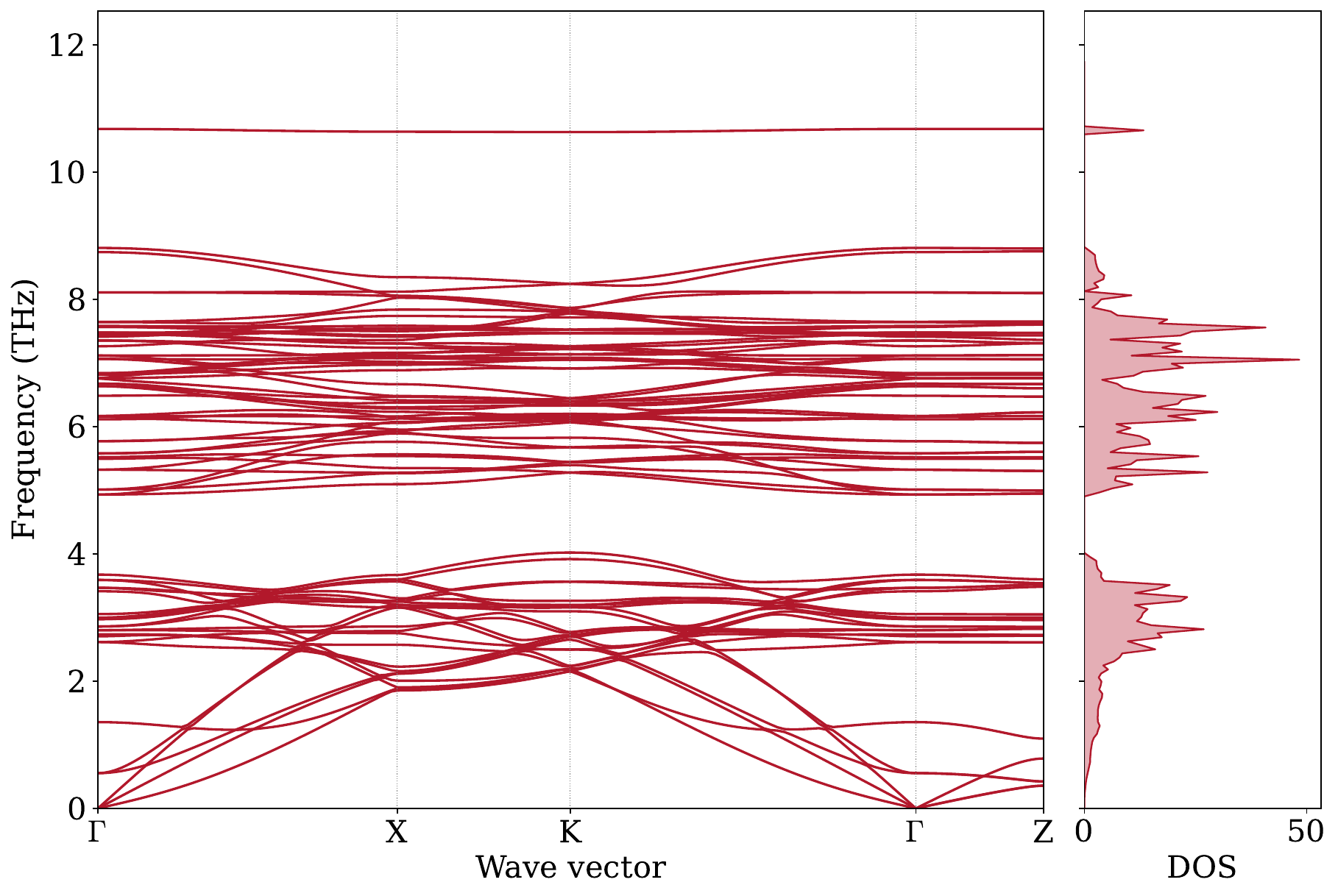}\label{subfig:DTMDC_Hc_DW_Se2_Li_D1}}
    \hfill
    \subfloat[]{\includegraphics[width=0.5\linewidth]{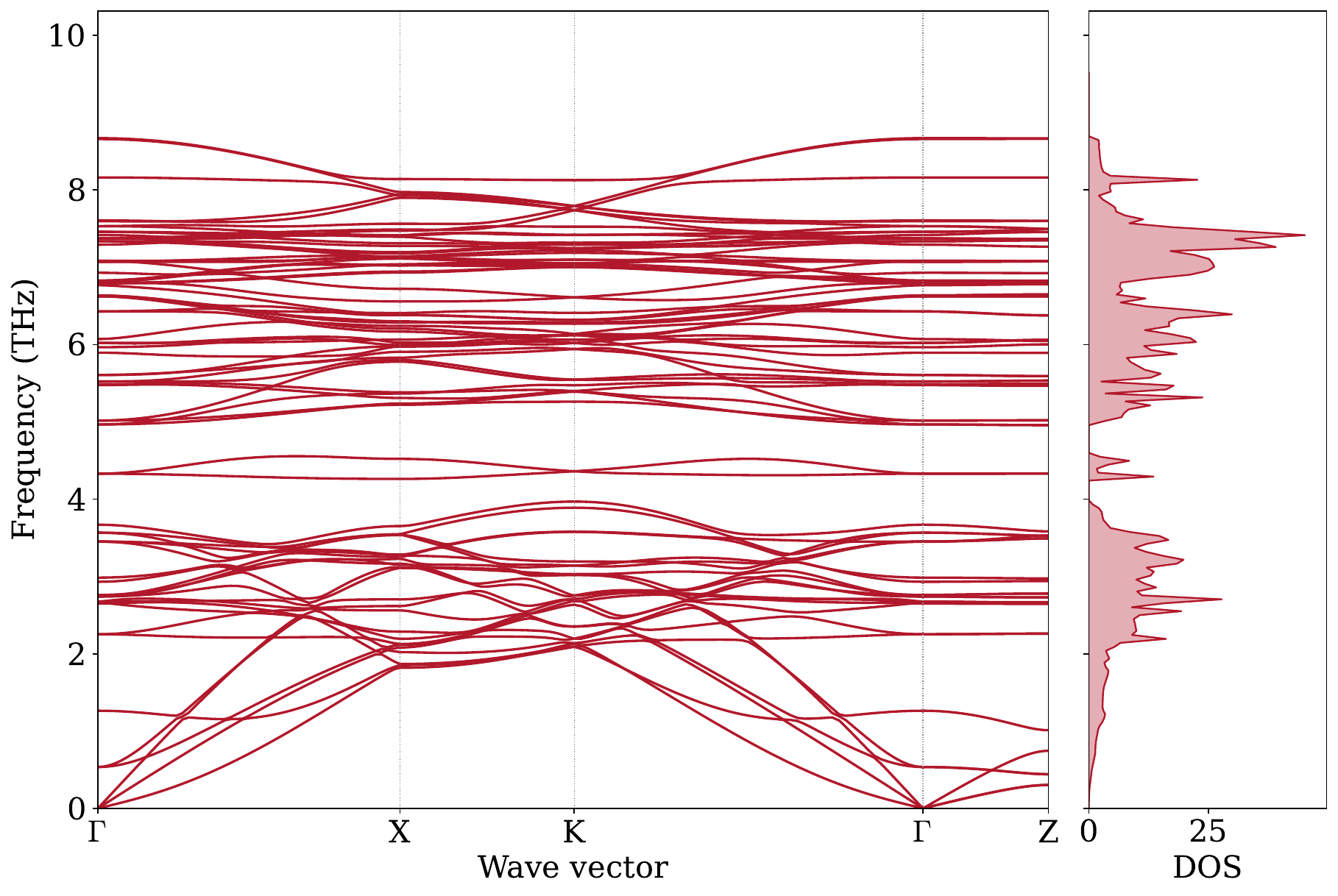}\label{subfig:DTMDC_Hc_DW_Se2_Mg_D1}}
    \caption{Phonon dispersion and total DOS for the WSe$_2$ family: (a)~bulk WSe$_2$; (b)~Li$_{0.125}$WSe$_2$; (c)~Mg$_{0.125}$WSe$_2$.}
    \label{suppfig:DTMDCHc_WSe}
\end{figure*}

\clearpage

\section{Intercalation voltage of Li$_a$ZrS$_2$}
\label{suppsec:voltage}

\begin{figure}[h!]
    \centering
    \includegraphics[width=0.5\linewidth]{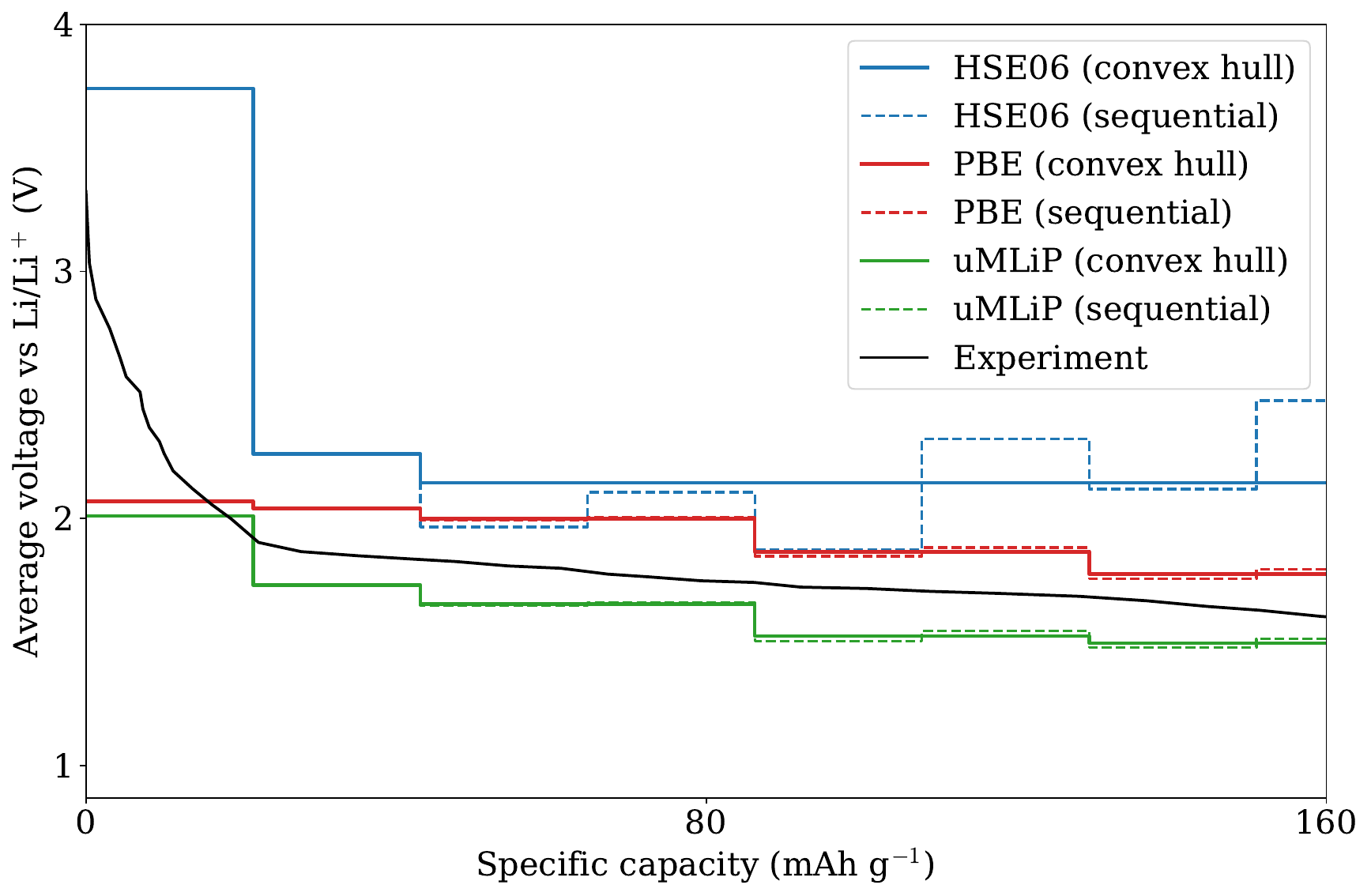}
    \caption{Sequential and convex-hull voltage profiles of Li$_a$ZrS$_2$. Intercalation voltage versus specific capacity calculated with the PBE~\cite{Perdew1996} and HSE06~\cite{Heyd2003,Heyd2006} functionals and with the uMLiP~\cite{NEQUIP}, compared with the experimental discharge curve~\cite{Kim2021}. Sequential profiles are drawn as dashed lines and convex-hull profiles as solid lines.}
    \label{suppfig:SequenceVoltage}
\end{figure}

\figref{suppfig:SequenceVoltage} compares the results of calculating the voltage dependency as a function of intercalation using either DFT ((PBE~\cite{Perdew1996} or HSE06~\cite{Heyd2003,Heyd2006})) or the uMLiP~\cite{NEQUIP}) with the experimental discharge curve~\cite{Kim2021}. For each approach, two profiles are plotted: a sequential profile (dashed lines), in which the voltage is evaluated between successive calculated Li contents, and a convex-hull profile (solid lines), which connects only the chemically stable states. For PBE and the uMLiP the two profiles are nearly coincident, whereas HSE06 identifies that only $a = 0$, $0.125$, $0.25$ and $1$ lie on the convex hull, as discussed in the main text.

\section{Volume change under intercalation}
\label{suppsec:volume}

\begin{figure*}[htbp]
    \centering
    \includegraphics[width=0.75\linewidth]{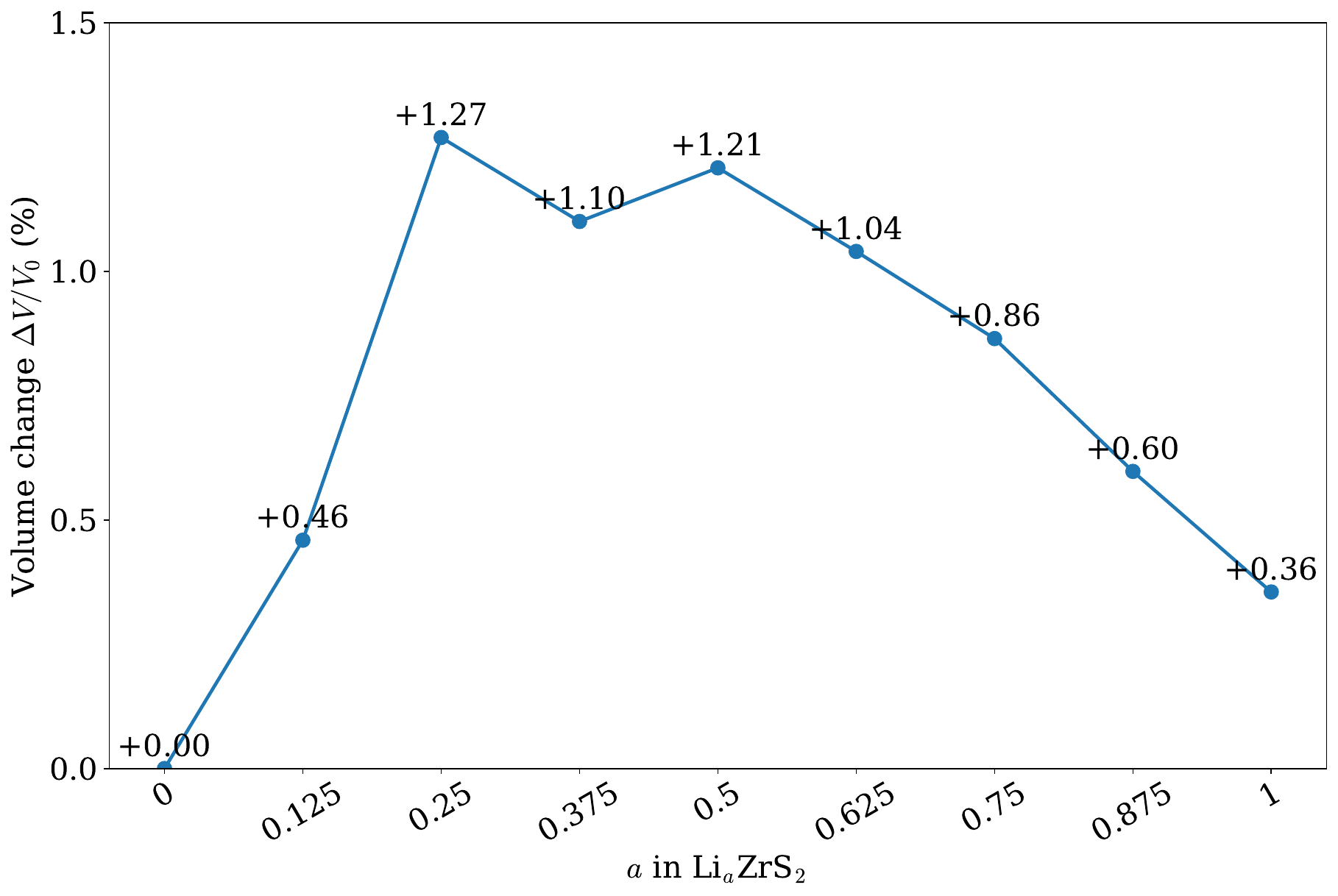}
    \caption{Volume change of Li$_a$ZrS$_2$ on intercalation. Relaxed cell volume relative to the pristine host as a function of Li content, from PBE.}
    \label{suppfig:VolumeExpansion}
\end{figure*}

\figref{suppfig:VolumeExpansion} shows the relaxed cell volume of Li$_a$ZrS$_2$ (PBE) as a function of Li content, expressed relative to the pristine host. The host expands only weakly: the maximum expansion is $\approx$1.3\% at $a = 0.25$, and the volume returns to within $\approx$0.4\% of the pristine value at full lithiation, so the net change across the whole intercalation range is $\approx$0.4\%. This anomalously small chemo-mechanical strain, consistent with the first-principles calculations of Ref.~\cite{Price2023}, is a practical advantage for a thermal switch because the host lattice is only minimally distorted as Li is inserted and extracted. Only the PBE volumes are shown: the uMLiP-relaxed cells used for the voltage profiles of \secref{suppsec:voltage} differ systematically (each cell is consistently larger than  the equivalent PBE result) and are not used in any phonon calculation.

\section{Phonon dispersions of Li$_a$ZrS$_2$}
\label{suppsec:band_dispersions}

We present here the dispersion relations and densities of states, calculated with DFT (solid lines) and with the uMLiP (dashed lines), which complement the $a=0$, $0.125$, $0.75$ and $1$ dispersions shown in Fig.~2 of the main text, so that all nine intercalation levels ($a=0$ to $1$ in steps of $0.125$) are covered. None of the dispersions shows imaginary (soft) modes, confirming that every composition considered here is dynamically stable. The two methods agree closely up to $a=0.5$, with a mean absolute deviation over the full $\Gamma$--X--K--$\Gamma$--Z path of $\approx$0.11~THz over all branches ($\approx$0.18~THz for modes below 3~THz at $a=0$, falling to $\approx$0.08~THz at $a=0.5$); at $a=0.75$ and $a=1$ the uMLiP does not reproduce the low-frequency acoustic branches, with the mean deviation below 3~THz rising to $\approx$0.28 and $\approx$0.40~THz, as discussed in the main text.

\paragraph{Li$_{0.125}$ZrS$_2$}

\begin{figure*}[h!]
    \centering
    \subfloat{\includegraphics[width=0.75\linewidth]{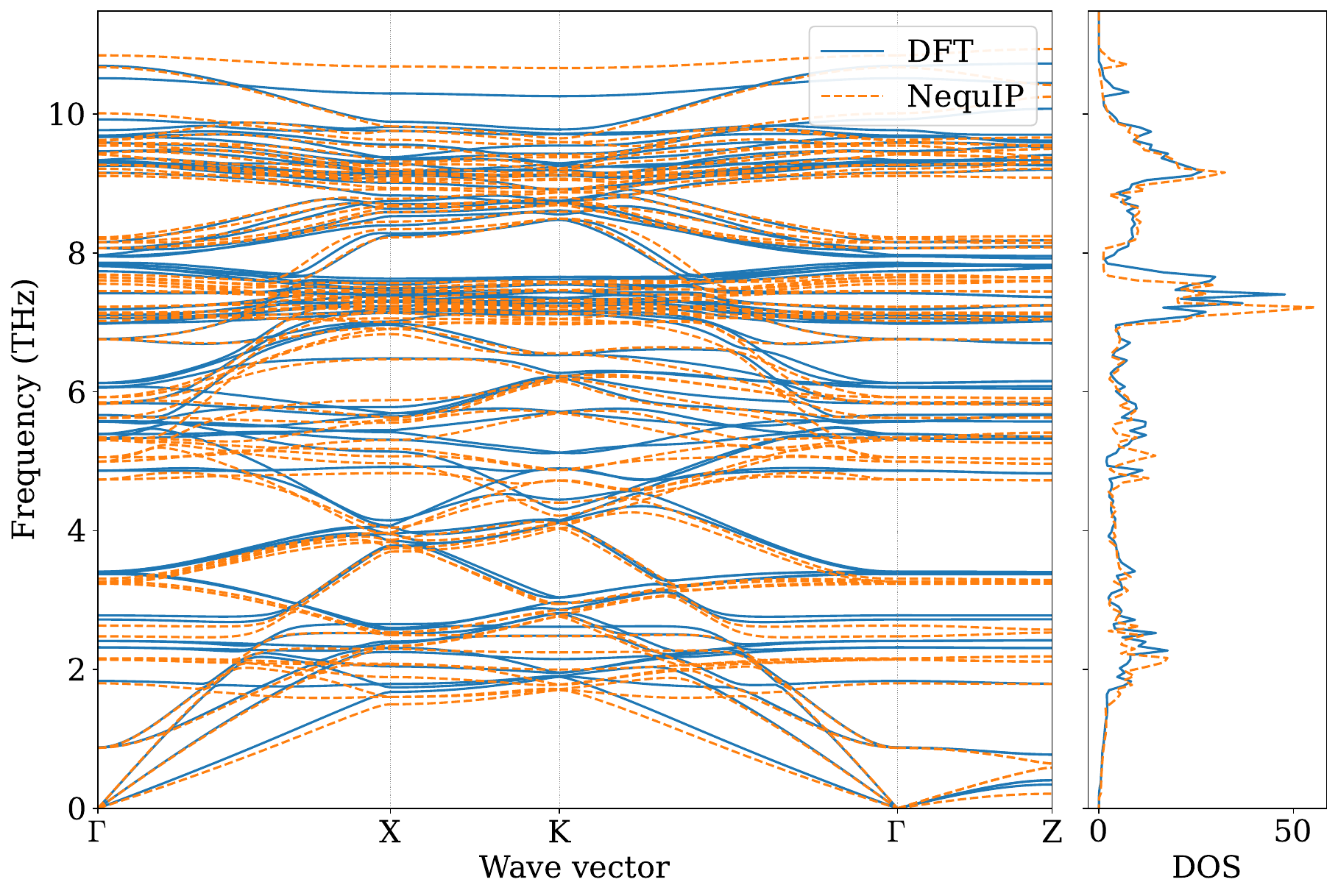}\label{subfig:DFT_100_Li1}}
    \caption{Phonon dispersion and density of states of Li$_{0.125}$ZrS$_2$. Band structure and total DOS calculated with DFT (solid lines) and with the uMLiP (dashed lines).}
    \label{suppfig:Li1BAND}
\end{figure*}

\paragraph{Li$_{0.25}$ZrS$_2$}

\begin{figure*}[h!]
    \centering
    \subfloat{\includegraphics[width=0.75\linewidth]{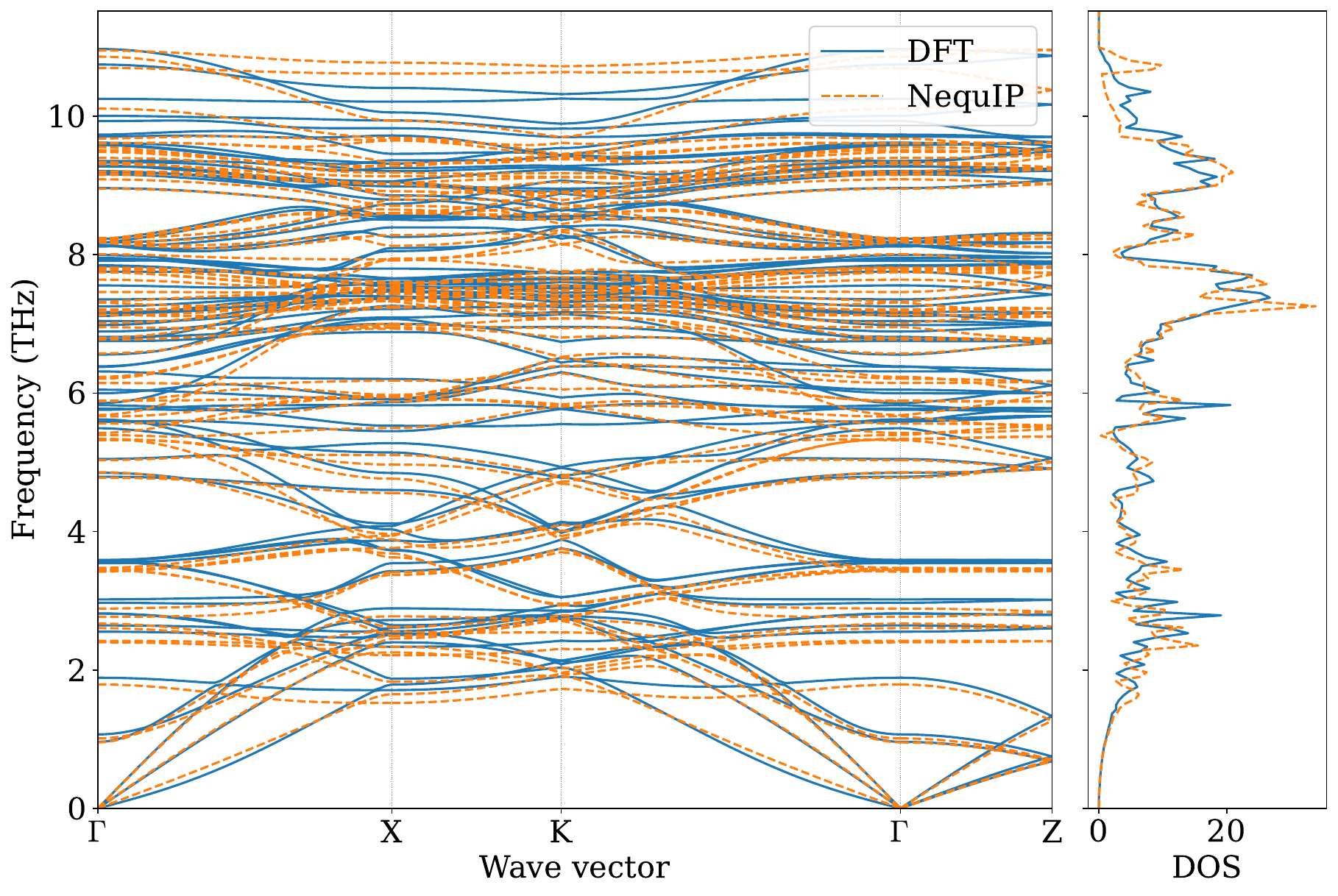}\label{subfig:DFT_100_Li2}}
    \caption{Phonon dispersion and density of states of Li$_{0.25}$ZrS$_2$. Band structure and total DOS calculated with DFT (solid lines) and with the uMLiP (dashed lines).}
    \label{suppfig:Li2BAND}
\end{figure*}

\paragraph{Li$_{0.375}$ZrS$_2$}

\begin{figure*}[h!]
    \centering
    \subfloat{\includegraphics[width=0.75\linewidth]{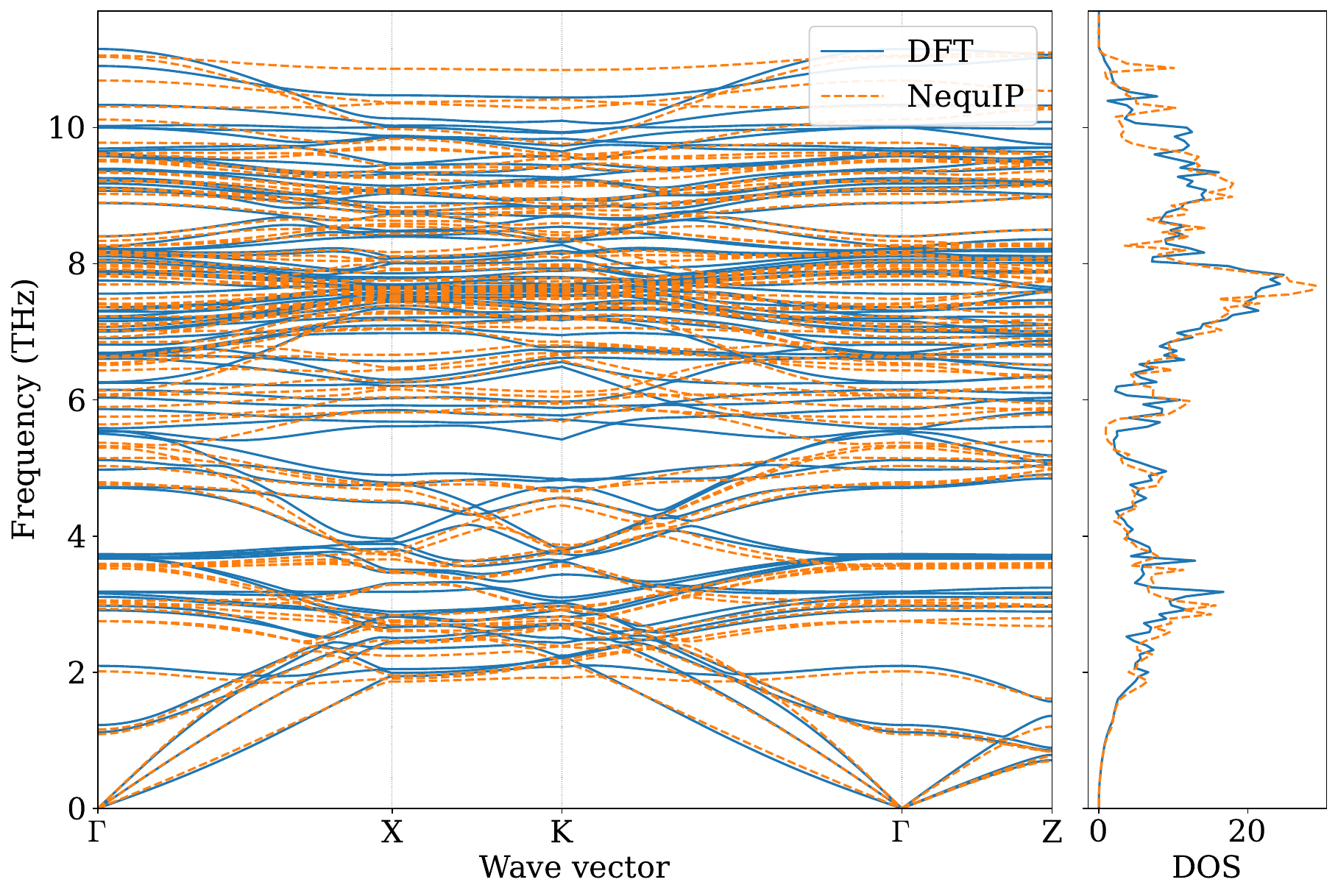}\label{subfig:DFT_100_Li3}}
    \caption{Phonon dispersion and density of states of Li$_{0.375}$ZrS$_2$. Band structure and total DOS calculated with DFT (solid lines) and with the uMLiP (dashed lines).}
    \label{suppfig:Li3BAND}
\end{figure*}

\paragraph{Li$_{0.5}$ZrS$_2$}

\begin{figure*}[h!]
    \centering
    \subfloat{\includegraphics[width=0.75\linewidth]{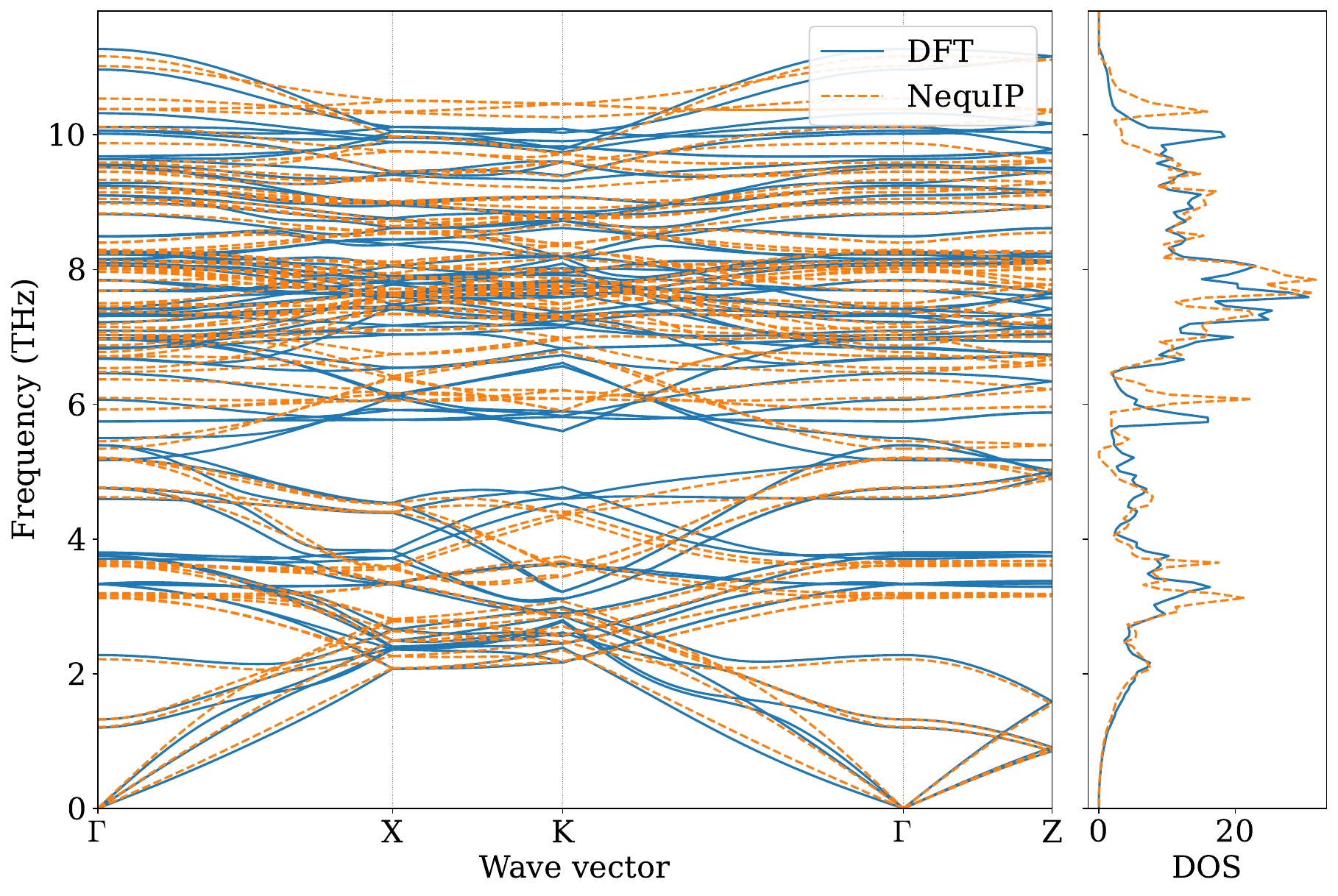}\label{subfig:DFT_100_Li4}}
    \caption{Phonon dispersion and density of states of Li$_{0.5}$ZrS$_2$. Band structure and total DOS calculated with DFT (solid lines) and with the uMLiP (dashed lines).}
    \label{suppfig:Li4BAND}
\end{figure*}

\paragraph{Li$_{0.625}$ZrS$_2$}

\begin{figure*}[h!]
    \centering
    \subfloat{\includegraphics[width=0.75\linewidth]{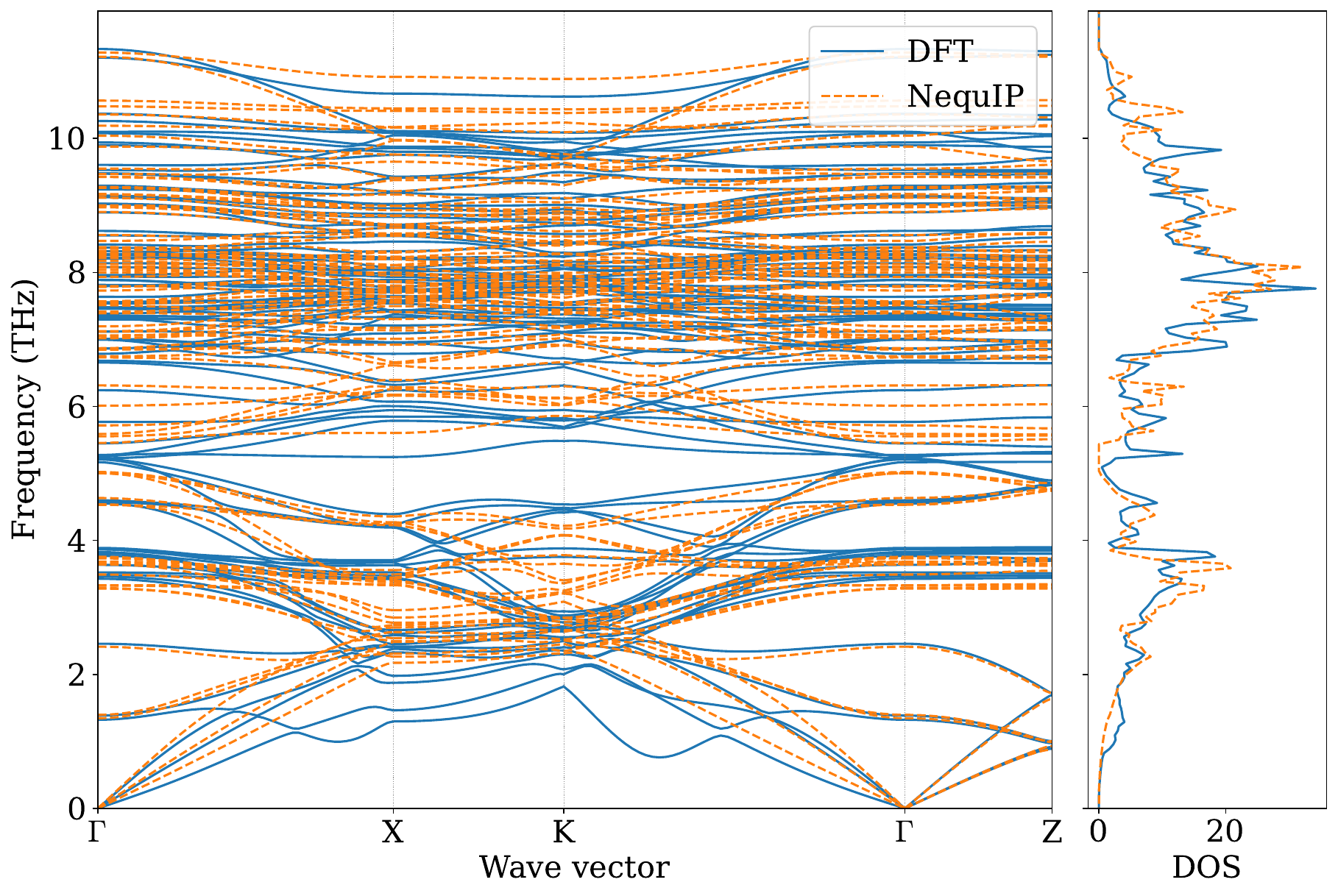}\label{subfig:DFT_100_Li5}}
    \caption{Phonon dispersion and density of states of Li$_{0.625}$ZrS$_2$. Band structure and total DOS calculated with DFT (solid lines) and with the uMLiP (dashed lines).}
    \label{suppfig:Li5BAND}
\end{figure*}

\paragraph{Li$_{0.875}$ZrS$_2$}

\begin{figure*}[h!]
    \centering
    \subfloat{\includegraphics[width=0.75\linewidth]{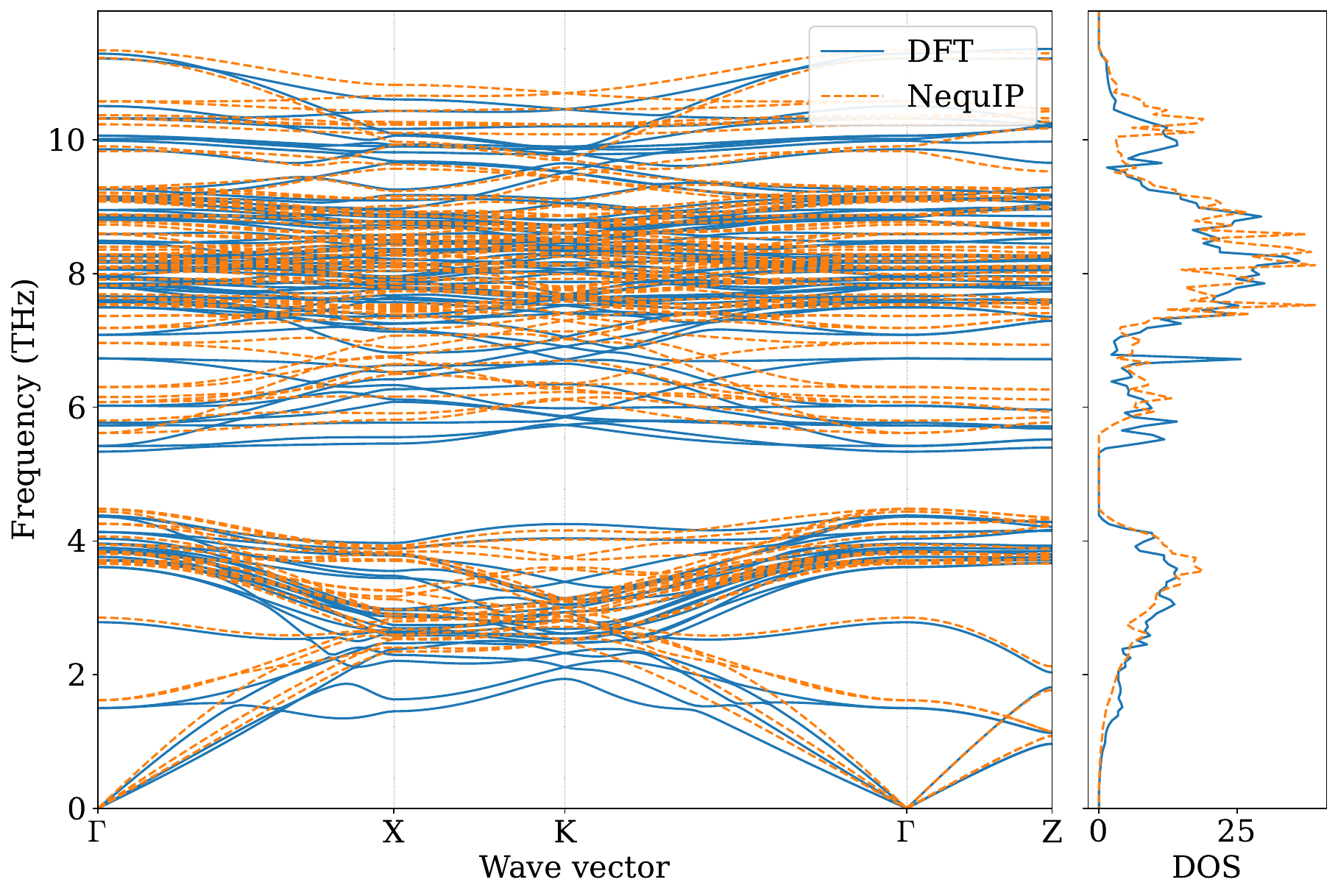}\label{subfig:DFT_100_Li7}}
    \caption{Phonon dispersion and density of states of Li$_{0.875}$ZrS$_2$. Band structure and total DOS calculated with DFT (solid lines) and with the uMLiP (dashed lines).}
    \label{suppfig:Li7BAND}
\end{figure*}

\clearpage

\section{Heat capacity of Li$_a$ZrS$_2$}
\label{suppsec:HeatCapacity}

\begin{figure*}[h!]
    \centering
    \includegraphics[width=\linewidth]{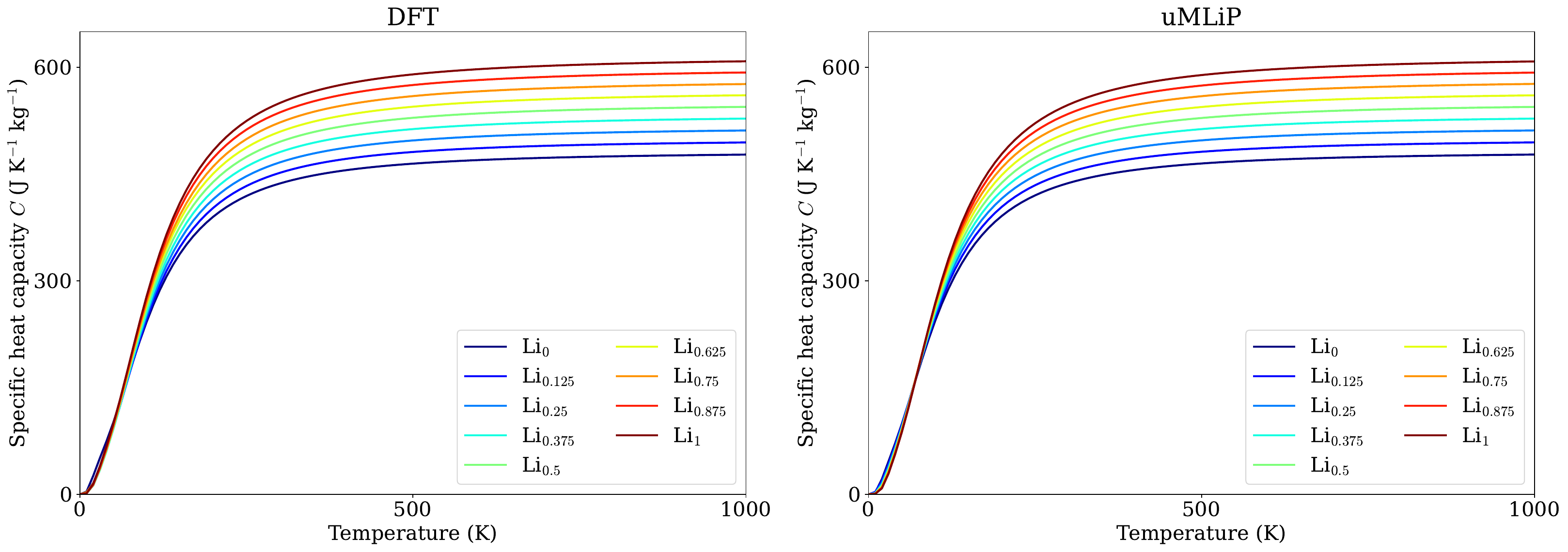}
    \caption{Specific heat capacity of Li$_a$ZrS$_2$. Specific heat capacity $C_v$ per unit mass versus temperature for $a = 0$ to $1$ in steps of $0.125$, from phonon calculations with DFT (left) and with the uMLiP (right). Each intercalation level is drawn in a different colour.}
    \label{suppfig:HeatCapacity}
\end{figure*}

\figref{suppfig:HeatCapacity} shows the phonon heat capacity as a function of temperature for every intercalation stage of Li$_a$ZrS$_2$ ($a = 0$ to $1$ in steps of $0.125$), computed from the previous phonon calculations with DFT (left) and with the uMLiP (right). The specific heat per unit mass rises monotonically with temperature towards the Dulong--Petit limit and increases with Li content, from $\approx$437 to $\approx$550~J~kg$^{-1}$~K$^{-1}$ at 300~K between $a=0$ and $a=1$, reflecting the light Li atoms added per formula unit. DFT and the uMLiP agree closely at every intercalation level across the full 0--1000~K range, with a mean absolute difference of $\approx$1~J~kg$^{-1}$~K$^{-1}$ that rises to $\approx$2.4~J~kg$^{-1}$~K$^{-1}$ at full lithiation.

\section{Force constants}
\label{suppsec:force_constants}

\begin{figure*}
    \centering
    \subfloat[]{\includegraphics[width=0.75\linewidth]{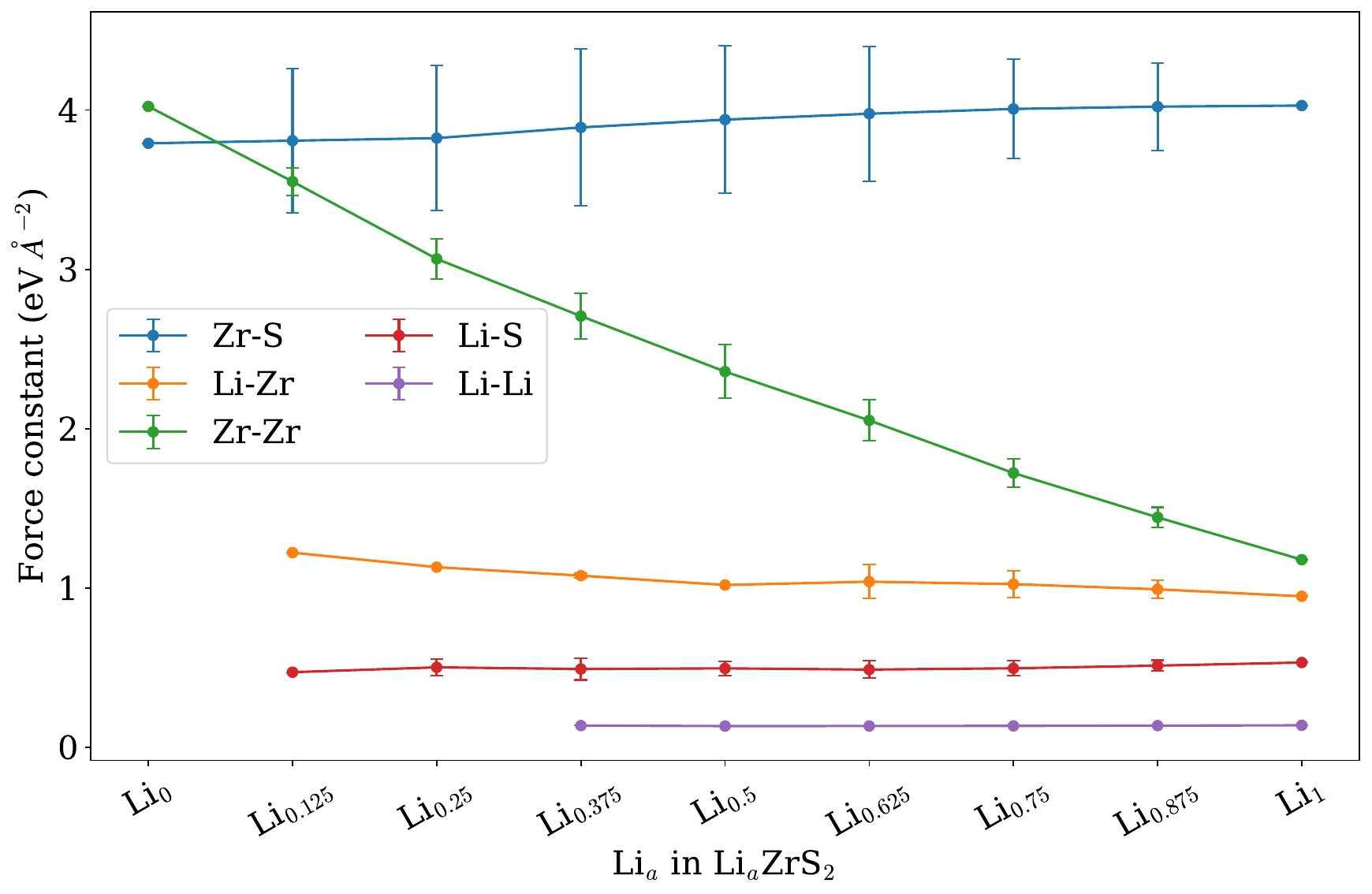}\label{subfig:BondFCVLiNequIP}}
    \caption{Bond-resolved force constants from the uMLiP. Mean second-order force constants of Li$_a$ZrS$_2$, resolved by bond type, as a function of Li content from the uMLiP.}
    \label{suppfig:BondFC}
\end{figure*}

\figref{suppfig:BondFC} shows the bond-resolved second-order force constants of Li$_a$ZrS$_2$ obtained with the uMLiP.
The trends mirror those found with DFT in the main text: the mean Zr--Zr force constant softens strongly with lithiation, the Zr--S bonds undergo slight hardening with a starting and ending in little variance but having strong variance between $a=0$ and $1$, and the newly formed Li--Zr, Li--S and Li--Li bonds are soft but remain relatively constant. The uMLiP reproduces the DFT values almost quantitatively across the full composition range ($4.0 \rightarrow 1.2$~eV\,\AA$^{-2}$ for the mean Zr--Zr constant, against $4.2 \rightarrow 1.2$~eV\,\AA$^{-2}$ from DFT), which supports the phonon and thermal-transport calculations.

\section{Phonon group velocities}
\label{suppsec:group_velocity}

\begin{figure*}[htbp]
    \centering
    \includegraphics[width=0.9\linewidth]{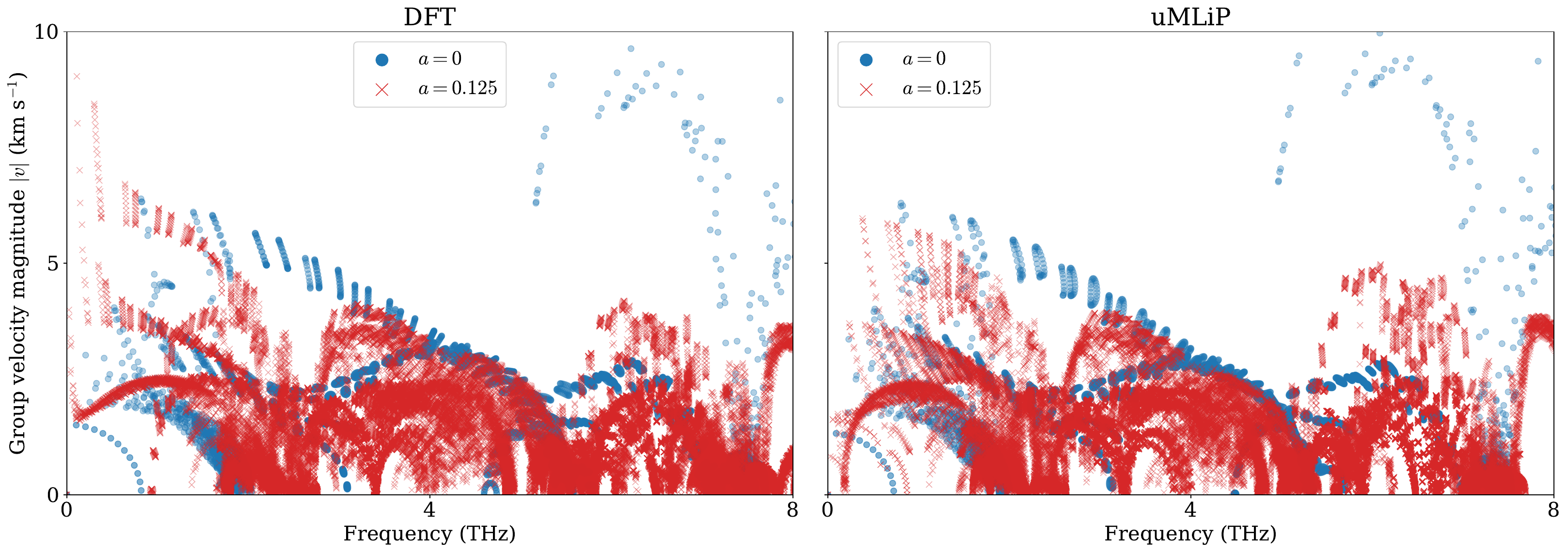}
    \caption{Phonon group velocities of ZrS$_2$ and Li$_{0.125}$ZrS$_2$. Group-velocity magnitude $|v|$ versus frequency at 300~K for pristine ZrS$_2$ ($a=0$) and Li$_{0.125}$ZrS$_2$ ($a=0.125$), from DFT (left) and from the uMLiP (right).}
    \label{suppfig:GroupVelocity}
\end{figure*}

\figref{suppfig:GroupVelocity} shows the phonon group-velocity magnitude $|v|$ as a function of frequency for pristine ZrS$_2$ and Li$_{0.125}$ZrS$_2$, computed with DFT (left) and the uMLiP (right). In the acoustic range below $\approx$3~THz that carries most of the heat, the mode velocities are consistently lower in the lithiated phase for both methods: the median velocity of the acoustic modes falls from $\approx$1.0 to $\approx$0.7~km~s$^{-1}$ in DFT and from $\approx$1.2 to $\approx$0.8~km~s$^{-1}$ in the uMLiP, and the mean out-of-plane component falls from $\approx$0.25 to $\approx$0.14~km~s$^{-1}$ and from $\approx$0.30 to $\approx$0.23~km~s$^{-1}$, respectively, providing direct evidence for the softening of the acoustic branches inferred from the force constants and used to interpret the reduction in thermal conductivity.

\section{Phonon lifetimes and thermal conductivity}
\label{suppsec:lifetimes}

\begin{figure*}
    \centering
    \subfloat[]{\includegraphics[width=0.5\linewidth]{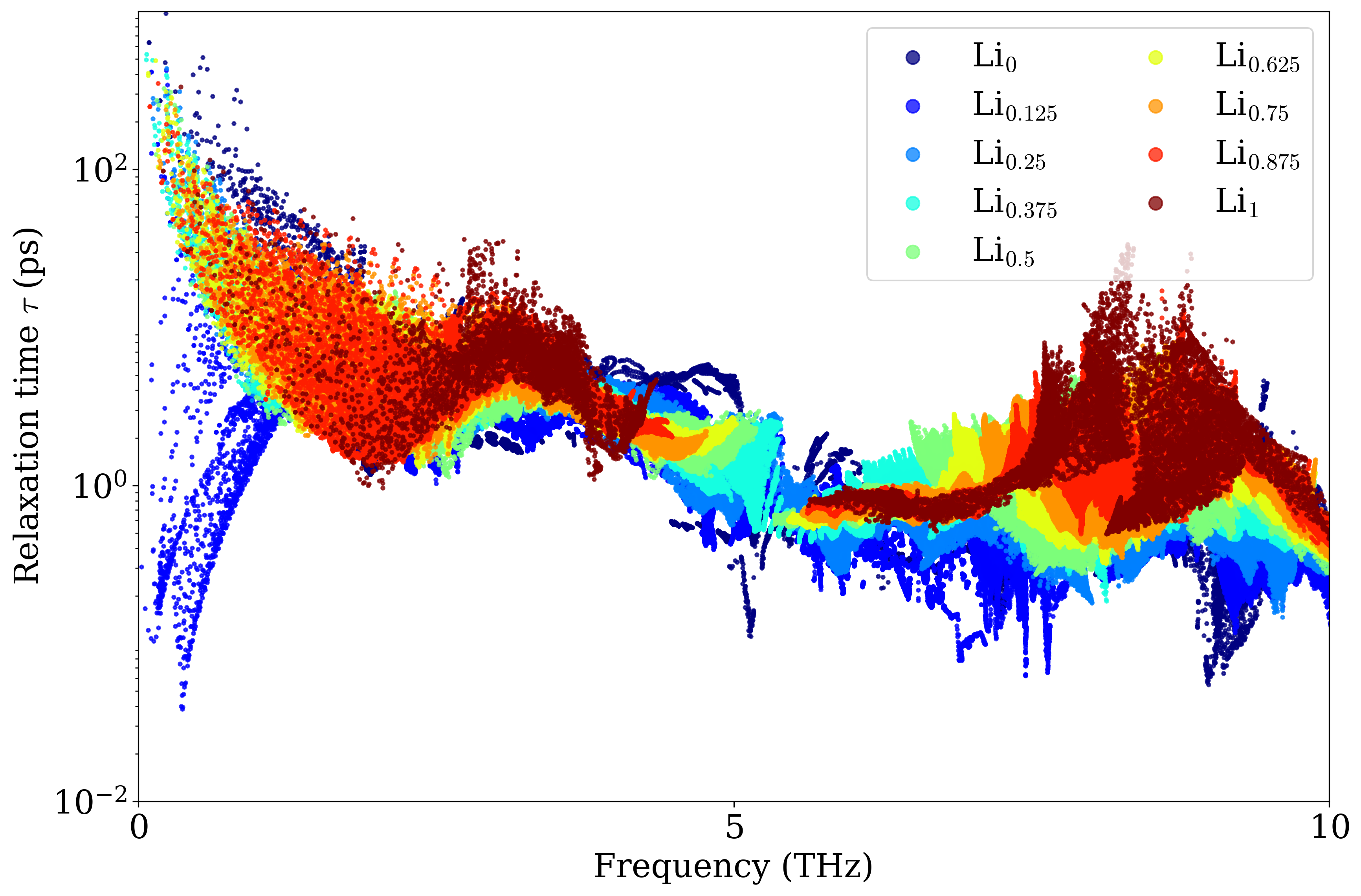}\label{subfig:RelaxationTimesVLiNequIP}}
    \subfloat[]{\includegraphics[width=0.5\linewidth]{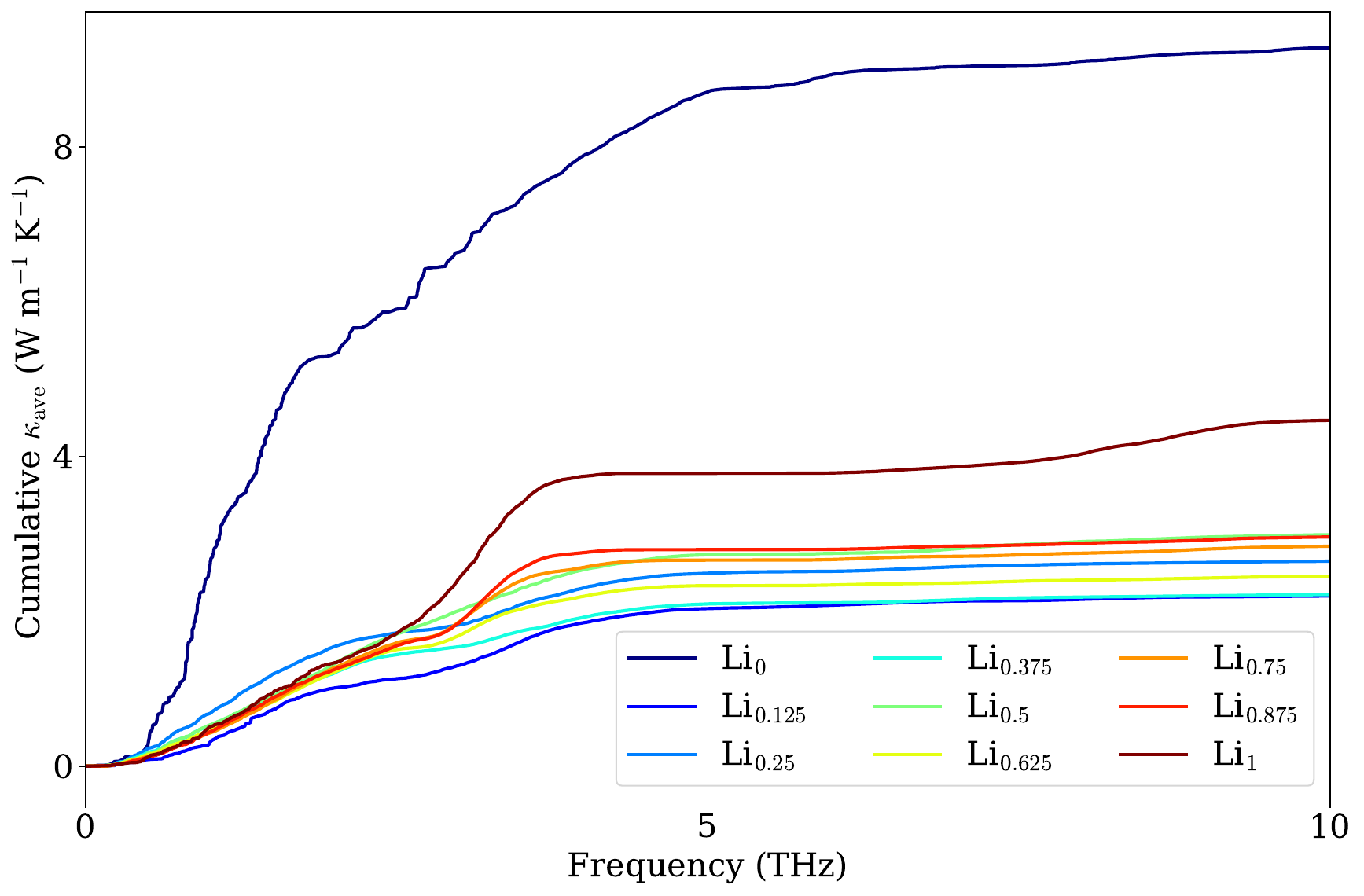}\label{subfig:CumulativeKappaNequIP}}
    \caption{Phonon lifetimes and cumulative thermal conductivity from the uMLiP. (a) RTA relaxation times $\tau$ at 300~K versus frequency for all Li concentrations. (b) Cumulative lattice thermal conductivity versus frequency at 300~K from the same RTA calculations.}
    \label{suppfig:NequIPTransport}
\end{figure*}

\figref{suppfig:NequIPTransport} shows the uMLiP (RTA) phonon relaxation times and cumulative thermal conductivity of $\mathrm{Li}_a\mathrm{ZrS}_2$. The relaxation-time distribution and the frequency-resolved contributions to $\kappa$ qualitatively follow the DFT results discussed in the main text: in every phase, the low-frequency acoustic modes carry the majority of the heat; the spectral midpoint is shifted to higher frequencies in the lithiated phases, and at low lithium content the population of long-lived acoustic modes is reduced.

\clearpage

\section{Direct solution of the Boltzmann transport equation}
\label{suppsec:dlbte}

Direct solutions of the linearized Boltzmann transport equation (DLBTE) for the four DFT compositions are described in the Methods of the main text; this section reports the resulting conductivities and switching metrics. The direct solution requires storing the full scattering matrix and, for the 25- and 28-atom cells, exceeds the available memory for meshes above a $10\times10\times10$ $q$-mesh, so the values reported here are $q$-mesh-limited and are given for completeness: the switching metrics of the main text are taken from the production $25\times25\times25$ RTA calculations.

The size of the mesh effect can be gauged from two comparisons. The same-mesh RTA contained in each DLBTE run differs from the production $25\times25\times25$ RTA by up to a factor of 1.5 at $a=0.125$ ($\kappa_{\mathrm{ave}}=1.31$ versus $0.88$~W~m$^{-1}$~K$^{-1}$). At $a=0$ the $10\times10\times10$ DLBTE result ($\kappa_{\mathrm{ave}}=7.46$~W~m$^{-1}$~K$^{-1}$) lies 22\% below its $15\times15\times15$ counterpart (9.56~W~m$^{-1}$~K$^{-1}$) and 23\% below the production-mesh RTA (9.67~W~m$^{-1}$~K$^{-1}$).

\begin{table}[h!]
\centering
\caption{DLBTE thermal conductivity of Li$_a$ZrS$_2$.
Lattice thermal conductivity at 300~K from the direct solution of the Boltzmann transport equation (DLBTE) on a $10\times10\times10$ $q$-mesh, with additional $15\times15\times15$ results for $a=0$ and $a=1$ listed separately as a convergence check, together with the relaxation-time-approximation (RTA) value obtained on the same mesh within the same run and the production $25\times25\times25$ RTA value used in the main text. All values are in W~m$^{-1}$~K$^{-1}$; $\kappa_{\mathrm{ave}}$ is the average of the three Cartesian components. Mesh labels $10^3$, $15^3$ and $25^3$ denote $10\times10\times10$, $15\times15\times15$ and $25\times25\times25$.}
\begin{tabular}{lcccc}
\hline
$a$ & component & DLBTE & RTA (same mesh) & RTA ($25^3$) \\
\hline
0 & $\kappa_{\parallel}$ & 9.22 & 8.93 & 12.64 \\
0 & $\kappa_{\perp}$ & 3.93 & 4.02 & 3.73 \\
0 & $\kappa_{\mathrm{ave}}$ & 7.46 & 7.30 & 9.67 \\
0 ($15^3$) & $\kappa_{\mathrm{ave}}$ & 9.56 & 9.25 & 9.67 \\
0.125 & $\kappa_{\parallel}$ & 3.26 & 1.93 & 1.30 \\
0.125 & $\kappa_{\perp}$ & 0.065 & 0.067 & 0.046 \\
0.125 & $\kappa_{\mathrm{ave}}$ & 2.20 & 1.31 & 0.88 \\
0.5 & $\kappa_{\parallel}$/$\kappa_{yy}$ & 2.73/3.54 & 2.44/3.04 & 2.48/3.11 \\
0.5 & $\kappa_{\perp}$ & 0.67 & 0.65 & 0.63 \\
0.5 & $\kappa_{\mathrm{ave}}$ & 2.31 & 2.04 & 2.07 \\
1 & $\kappa_{\parallel}$ & 3.17 & 2.78 & 2.67 \\
1 & $\kappa_{\perp}$ & 1.33 & 1.33 & 1.08 \\
1 & $\kappa_{\mathrm{ave}}$ & 2.56 & 2.30 & 2.14 \\
1 ($15^3$) & $\kappa_{\mathrm{ave}}$ & 2.47 & 2.20 & 2.14 \\
\hline
\end{tabular}
\label{supptab:DLBTEkappa}
\end{table}

The resulting DLBTE switching metrics at $10\times10\times10$ are a cross-plane ratio of 60 (score 0.983), an in-plane ratio of 2.8 (0.646), and an orientation-averaged ratio of 3.4 (0.705), to be compared with 81 (0.988), 9.8 (0.897), and 11 (0.909) from the production-mesh RTA. The DLBTE values are shown alongside the RTA and uMLiP results in the figures below: the principal $\kappa$ components (\figref{suppfig:ComponentsDLBTE}), the switch score and on/off ratio versus temperature with the DLBTE values at 300~K overlaid (\figref{suppfig:ScoreTDLBTE}), the cumulative thermal conductivity with the same-mesh and production-mesh RTA and the DLBTE totals (\figref{suppfig:CumulativeDLBTE}), the convergence of the orientation-averaged conductivity with method and mesh (\figref{suppfig:DLBTEConvergence}), and the on/off conductivity and switch-score comparisons with the literature switches (\figref{suppfig:KappaComparisonDLBTE} and \figref{suppfig:ScoreComparisonDLBTE}).

\begin{figure*}[h!]
    \centering
    \subfloat{\includegraphics[width=0.75\linewidth]{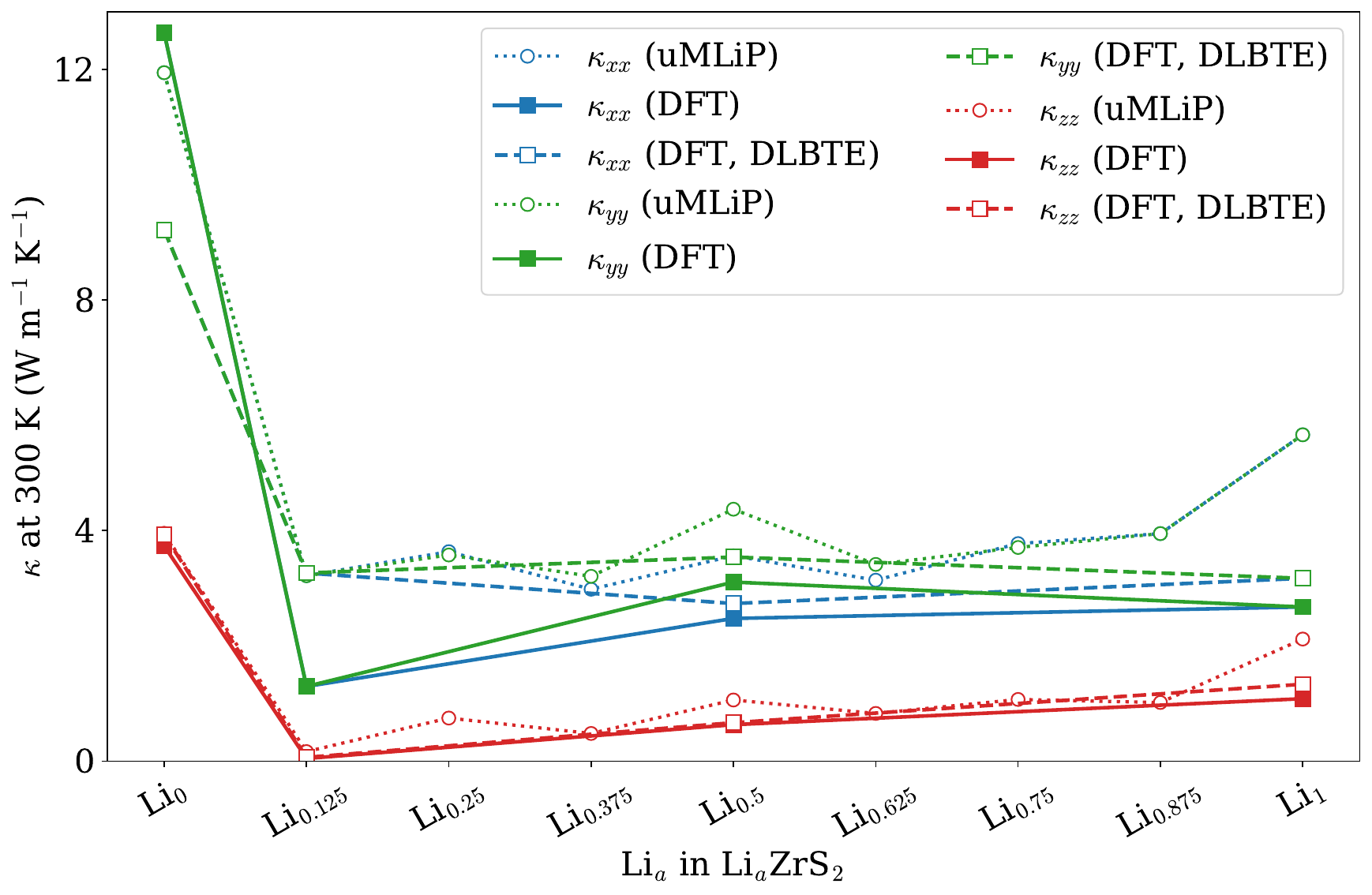}}
    \caption{Thermal conductivity components including the DLBTE. Principal Cartesian components of the lattice thermal conductivity at 300~K versus intercalation, from the DFT RTA, the DFT DLBTE ($10\times10\times10$) and the uMLiP RTA.}
    \label{suppfig:ComponentsDLBTE}
\end{figure*}

\begin{figure*}[h!]
    \centering
    \subfloat{\includegraphics[width=0.75\linewidth]{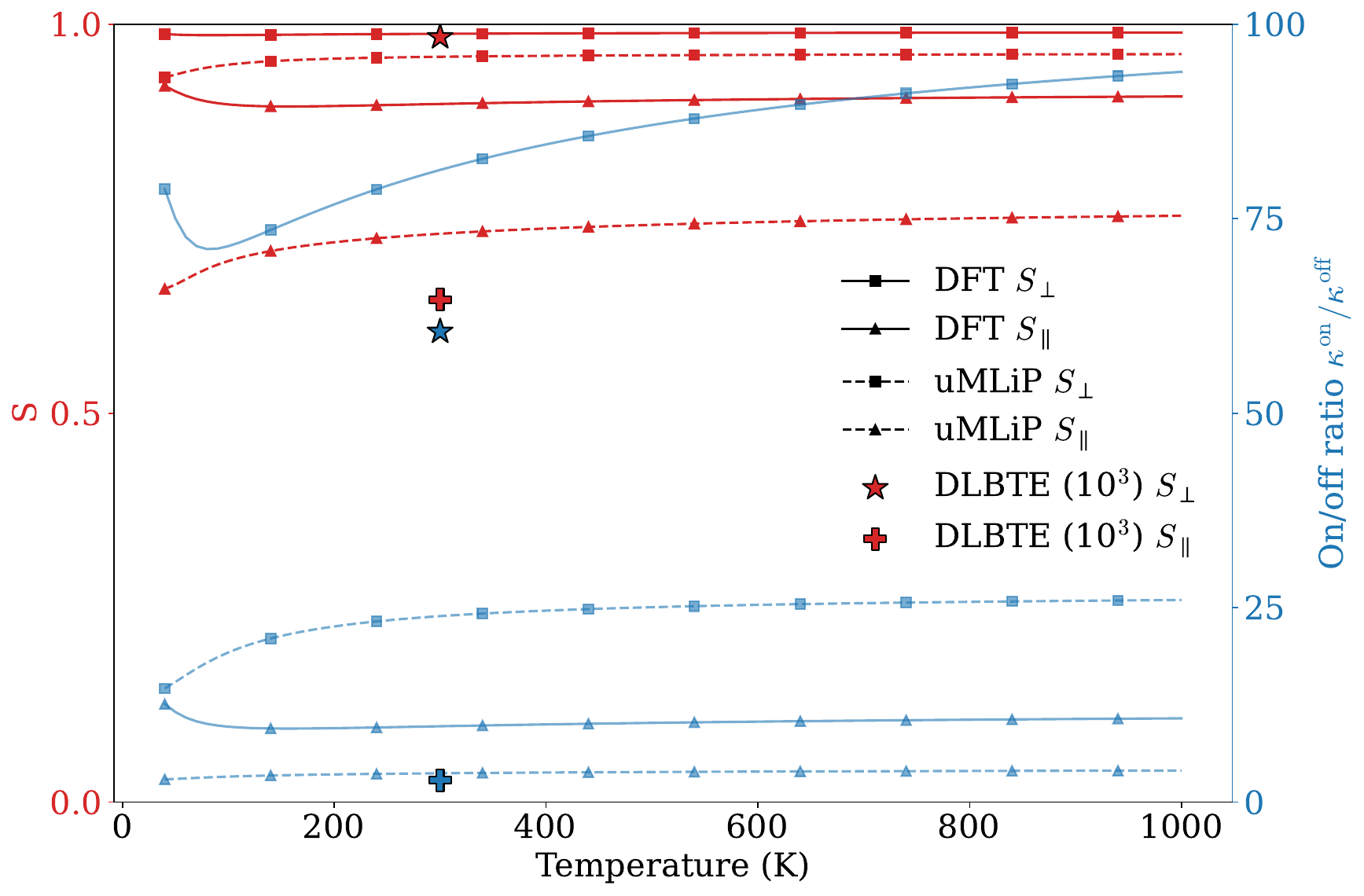}}
    \caption{Temperature dependence of the switch score including the DLBTE. Switch score (left axis) and on/off ratio (right axis) versus temperature for the cross-plane (solid lines) and in-plane (dashed lines) channels from DFT and from the uMLiP (RTA), with the DLBTE ($10\times10\times10$) values at 300~K overlaid as markers.}
    \label{suppfig:ScoreTDLBTE}
\end{figure*}

\begin{figure*}[h!]
    \centering
    \subfloat{\includegraphics[width=0.75\linewidth]{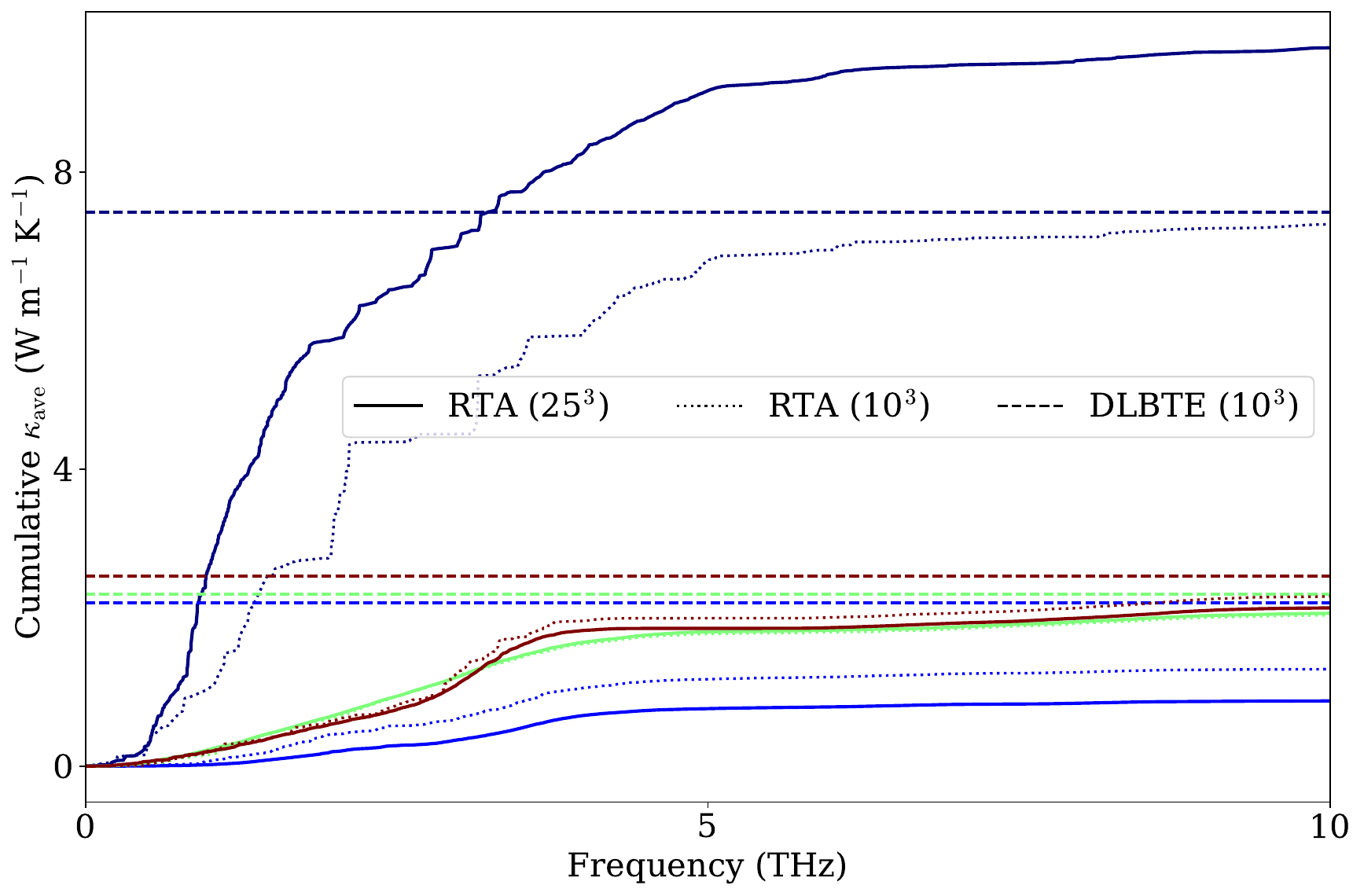}}
    \caption{Cumulative thermal conductivity including the DLBTE. Cumulative orientation-averaged thermal conductivity versus frequency for $a=0$, $0.125$, $0.5$ and $1$, from the production $25\times25\times25$ RTA (solid lines), the same-mesh $10\times10\times10$ RTA (dotted lines) and the DLBTE total (dashed horizontal lines).}
    \label{suppfig:CumulativeDLBTE}
\end{figure*}

\begin{figure*}[h!]
    \centering
    \subfloat{\includegraphics[width=0.75\linewidth]{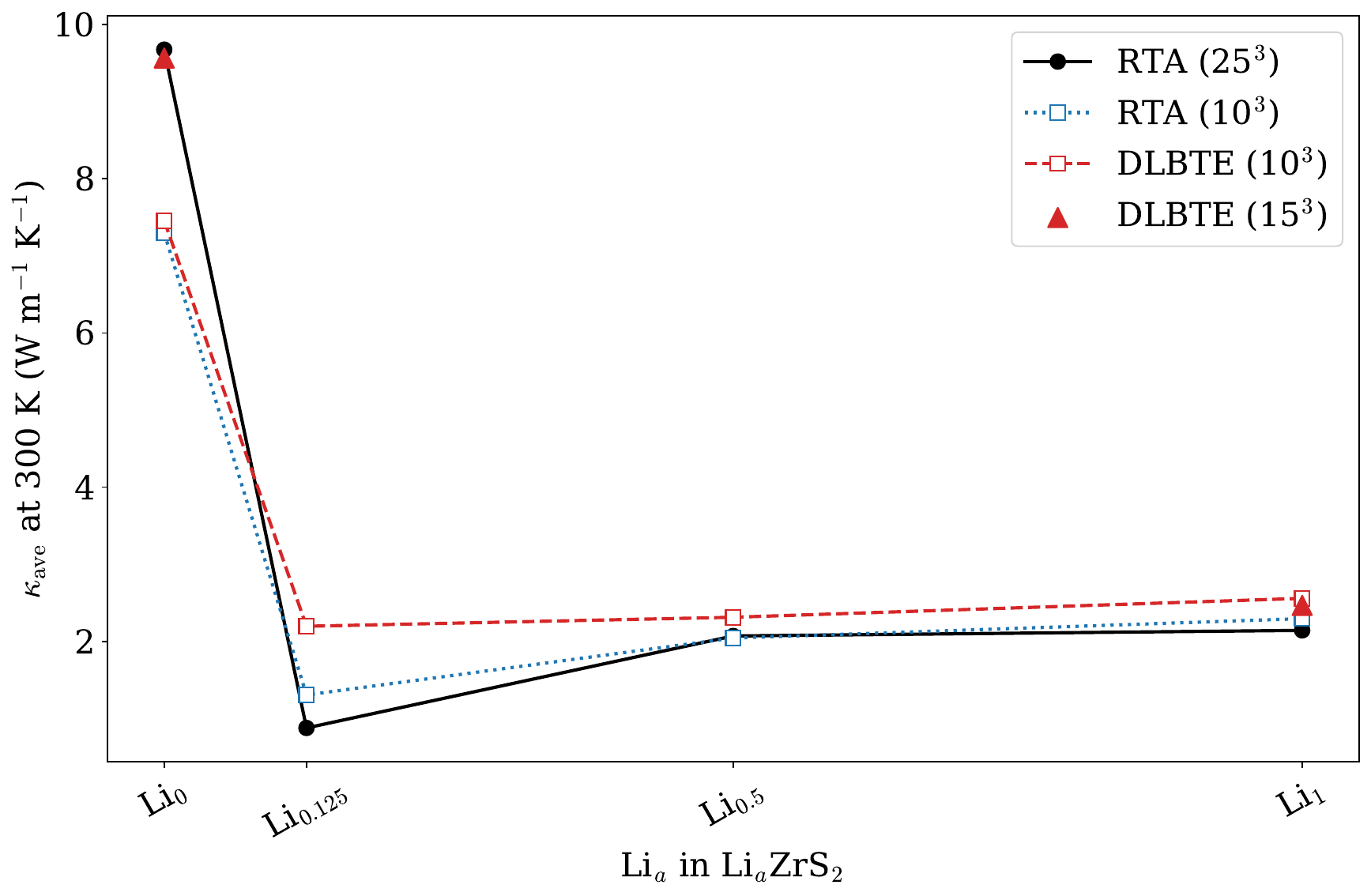}}
    \caption{$q$-mesh convergence of the orientation-averaged conductivity. Orientation-averaged lattice thermal conductivity at 300~K versus intercalation from the production-mesh RTA ($25^3$), the same-mesh RTA ($10^3$) and the DLBTE ($10^3$), with the DLBTE $15^3$ values for $a=0$ and $a=1$.}
    \label{suppfig:DLBTEConvergence}
\end{figure*}

\begin{figure*}[h!]
    \centering
    \subfloat{\includegraphics[width=0.75\linewidth]{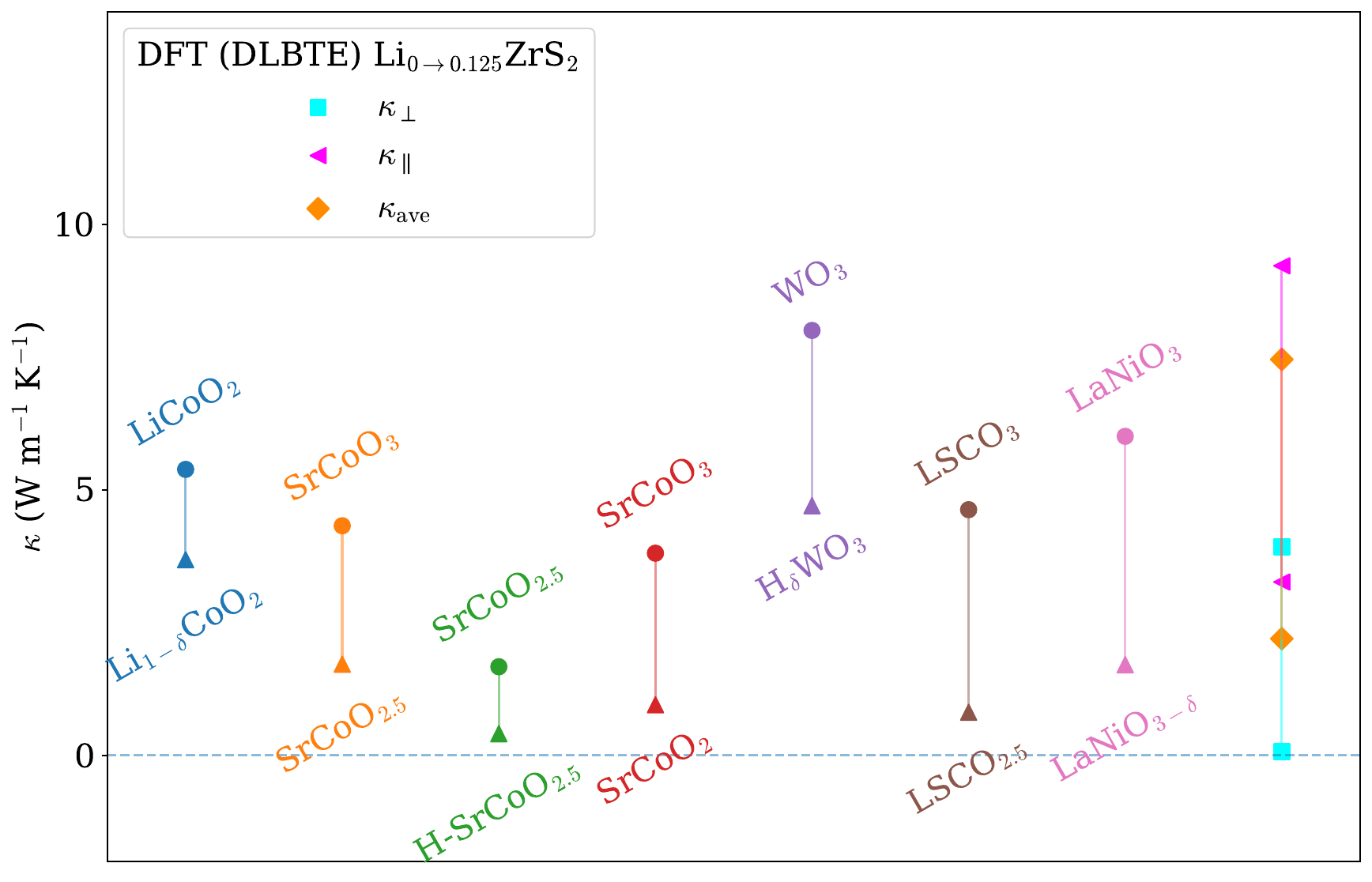}}
    \caption{On/off-state conductivities compared with other electrochemical switches (DLBTE). On- and off-state thermal conductivity of ZrS$_2$ from the DLBTE ($10\times10\times10$; cross-plane, in-plane and orientation-averaged) compared with other solid-state electrochemical thermal-switch materials.}
    \label{suppfig:KappaComparisonDLBTE}
\end{figure*}

\begin{figure*}[h!]
    \centering
    \subfloat{\includegraphics[width=0.75\linewidth]{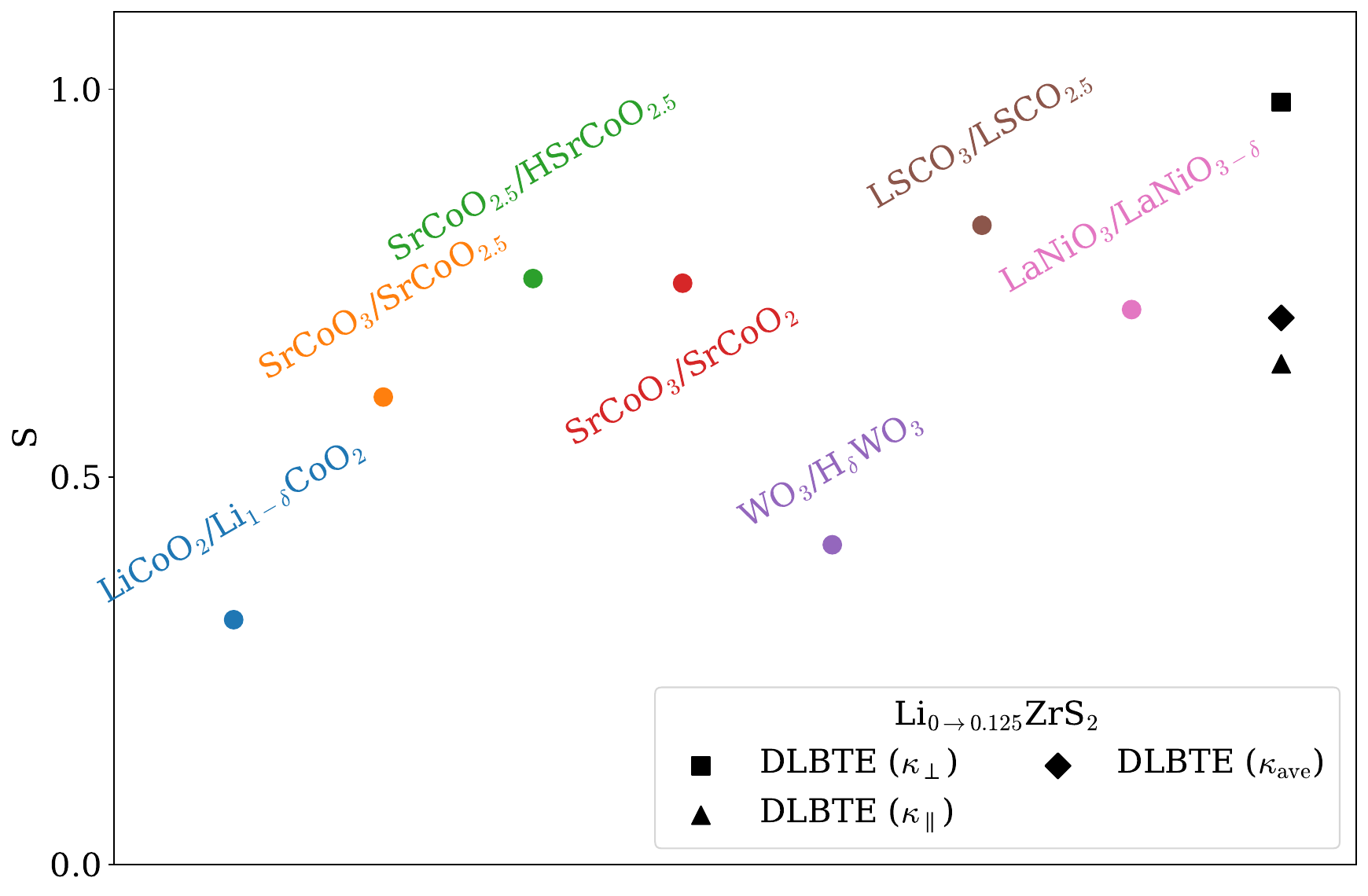}}
    \caption{Switch scores compared with other electrochemical switches (DLBTE). Switch score of Li$_a$ZrS$_2$ from the DLBTE ($10\times10\times10$; cross-plane, in-plane and orientation-averaged) compared with other electrochemical thermal-switch materials.}
    \label{suppfig:ScoreComparisonDLBTE}
\end{figure*}

\begin{table}[h!]
\centering
\caption{Waste-heat recovery for the DLBTE switch.}
Recovery after 20 pulses for the DLBTE ($10\times10\times10$) switch, using the same device model and pulse schedule as \tabref{supptab:WHRresults}. $\eta$ is the capture efficiency, $\eta_{\mathrm{rel}}$ its ratio to the ideal switch, backflow the off-state leakage as a fraction of the pulse energy and duty the on-time fraction of the gate; the ideal-switch benchmark uses the DFT RTA cross-plane on-state conductivity, $\kappa^{\mathrm{on}}_{\perp}$. Thermal conductivities are in W~m$^{-1}$~K$^{-1}$.
\begin{tabular}{lcccccc}
\hline
System & $\kappa^{\mathrm{on}}$ & $\kappa^{\mathrm{off}}$ & $\eta$ (\%) & $\eta_{\mathrm{rel}}$ & Backflow (\%) & Duty (\%) \\
\hline
ZrS$_2$/Li$_{0.125}$ZrS$_2$ (DLBTE, $\kappa_{\perp}$) & 3.93 & 0.06 & 5.40 & 0.82 & 1.9 & 8.1 \\
ZrS$_2$/Li$_{0.125}$ZrS$_2$ (DLBTE, $\kappa_{\text{ave}}$) & 7.46 & 2.20 & 1.68 & 0.26 & 13.9 & 10.4 \\
ZrS$_2$/Li$_{0.125}$ZrS$_2$ (DLBTE, $\kappa_{\parallel}$) & 9.22 & 3.26 & 1.51 & 0.23 & 15.1 & 9.3 \\
\hline
\end{tabular}
\label{supptab:DLBTEWHR}
\end{table}

\begin{figure*}[h!]
    \centering
    \subfloat{\includegraphics[width=0.75\linewidth]{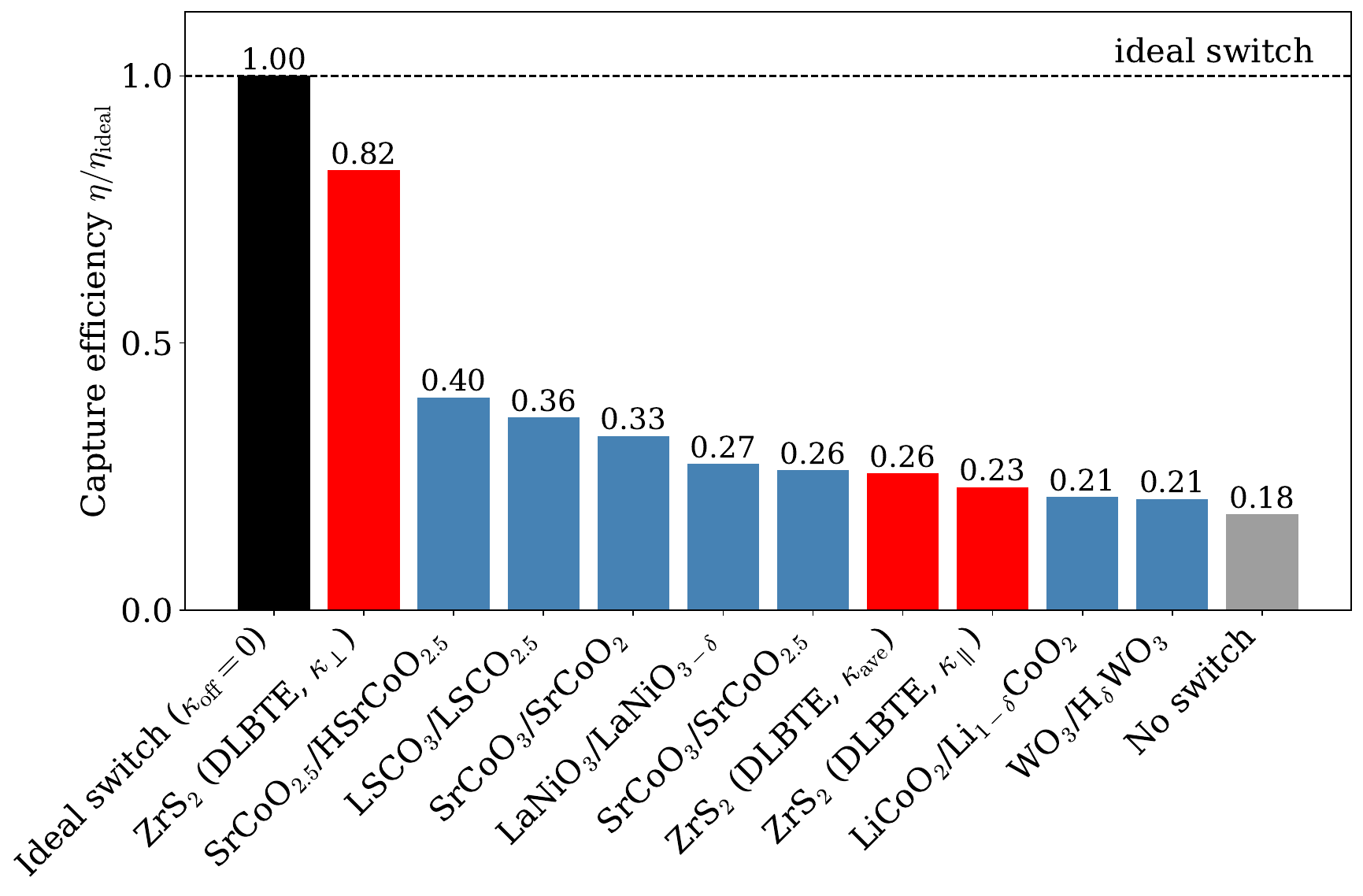}}
    \caption{Capture efficiency of the DLBTE switch. Capture efficiency relative to the ideal switch for the DLBTE ($10\times10\times10$) switch of this work, the literature switches and the reference systems.}
    \label{suppfig:WHREfficiencyDLBTE}
\end{figure*}

\begin{figure*}[h!]
    \centering
    \subfloat{\includegraphics[width=0.75\linewidth]{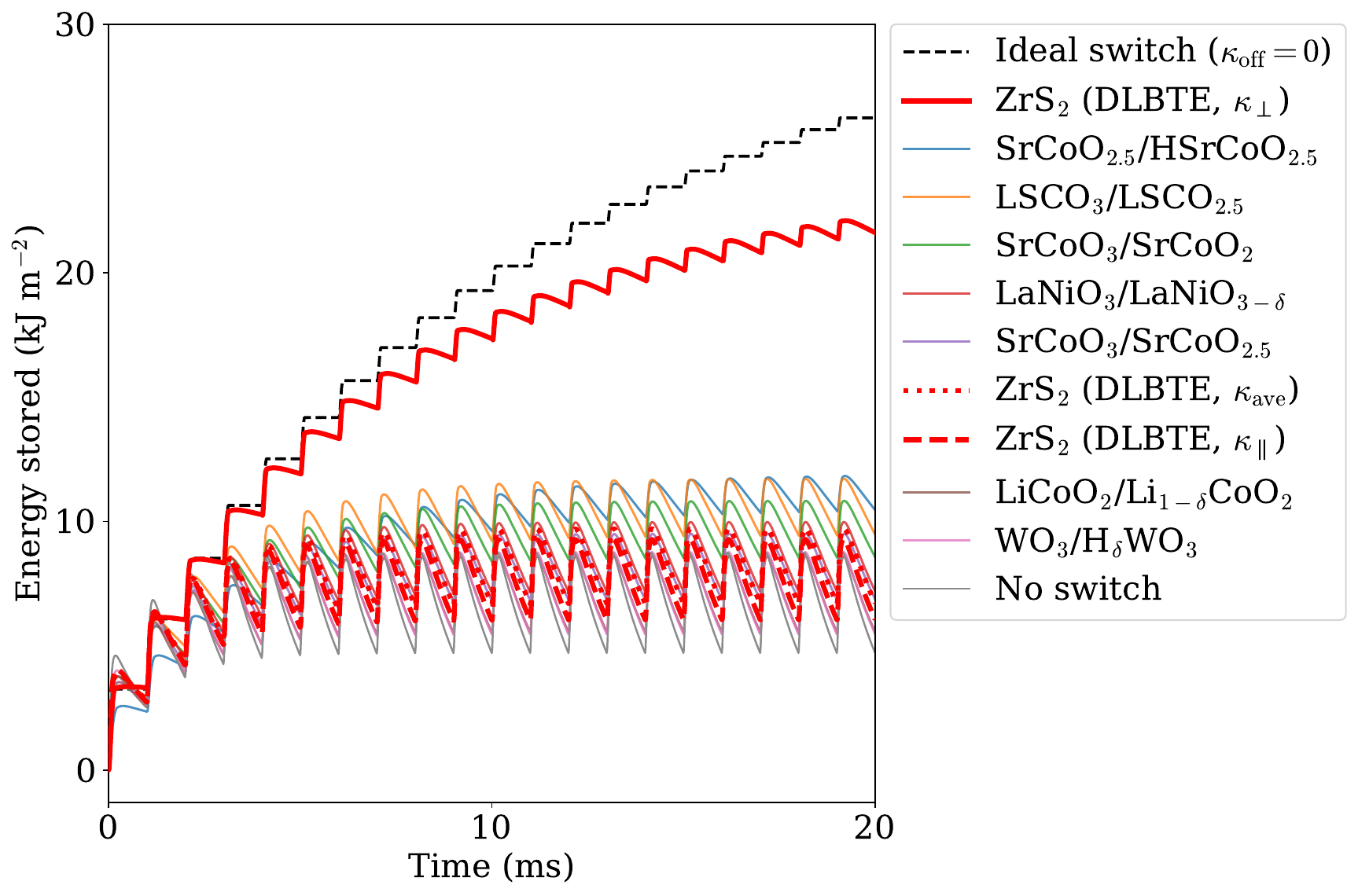}}
    \caption{Storage evolution of the DLBTE switch. Energy stored in the insulated storage over 20 heat pulses for the same systems as \figref{suppfig:WHREfficiencyDLBTE}.}
    \label{suppfig:WHREvolutionDLBTE}
\end{figure*}

\begin{figure*}[h!]
    \centering
    \subfloat{\includegraphics[width=0.8\linewidth]{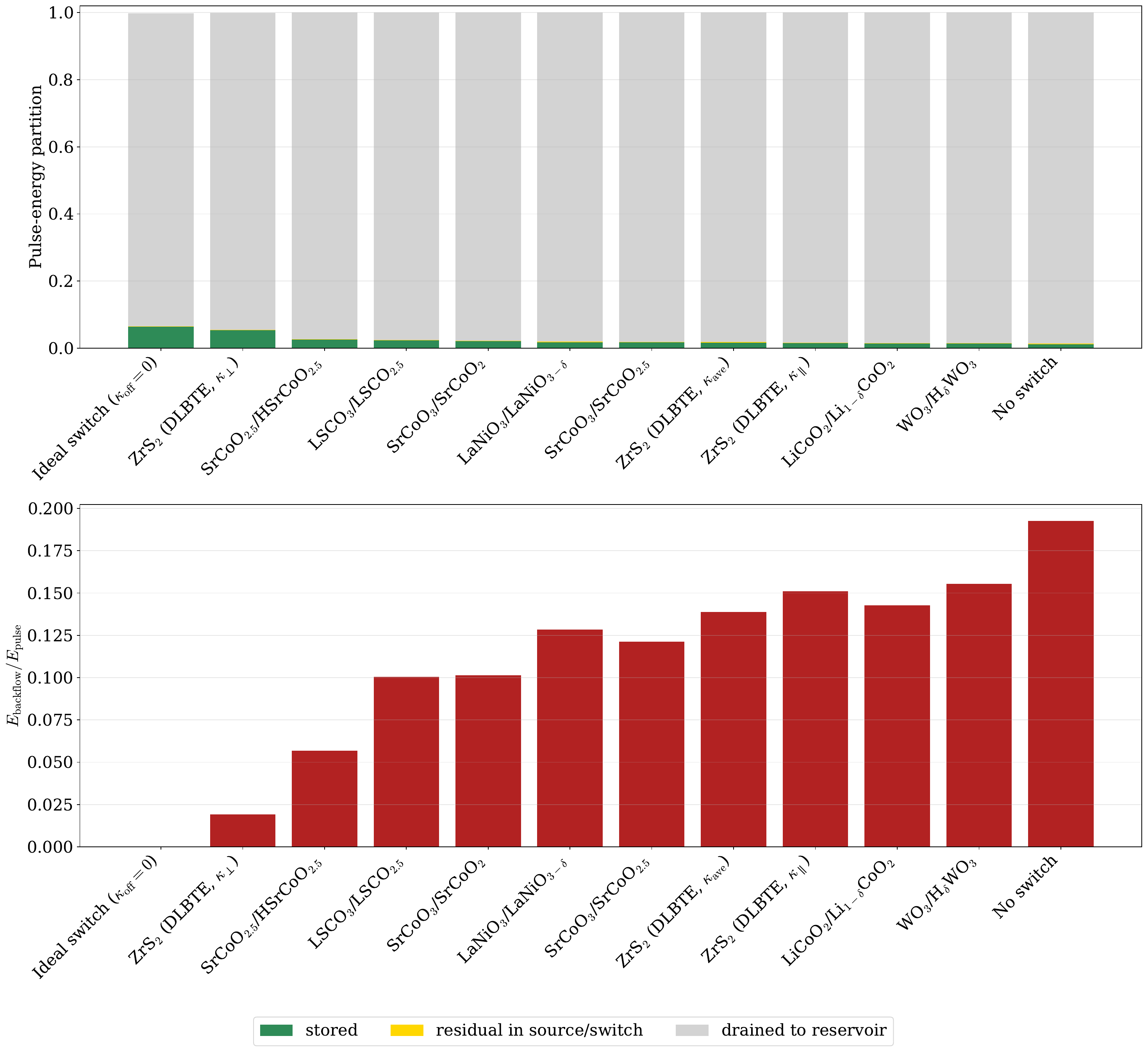}}
    \caption{Energy partition of the DLBTE switch. Pulse-energy partition (left) and off-state backflow leakage (right) after 20 pulses for the DLBTE switch of this work, the literature switches~\cite{Cho2014,Lu2020,Yang2023,Ning2019,Zhang2023,Bian2024} and the reference systems.}
    \label{suppfig:WHRAuditDLBTE}
\end{figure*}

\begin{figure*}[h!]
    \centering
    \subfloat{\includegraphics[width=0.8\linewidth]{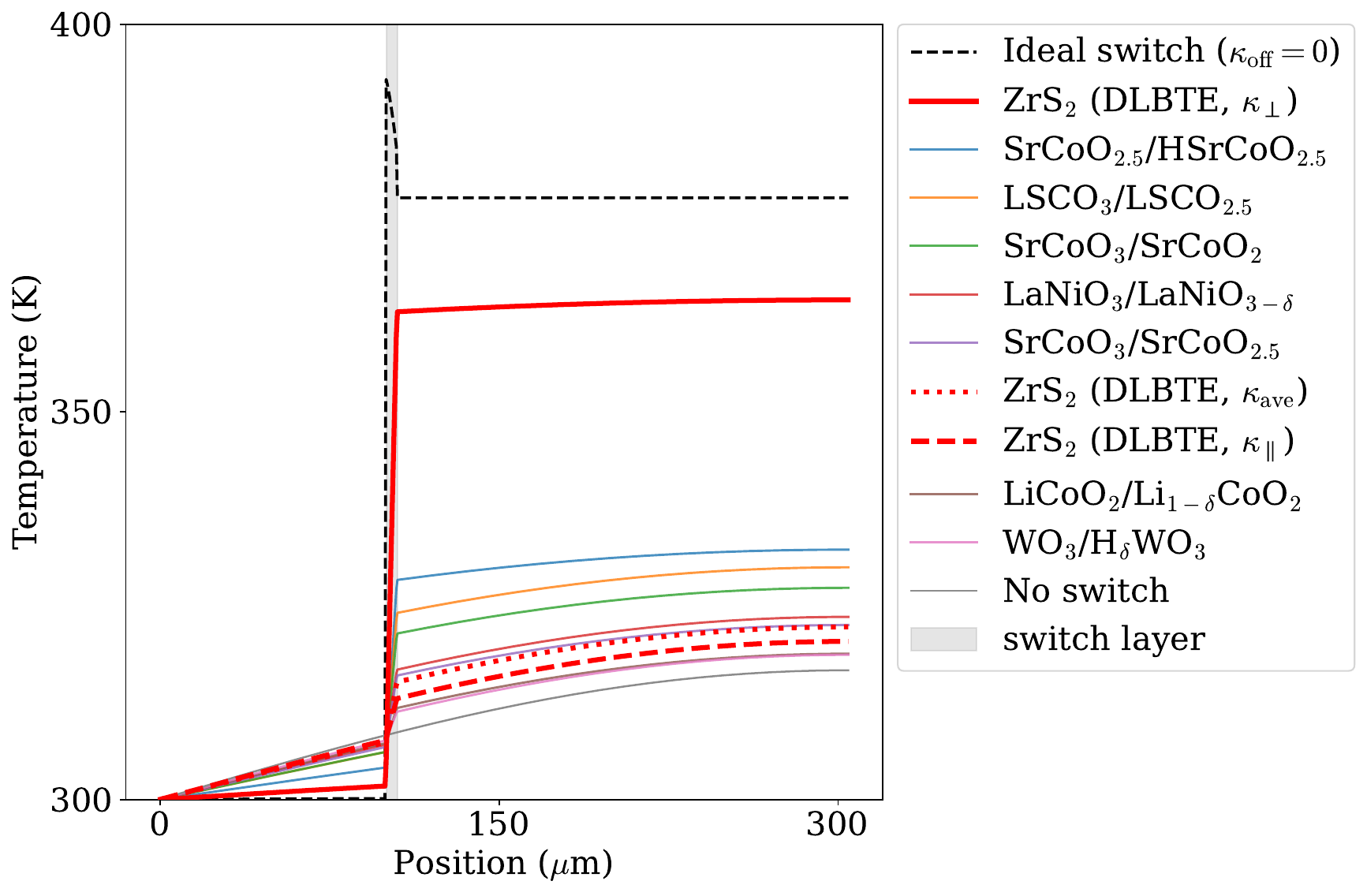}}
    \caption{Temperature profiles of the DLBTE switch. Temperature profiles after 20 heat pulses for the DLBTE switch of this work, the literature switches~\cite{Cho2014,Lu2020,Yang2023,Ning2019,Zhang2023,Bian2024} and the reference systems.}
    \label{suppfig:WHRProfilesDLBTE}
\end{figure*}

\clearpage

\section{Waste-heat-recovery results}
\label{suppsec:WHRresults}

For our continuum level simulations, the device model, its discretization and all of its parameters are given in the Methods of the main text. This section collects the complete numerical results for every switch material simulated there, using the identical geometry, pulse schedule and volumetric heat capacity. \tabref{supptab:WHRresults} lists, for each system, the on- and off-state conductivities of the switch layer, the capture efficiency $\eta$, its ratio $\eta_{\mathrm{rel}}$ to the ideal switch ($\kappa^{\mathrm{off}}=0$), the off-state backflow as a fraction of the injected pulse energy, and the on-time fraction (duty) of the gate after 20 pulses.

\begin{table}[h!]
\centering
\caption{Waste-heat recovery for every switch material.}
Recovery after 20 pulses for identical devices that differ only in the switch material. $\eta$ is the capture efficiency, $\eta_{\mathrm{rel}}$ its ratio to the ideal switch ($\kappa^{\mathrm{off}}=0$), backflow the off-state leakage as a fraction of the pulse energy, and duty the on-time fraction of the gate; the total injected energy is $4.00\times10^{5}$~J~m$^{-2}$ for every system. On- and off-state thermal conductivities are in W~m$^{-1}$~K$^{-1}$; the ``no switch'' row is an unswitched bar of the host conductor.
\begin{tabular}{lcccccc}
\hline
System & $\kappa^{\mathrm{on}}$ & $\kappa^{\mathrm{off}}$ & $\eta$ (\%) & $\eta_{\mathrm{rel}}$ & Backflow (\%) & Duty (\%) \\
\hline
Ideal switch ($\kappa^{\mathrm{off}}=0$) & 3.73 & 0.00 & 6.56 & 1.00 & 0.0 & 7.6 \\
ZrS$_2$/Li$_{0.125}$ZrS$_2$ (DFT, $\perp$) & 3.73 & 0.05 & 5.65 & 0.86 & 1.3 & 8.0 \\
ZrS$_2$/Li$_{0.125}$ZrS$_2$ (uMLiP, $\perp$) & 3.95 & 0.17 & 4.35 & 0.66 & 4.2 & 9.1 \\
SrCoO$_{2.5}$/HSrCoO$_{2.5}$ & 1.67 & 0.41 & 2.61 & 0.40 & 5.7 & 14.2 \\
ZrS$_2$/Li$_{0.125}$ZrS$_2$ (DFT, average) & 9.67 & 0.88 & 2.46 & 0.38 & 11.4 & 8.4 \\
LSCO$_3$/LSCO$_{2.5}$ & 4.63 & 0.81 & 2.36 & 0.36 & 10.0 & 10.5 \\
SrCoO$_3$/SrCoO$_2$ & 3.81 & 0.95 & 2.14 & 0.33 & 10.1 & 11.5 \\
ZrS$_2$/Li$_{0.125}$ZrS$_2$ (DFT, $\parallel$) & 12.64 & 1.30 & 2.13 & 0.32 & 13.1 & 7.8 \\
LaNiO$_3$/LaNiO$_{3-\delta}$ & 6.01 & 1.71 & 1.80 & 0.27 & 12.8 & 10.1 \\
SrCoO$_3$/SrCoO$_{2.5}$ & 4.32 & 1.72 & 1.72 & 0.26 & 12.1 & 12.0 \\
ZrS$_2$/Li$_{0.125}$ZrS$_2$ (uMLiP, average) & 9.28 & 2.20 & 1.71 & 0.26 & 14.2 & 8.6 \\
ZrS$_2$/Li$_{0.125}$ZrS$_2$ (uMLiP, $\parallel$) & 11.94 & 3.22 & 1.54 & 0.24 & 15.4 & 8.2 \\
LiCoO$_2$/Li$_{1-\delta}$CoO$_2$ & 5.38 & 3.68 & 1.39 & 0.21 & 14.3 & 11.1 \\
WO$_3$/H$_\delta$WO$_3$ & 8.00 & 4.70 & 1.37 & 0.21 & 15.5 & 9.9 \\
No switch (uniform host) & 50.0 & 50.0 & 1.18 & 0.18 & 19.3 & 0.0 \\
\hline
\end{tabular}
\label{supptab:WHRresults}
\end{table}

\begin{figure*}[h!]
    \centering
    \subfloat{\includegraphics[width=0.95\linewidth]{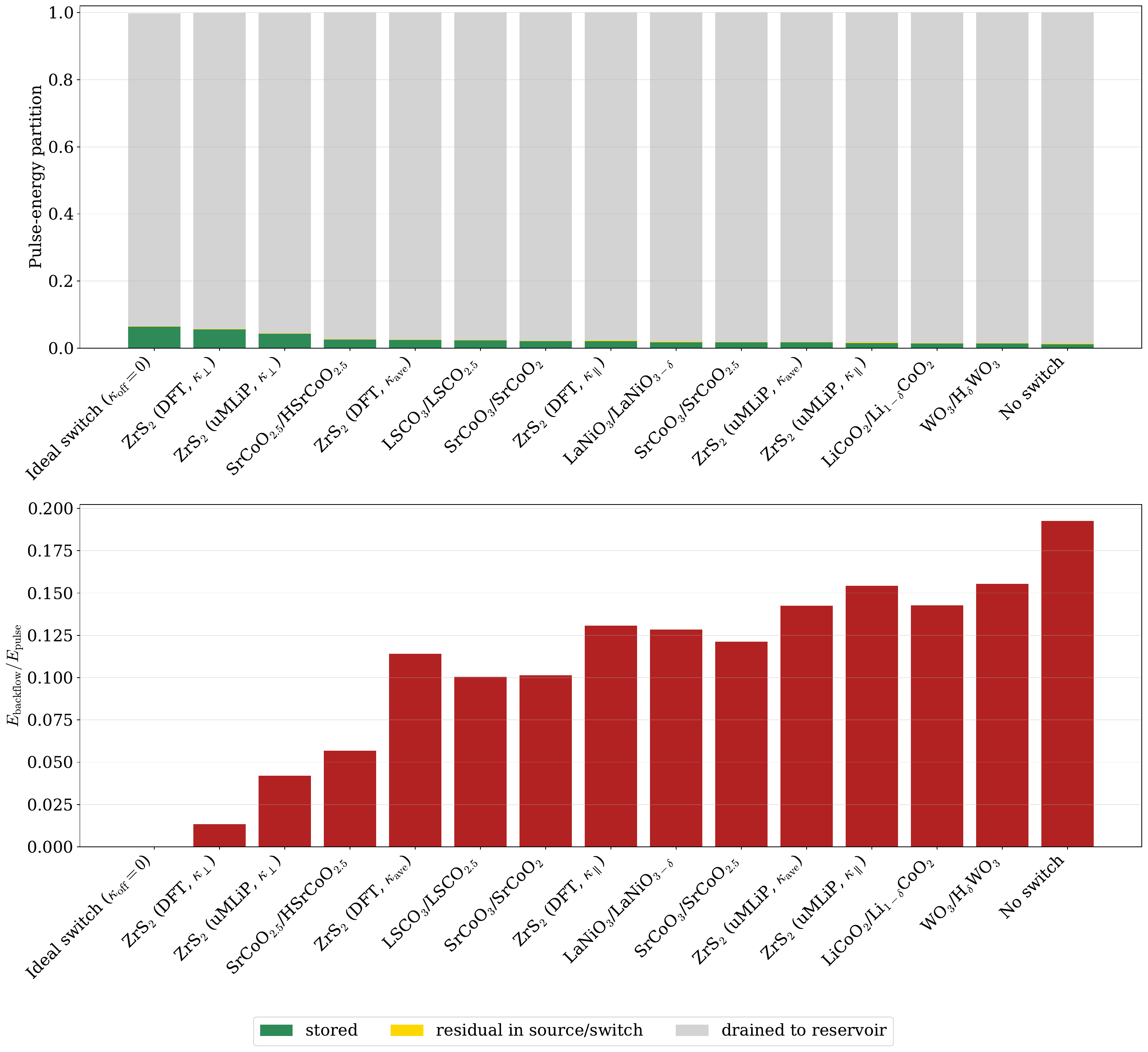}}
    \caption{Energy partition of the waste-heat-recovery cycle. Pulse-energy partition (left) and off-state backflow leakage (right) after 20 pulses, for the DFT and uMLiP switches of this work, the literature switches and the reference systems, all on identical geometry and an identical pulse schedule.}
    \label{suppfig:WHRAudit}
\end{figure*}

\begin{figure*}[h!]
    \centering
    \subfloat{\includegraphics[width=0.95\linewidth]{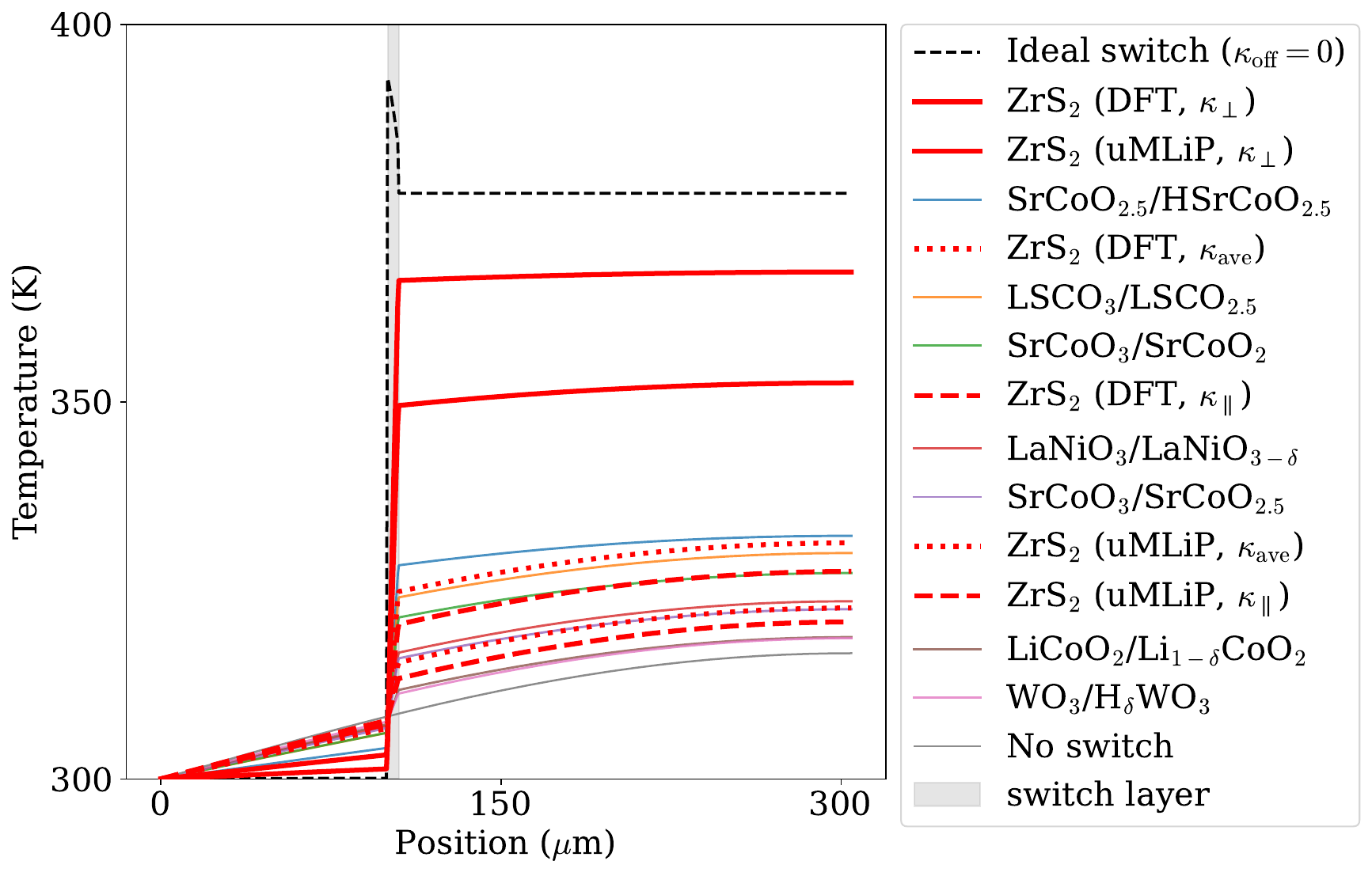}}
    \caption{Temperature profiles of the waste-heat-recovery device. Temperature profiles after 20 heat pulses for the DFT and uMLiP switches of this work, the literature switches and the reference systems. The shaded band marks the switch layer and the dotted horizontal line the reservoir temperature.}
    \label{suppfig:WHRProfiles}
\end{figure*}

\begin{figure*}
    \centering
     \subfloat[]{\includegraphics[width=0.5\linewidth]{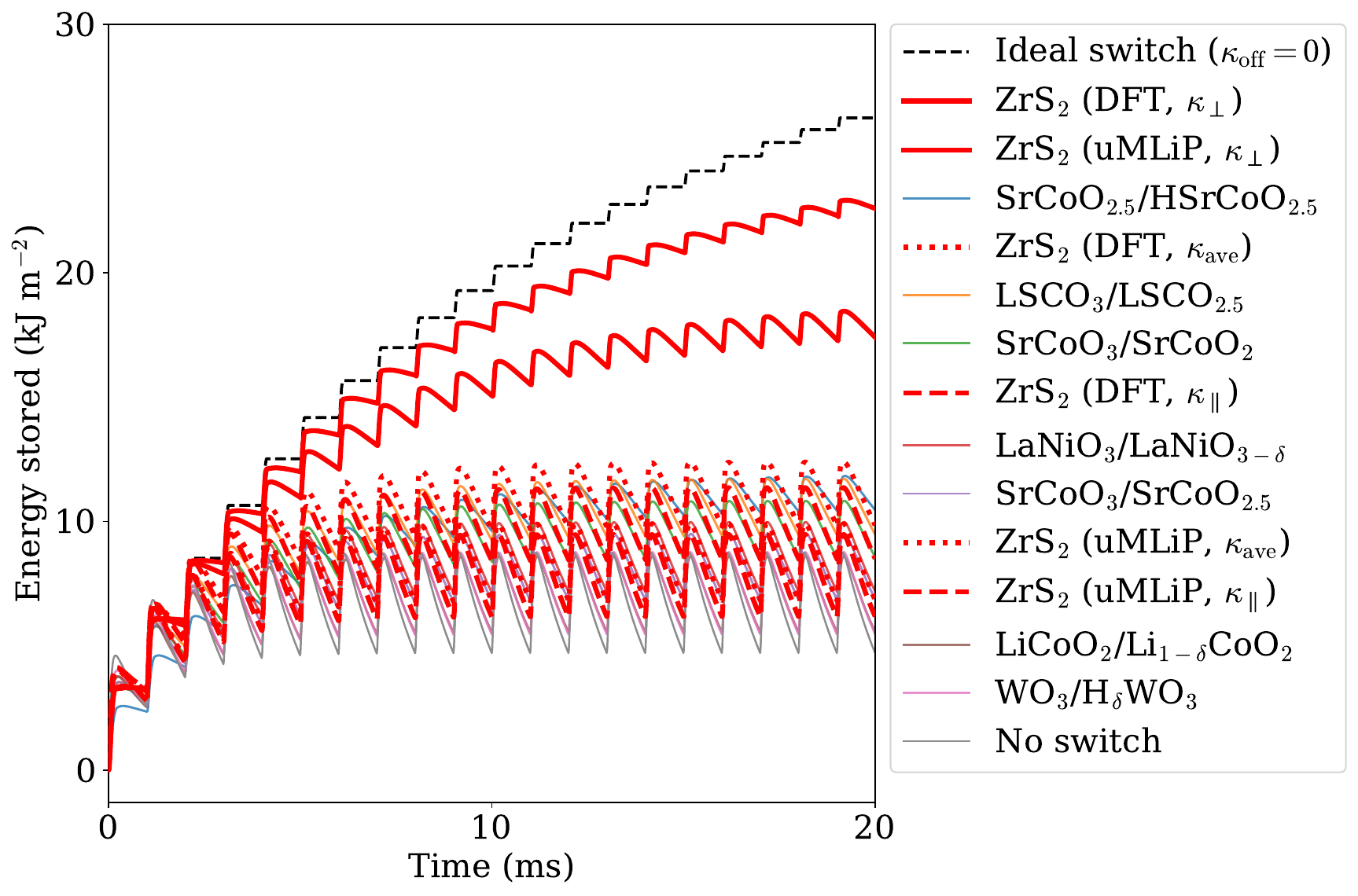}}
    \caption{Energy stored in the sink over 20 pulses: low-contrast switches cycle their storage, whereas the cross-plane ZrS$_2$ switch retains each capture.}
    \label{suppfig:WHR_stored}
\end{figure*}

\figref{suppfig:WHRAudit} shows the pulse-energy partition for the DFT and uMLiP switches of this work, the literature switches and the reference systems: the fraction stored in the sink, the residual heat still in transit in the source and switch layer, the heat drained to the reservoir, and the off-state backflow. \figref{suppfig:WHRProfiles} shows the final temperature profiles after 20 pulses for the same systems. The sink temperature reached is ordered by the on/off ratio rather than by the on-state conductivity: 378~K for the ideal switch, 367~K for the cross-plane DFT ZrS$_2$ switch and 352~K for the cross-plane uMLiP ZrS$_2$ switch, falling to 314--331~K for the literature switches, the in-plane and orientation-averaged ZrS$_2$ switches and the unswitched host. The large cross-plane contrast of ZrS$_2$, rather than the absolute magnitude of its thermal conductivity, is therefore what maximises the recovered energy.

\end{document}